\documentclass[12pt]{article}
\usepackage[dvips]{epsfig}
\usepackage{amsfonts}
\usepackage{graphicx}
\usepackage{amsmath}
\long\def\commabs #1\commabsend{#1}
\long\def\commful #1\commfulend{}
\long\def\commEREZDEL #1\commEREZDELend{}

\commabs
\commabsend

\newtheorem{theorem}{Theorem}[section]
\newtheorem{lemma}[theorem]{Lemma}
\newtheorem{observation}[theorem]{Observation}
\newtheorem{corollary}[theorem]{Corollary}

\newtheorem{claim}[theorem]{Claim}
\newtheorem{fact}[theorem]{Fact}

\newcommand{\R}{\mathbb R}
\def\deg{\mbox{\tt deg}}

\def\cC{{\cal C}}

\commful
\def\dnsparagraph{\vspace{-10pt}\paragraph}
\commfulend
\commabs
\def\dnsparagraph{\paragraph}
\commabsend
\def\inline#1:{\par\vskip 7pt\noindent{\bf #1:}\hskip 10pt}
\def\Proof{\par\noindent{\bf Proof:~}}
\def\blackslug{\hbox{\hskip 1pt \vrule width 4pt height 8pt
    depth 1.5pt \hskip 1pt}}
\def\QED{\quad\blackslug\lower 8.5pt\null\par}

\newcommand{\PowerRatio}[0]{\eta}
\def\SturmCell{\mbox{\sf SturmCell}}

\def\SturmCellB{\mbox{\sf SturmCellB}}

\def\TagCell{\mbox{\sf TagCell}}
\def\SegTest{\mbox{\sf SegTest}}
\def\cJ{{\cal J}}
\def\dhat{{\hat d}}
\def\betahat{{\hat\beta}}

\newcommand{\Station}[0]{\mathit{s}}
\newcommand{\ZStation}[0]{\mathit{s}}
\newcommand{\ZStationSet}[0]{\mathit{S}}
\newcommand{\ZStationC}[0]{\mathit{s}_{0}}
\newcommand{\cA}{\mathcal{A}}
\newcommand{\Power}[0]{\psi}
\newcommand{\MaxPower}[0]{\Psi}
\newcommand{\Noise}[0]{\mathit{N}}
\newcommand{\Reals}[0]{\mathbb{R}}
\newcommand{\Integers}[0]{\mathbb{Z}}
\newcommand{\Grid}[0]{\mathit{G}}
\newcommand{\Area}[0]{\mathrm{area}}
\newcommand{\Perimeter}[0]{\mathrm{per}}
\newcommand{\SmallRadius}[0]{\delta}
\newcommand{\LargeRadius}[0]{\Delta}
\newcommand{\Boundary}[0]{\Phi}
\newcommand{\Circumference}[0]{L}
\newcommand{\DataStructure}[0]{\ensuremath{\mathtt{DS}}}
\newcommand{\QuerDS}[0]{\ensuremath{\mathtt{QDS}}}

\newcommand{\Energy}[0]{\mathrm{E}}
\newcommand{\GridSpace}[0]{\gamma}
\newcommand{\ReceptionZone}[0]{\mathcal{H}}
\newcommand{\NZones}[0]{\mathcal{\tau}}
\newcommand{\Segment}[2]{\overline{{#1} \, {#2}}}
\newcommand{\arc}[2]{\widehat{#1 \, #2}}
\newcommand{\Interference}[0]{\mathrm{I}}
\newcommand{\SINR}[0]{\mathrm{SINR}}

\newcommand{\Ball}[0]{\mathit{B}}

\newcommand{\MinDist}[0]{\kappa}
\newcommand{\dist}[1]{\mathrm{dist} ({#1})}

\newcommand{\LargeRadiusBound}[0]{\LargeRadius}
\newcommand{\SmallRadiusBound}[0]{\SmallRadius}
\newcommand{\FatnessParameterBound}[0]{\varphi}
\newcommand{\MaxFatnessParameterBound}[0]{\varphi_{\rm max}}
\newcommand{\SumFatnessSquares}[0]{\varphi_{\rm sum}}

\newcommand{\FatnessParameter}[0]{\varphi}

\newcommand{\UPN}[0]{uniform power network}
\newcommand{\NUPN}[0]{non-uniform power network}

\newcommand{\vor}[0]{\mbox{\sc Vor}}
\newcommand{\wvorsys}[0]{V}
\newcommand{\wvor}[0]{\mbox{\sc WVor}}
\newcommand{\Vweight}[0]{w}
\newcommand{\Vweights}[0]{w}

\newcommand{\cS}[0]{{\cal S}}
\newcommand{\WPower}[0]{{\Theta}}
\newcommand{\wPower}[0]{{\theta}}
\newcommand{\segment}[0]{{\sigma}}
\newcommand{\hypergeodesic}[1]{\mathrm{h} ({#1})}

\newcommand{\Center}[0]{q}

\newcommand{\diffEnergyp}[0]{g_{(1)}}
\newcommand{\diffEnergym}[0]{g_{(-1)}}

\newcommand{\rootp}[0]{r_{(1)}}
\newcommand{\rootm}[0]{r_{(-1)}}

\newcommand{\sqS}[1]{{\hat S}_{#1}}

\newcommand{\thetaci}[0]{\theta^\chi_i}
\newcommand{\thetacj}[0]{\theta^\chi_j}

\def\cC{{\cal C}}

\def\epsilonhat{\eta}

\begin{document}


\begin{titlepage}
\def\thepage{}

\title{The Topology of Wireless Communication
\thanks{Supported in part by a grant from the Israel Science Foundation.}
}

\author{
Erez Kantor
\thanks{The Technion, Haifa, Israel.
E-mail: {\tt erez.kantor@gmail.com}.}
\and
Zvi Lotker
\thanks{
Ben Gurion University, Beer-Sheva, Israel.
E-mail: {\tt zvilo@cse.bgu.ac.il}.}
\and
Merav Parter
\thanks{
The Weizmann Institute of Science, Rehovot, Israel.
E-mail: {\tt \{merav.parter,david.peleg\}@ weizmann.ac.il}.}
\and
David Peleg $^\S$
}

\date{\today}

\maketitle
\begin{abstract}
In this paper we study the topological properties of wireless communication
maps and their usability in algorithmic design.  We consider the SINR model,
which compares the received power of a signal at a receiver against the sum
of strengths of other interfering signals plus background noise.
To describe the behavior of a multi-station network,
we use the convenient representation of a \emph{reception map}. In the SINR
model, the resulting \emph{SINR diagram} partitions the plane into
reception zones, one per station, and the complementary region of the plane
where no station can be heard.
\emph{SINR diagrams} have been studied in \cite{Avin2009PODC}
for the specific case where all stations use the same power.
It is shown that the reception zones are convex (hence connected) and fat,
and this is used to devise an efficient algorithm for the fundamental
problem of point location. Here we consider the more general (and common)
case where transmission energies are arbitrary (or non-uniform).
Under that setting, the reception zones are not necessarily convex
or even connected. This poses the algorithmic challenge of designing
efficient point location techniques for the non-uniform setting,
as well as the theoretical challenge of understanding the geometry
of SINR diagrams (e.g., the maximal number of connected components
they might have). We achieve several results in both directions.
We establish a form of weaker convexity
in the case where stations are aligned on a line and use this to derive
a tight bound on the number of connected components in this case.
In addition, one of our key results concerns the behavior
of a $(d+1)$-dimensional map, i.e., a map in one dimension higher
than the dimension in which stations are embedded.
Specifically, although the $d$-dimensional map might be highly fractured,
drawing the map in one dimension higher ``heals'' the zones,
which become connected (in fact hyperbolically connected).
In addition, as a step toward establishing a weaker form of convexity
for the $d$-dimensional map, we study the interference function
and show that it satisfies the maximum principle. This is done through
an analysis technique based on looking at the behavior of systems composed
on lines of densely placed weak stations, as the number
of stations tends to infinity, keeping their total transmission energy fixed.
Finally, we turn to consider algorithmic applications, and
propose a new variant of approximate point location.
\end{abstract}

\end{titlepage}

\section{Introduction}
\label{sec:introduction}

\paragraph{Background and motivation:}
The use of wireless technology in communication networks is rapidly
\commful
growing, imposing
\commfulend
\commabs
growing. This trend imposes
\commabsend
increasingly heavy loads on the resources required by wireless networks.
One of the main resources required for such communication is radio spectrum,
which is limited by nature. Hence careful design of all aspects of
the network is crucial to efficient utilization of its resources.
Good planning of radio communication networks must take advantage of all
its features, including both physical properties of
the channels and structural properties of the entire network.
While the physical properties of channels have been thoroughly studied,
see \cite{G05,TV05}. Relatively little is known about the topology and geometry
of the wireless network structure and their influence on performance issues.

There is a wide range of challenges in wireless communication for which
better organization of the
\commabs
communication
\commabsend
network may become useful.
Specifically, understanding the topology of the underlying
communication network may lead to more sophisticated algorithms
for problems such as scheduling, topology control and connectivity.
We study wireless communication in free space; this is simpler than
the irregular environment of radio channels in a general setting,
which involves reflection and shadowing.
We use the \emph{Signal to Interference-plus-Noise Ratio (SINR)} model
which is widely used by the Electrical Engineering community,
and is recently being explored by Computer Scientists as well.
Let
$$\SINR(\Station_{i},p)=\frac{\Power_i \cdot \dist{\Station_i, p}^{-\alpha}}{\sum_{j \neq i}
\Power_j \cdot \dist{\Station_j, p}^{-\alpha} + \Noise}.$$
In this model, a receiver at point $p \in \R^{d}$ successfully receives a message from the sender $\Station_{i}$ if and only if $\SINR(\Station_{i},p)\geq \beta$,
where $\Noise$ is the environmental noise, the constant $\beta \geq 1$ denotes the minimum SINR required for a message to be successfully received, $\alpha$ is the path-loss parameter and $S=\{\Station_1, \ldots, \Station_n\}$ is the set of concurrently transmitting stations using power assignment $\Power$.
Within this context, we focus on one specific algorithmic challenge,
namely, the point location problem, defined as follows.
Given a query point $p$, it is required to identify which of the $n$
transmitting stations is heard at \(p\), if any,
under interference from all other $n-1$ transmitting stations
and background noise \(\Noise\). Obviously, one can directly
compute \(\SINR_{\cA}(\Station_i, p)\) for every $i \in \{1, \ldots ,n\}$
in time \(\Theta (n)\) and answer the above question accordingly.
Yet, this computation may be too expensive, if the query is asked
for many different points \(p\).
Avin et al. \cite{Avin2009PODC} initiated the study of the topology
and geometry of wireless communication in the SINR model, and its application
to the point location problem, in the relatively simple setting of
{\em uniform powers}, namely, under the assumption that all stations transmit
with the same power level. They show that in this setting,
the SINR diagram assumes a particularly convenient form:
the reception zones of all senders are convex and ``fat''.
They later exploit these properties to devise an efficient data structure
for point location queries, resulting in a logarithmic query time complexity.

In actual wireless communication systems, however, most wireless
communication devices can modify their transmission power.
Moreover, it has been demonstrated convincingly that allowing transmitters
to use different power levels increases the efficiency of various
communication patterns in terms of resource utilization (particularly,
energy consumption and communication time).
Hence it is important to develop both a deep understanding of the underlying
structural properties and suitable algorithmic techniques for handling
various communication-related problems in non-uniform wireless networks
as well. In particular, it may be useful to develop algorithms for
solving the problem of point location in such networks.
Unfortunately, it turns out that once we turn to the more general case of
non-uniform wireless networks, the picture becomes more involved,
and the topological features of the SINR diagram
are more complicated than
in the uniform case. In particular, simple examples (with as few as
five stations, as illustrated later on)
show that the reception zone of a station is not necessarily connected,
and therefore is not convex. Other ``nice'' features of the problem
in the uniform setting, such as fatness, are no longer satisfied as well.
Subsequently, algorithmic design problems become more difficult.
In particular, the point location problem becomes harder,
and cannot be solved directly
via the techniques developed in \cite{Avin2009PODC} for the uniform case.

In this paper we aim to improve our understanding of the topological
and geometric structure of the reception zones of SINR diagrams
in the general (non-uniform) case.
The difficulty in point location with variable power follows from
several independent sources. First, one must overcome the fact that the number
of connected cells is not always known (and there are generally
several connected cells). A second problem is that the shape of each
connected cell is no longer as simple as in the uniform case.
Yet another problem is the possibility of singularity points
on the boundaries of the reception zones.
(Typically, those problems become harder in higher dimensions, but
as seen later, this is not always the case for wireless networks.)

Nevertheless, we manage to establish several properties of SINR diagrams
in non-uniform networks that are slightly weaker than convexity,
but are still useful for tackling our algorithmic problems,
such as satisfying the maximum principle of the interference function and enjoying hyperbolic convexity.
%
%
To illustrate these properties, let us take a look
at the simplest example where a problem already occurs.
When we look at two stations in one dimension,
the reception zones are not connected.
Surprisingly, when we look at the same example in two dimensions
(instead of one), the reception zones of both stations become connected.
As shown later on, this is no coincidence.
Moreover, when we examine closely the two-dimensional case,
we see that the reception zones are no longer convex but actually
hyperbolic convex (as opposed to non convex in the one dimensional case).
We use this strategy of adding a dimension to the original problem
and moving from Euclidean geometry to hyperbolic geometry
to solve the point location problem.

\dnsparagraph{Contributions:}
\commabs
In this paper we aimed toward gaining better understanding of SINR maps
with non-uniform power. Better characterization of reception map
has a theoretical as well as practical motivation.
\commabsend
The starting point of our work is the following observation:
in non-uniform setting, reception zones are neither convex nor fat.
In addition, they are not connected.
The loss of these ``niceness" properties, previously established
for the uniform power setting \cite{Avin2009PODC}, appears even for
the presumably simple case where all stations are aligned on a line.

This raises several immediate questions.  The first is a simple ``counting''
question that has strong implications on our algorithmic question:
What is the maximal number of reception cells that may occur
in an SINR diagram of a wireless network on $n$ transmitters.
The second question has a broader scope: Are there any ``niceness" properties
that can be established in non-uniform setting.
Specifically, we aim toward finding other (weaker but still useful)
forms of convexity that are satisfied by cells in non-uniform reception maps.
Apart from their theoretical interest, these questions are also
of considerable practical significance, as
obviously, having reception zones with some form of convexity
might ease the development of protocols for various design and
communication tasks.

We establish two weaker forms of convexity and show their theoretical
as well as algorithmic implications.
Starting with the one-dimensional case, where stations are aligned on a line,
we show that although the zones are not convex, they are
convex in a region that is free from stations.
We then use this ``No-Free-Hole'' (NFH) property
to establish the fact that in one dimension, the number of
reception cells generated by $n$ stations is bounded above by $2n-1$
(and this can be realized).
For the general setting where stations are embedded in $\R^{d}$,
the problem of bounding the number of connected cells seems to be harder,
even for $d=2$. We are able to show that the number of reception cells
is no more than $O(n^{d+1})$ and provide examples with $\Omega(n)$
reception cells for a single station.  Do $d$-dimensional zones enjoy
the NFH property? Although this remains an open question,
we make two major advances in this context.

First, we consider the $(d+1)$-dimensional SINR map
of a wireless network whose stations are embedded in $d$-dimensional space,
and establish a much stronger property.
It turns out, that while in the $d$-dimensional space the network's SINR map
might be highly fractured,
going one dimension higher miraculously ``heals'' the reception zones,
which become connected (in fact, hyperbolically connected or
hyperbolically convex). This may have practical ramifications.
For instance, considering
stations located in the 2-dimensional plane, one realizes that
their reception zones in 3-dimensional space are connected,
which aids in answering point location queries in this realistic setting.

Turning back to the $d$-dimensional map, we consider a well known property
of harmonic functions, namely, the maximum principle.
Generally speaking, the maximum principle refers to the case where
the maximum value of the function in a given domain,
is attained at the circumference of that domain.
Does the SINR function follow the maximum principle?
This is yet another open question.  If so, NFH property
is followed.  As a step toward achieving this goal,
we then examine the properties of the interference function
(appearing in the denominator of the SINR function),
and establish the fact that this function satisfies the maximum principle.
This is done through an analysis technique based on looking at the behavior
of systems composed on lines of densely placed weak stations, as the number
of stations tends to infinity, keeping their total transmission energy fixed.

Finally, we consider the point location task, defined as follows.
Given a set of broadcasting stations $S$ and a point $p$,
we are interested in knowing whether the transmission of station $\Station$
is correctly received at $p$.  We present a construction scheme
of a data structure (per station) that maintains a partition of the plane
into three zones: a zone of all points that correctly receive
the transmissions of $\Station$, i.e., points $p$ with
$\SINR(\Station,p) \geq \beta$;  a zone where the transmission
of $\Station$ cannot be correctly received, i.e., points $p$ with
$\SINR(\Station,p) < \beta$; and a zone of uncertainty corresponding
to points that might receive the transmission in a somewhat lower quality,
i.e., points $p$ with
$\SINR(\Station,p) \geq (1-\epsilon)^{2\alpha}\cdot\beta$,
where $\epsilon$ is predefined performance parameter.
Using this data structure, a point location query can be answered
in logarithmic time.

\begin{figure}[htb]
\begin{center}
\includegraphics[scale=0.8]{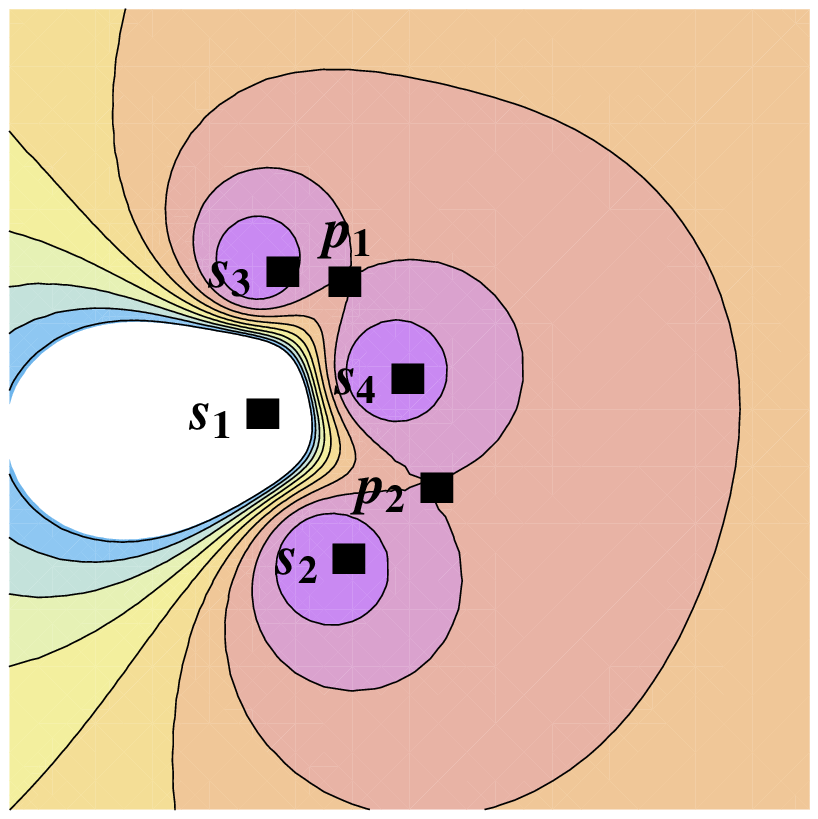}
\hfill
\includegraphics[scale=0.8]{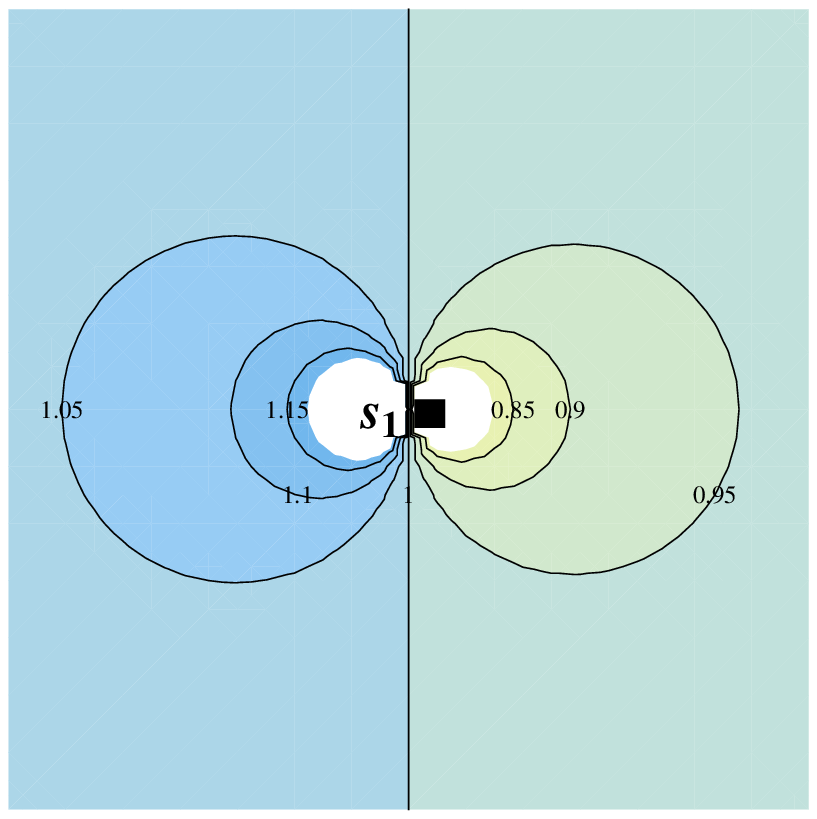}
\caption{ \label{figure:topological_complex}
\sf Topographic view of $\ReceptionZone_{1}$ (with no noise).
Heights indicate SINR thresholds.
(a) SINR map of 4-station non-uniform power network: singular points
($p_{1}$ and $p_{2}$) and contour lines of $\SINR(\Station_1,(x,y))$.
(b) Low gradient regions in SINR map in 2-station uniform power network.
$\ReceptionZone_{1}$ is unbounded when $\beta \leq 1$ but finite for $\beta>1$,
illustrating the impossibility of getting uniformly bounds of the area
between two SINR curves corresponding to two different threshold levels.
}
\end{center}
\end{figure}

\dnsparagraph{Related work:}
\commabs
Our starting point is the work of Avin at. el. \cite{Avin2009PODC},
where it is proven that if all transmitters use the same power
then the reception zones are convex and fat.
\commabsend
Several papers have shown
that the capacity of wireless networks increases when transmitters
can adapt their transmission power. In their seminal paper \cite{AE99},
Gupta and Kumar analyzed the capacity
of wireless networks in the physical and protocol models.
Moscibroda \cite{Mo07} analyzed the worst-case capacity of wireless networks,
without any assumption on the deployment of nodes in the plane,
as opposed to almost all previous works on this problem.
Non-uniform power assignments can clearly outperform a uniform assignment
\cite{Moscibroda2006Protocol,MoWa06} and increase the capacity of a network.
Therefore the majority of the literature on capacity and scheduling
addresses non-uniform power.
\commabs
In the engineering community, the physical interference (SINR) model
has been scrutinized for almost four decades.
\commabsend
Assuming that the power of all transmitters is uniform,
we know from \cite{Avin2009PODC} that the reception zones are convex
and fat. Therefore the singularity points of a zone can be easily handled.
Yet when power is not uniform, handling the singularity points
becomes a major challenge. We remark that recently,
Gabrielov, Novikov, and Shapiro have shown that the number of singular points
of functions similar to the interference function is
finite, see \cite{GNS07}.
Maxwell conjectured that the number of singularity points
in the interference function is bound by $(n-1)^2$
where $n$ is the number of transmitters; see \cite{Max54} for more details.
For illustration see Figure \ref{figure:topological_complex}a.

Another challenge that one has to deal with in non-uniform networks is
the possible existence of regions with very small gradient in the $\SINR$
function, as exemplified in Figure \ref{figure:topological_complex}b,
which reflects the fact that the area containing all points $p$ such that
$\SINR(\Station_i,p) \in [\beta,\beta+\epsilon]$ cannot be bounded even for small $\epsilon>0$.
\commabs
\commabsend

\commful
Understanding the geometry of SINR diagrams is relevant for the joint problem
of scheduling and power control.
\commfulend
\commabs
It is hoped that a better understanding of the topology of the SINR diagram
will improve our understanding of the joint problem of scheduling and
power control.
\commabsend
The complexity of this problem in the physical model,
taking into account the geometry of the problem, is unknown.
Nevertheless, many algorithms and heuristics have been suggested, e.g.,
\cite{BE06a,cruz03,Lee95,MoWa06,Wang05,Zander92b}.
See \cite{Moscibroda2007How} for a more detailed discussion
of these approaches.
Recently, Kesselheim \cite{K11} has shown how to achieve a constant
approximation
for the capacity problem with power control, for doubling metric spaces.
His algorithm yields $O(\log n)$ approximation for general metrics.
Halldórsson and Mitra \cite{HM11} show tight characterizations of
capacity maximization under power control, using oblivious power assignments
in general metrics.

\section{Preliminaries}
\label{section:Preliminaries}

\subsection{Geometric notions}
\label{section:GeometricNotions}
Throughout, we consider the $d$-dimensional Euclidean space \(\Reals^d\) (for
$d \in \Integers_{\geq 1}$).
The \emph{distance} between points \(p\) and point \(q\) is denoted by
\( \dist{p,q}
= \| q - p \| \).
A \emph{ball} of radius \(r\) centered at point \(p \in \R^{d}\)
is the set of all points
at distance at most \(r\) from \(p\), denoted by \( \Ball^{d}(p, r) = \{ q \in
\Reals^d \mid \dist{p, q} \leq r \} \).
Unless stated otherwise, we assume the 2-dimensional Euclidean plane,
and omit $d$.
The basic notions of open, closed, bounded, compact and connected
sets of points are defined in the standard manner.
\commabs
A point set \(P\) is said to be \emph{open} if all points \( p \in P \) are
internal points, and \emph{closed} if its complement \({\bar P}\) is open.
If there exists some real \(r\) such that \( \dist{p, q} \leq r \) for every
two points \( p, q \in P \), then \(P\) is said to be \emph{bounded}.
A \emph{compact} set is a set that is both closed and bounded.
The \emph{closure} of \(P\), denoted \(cl(P)\),
is the smallest closed set containing \(P\).
\commabsend
The \emph{boundary} of a point set \(P\), denoted by \(\Boundary (P)\),
\commabs
is the intersection of the closure of \(P\) and the closure of its
complement, i.e., \(\Boundary (P) = cl(P) \cap cl(\bar{P}) \).
Let $\Circumference(\Boundary (P))$ denote the length of \(\Boundary (P)\).
A \emph{connected set} is a point set \(P\) that cannot be partitioned to two
non-empty subsets \(P_1,P_2\) such that each of the subsets has no point
in common with the closure of the other (i.e., $P$ is connected if
for every $P_1,P_2\ne\emptyset$ such that $P_1\cap P_2=\emptyset$
and $P_1\cup P_2=P$, either $P_1\cap cl(P_2)\ne\emptyset$ or
$P_2\cap cl(P_1)\ne\emptyset$.).
\commabsend
A \emph{maximal connected subset} \(P_1 \subseteq P\) is a
connected point set such that $P_1 \cup \{p\}$ is no longer connected
for every $p \in P \setminus P_1$.

We use the term \emph{zone} to describe a point set with some
``niceness'' properties.
Unless stated otherwise, a zone refers to the union of an open connected set
and some subset of its boundary.
It may also refer to a single point or to the finite union of zones.
\commabs
\par
\commabsend
Let $F: \R^{d} \to \R^{d}$ and let $p\in \R^{d}$. Then $F$ is the
\emph{characteristic polynomial} of a zone \(Z\) if
$p \in Z ~\Leftrightarrow~ F(p) \leq 0$.
\commabs
\par
\commabsend
Denote the \emph{area} of a bounded zone \(Z\)
(assuming that it is well-defined) by \(\Area(Z)\).
For a non-empty bounded zone \(Z\ne\emptyset\) and an internal \(p\in Z\),
denote the maximal and minimal radii of $Z$ w.r.t. $p$ by
\commful
$\SmallRadius(p, Z) = \sup \{ r > 0 \mid Z \supseteq \Ball(p, r) \}$ and
$\LargeRadius(p, Z) = \inf \{ r > 0 \mid Z \subseteq \Ball(p, r) \}$,
\commfulend
\commabs
$$\SmallRadius(p, Z) ~=~ \sup \{ r > 0 \mid Z \supseteq \Ball(p, r) \} ~ ,
\qquad
\LargeRadius(p, Z) ~=~ \inf \{ r > 0 \mid Z \subseteq \Ball(p, r) \} ~ ,$$
\commabsend
and define the \emph{fatness parameter} of \(Z\) \emph{with respect to} \(p\)
to be
\( \FatnessParameter(p, Z) = \LargeRadius(p, Z) / \SmallRadius(p, Z) \).
The zone \(Z\) is said to be \emph{fat} with respect to \(p\) if
\(\FatnessParameter(p, Z)\) is bounded by some constant.

\subsection{Wireless networks}
\label{section:WirelessNetworks}
We consider a wireless network \( \cA = \langle d, S, \Power, \Noise, \beta,
\alpha \rangle \), where
$d \in \Integers_{\geq 1}$ is the dimension,
\( S = \{ \Station_1, \Station_2, \dots, \Station_{n} \} \) is a set of
transmitting \emph{radio stations} embedded in the $d$-dimensional space,
\(\Power\) is an assignment of a positive real \emph{transmitting power}
\(\Power_i\) to each station \(\Station_i\),
\( \Noise \geq 0 \) is the \emph{background noise},
\( \beta \geq 1 \) is a constant that serves as the \emph{reception threshold}
(to be explained soon), and
$\alpha>0$ is the {\em path-loss parameter}.
We sometimes wish to consider a network obtained from \(\cA\) by modifying
one of the parameters while keeping all other parameters unchanged.
To this end we employ the following notation.
Let \(\cA_{d'}\) be a network identical to \(\cA\) except its dimension
is $d' \neq d$. \(\cA_{\beta'}\) and \(\cA_{\alpha'}\) are defined
in the same manner.
For notational simplicity, \(\Station_i\) also refers to the point
$(x_1^{\Station_i},...,x_d^{\Station_i})$ in the $d$-dimensional space $\Reals^{d}$ where the station
\(\Station_i\) resides, and moreover,
when $d=2$, the point \(\Station_i\) in the Euclidean plane is denoted \((x_i,
y_i)\).
The network is assumed to contain at least two stations, i.e., \( n \geq 2 \).
The \emph{energy} of station \(\Station_i\) at point \( p \neq \Station_i \)
is defined to be \( \Energy_{\cA}(\Station_i, p) = \Power_i \cdot
\dist{\Station_i, p}^{-\alpha} \).
The \emph{energy} of a set of stations \( T \subseteq S \) at a point
\(p \not\in T\)
is defined to be
\(\Energy_{\cA}(T, p) =
\sum_{\Station_i \in T} \Energy_{\cA}(\Station_i, p) \).
Fix some station \(\Station_i\) and consider some point \( p \notin S \).
We define the \emph{interference} of \(\Station_j\)
to be the energy of \(\Station_j\) at \(p\), $j \neq i$ denoted
\( \Interference_{\cA}(\Station_j, p) = \Energy_{\cA}(\Station_j, p)\).
The \emph{interference} of a set of stations \( T \subseteq S \setminus \{ \Station_i\}\)
at a point \(p \not\in S\) is defined to be
\(\Interference_{\cA}(T,p) = \Energy_{\cA}(T, p)\).
The \emph{signal to interference \& noise ratio (SINR)} of \(\Station_i\)
at point \(p\) is defined as
\begin{equation}
\label{eq:Def-SINR}
\SINR_{\cA}(\Station_i, p)
~=~ \frac{\Energy_{\cA}(\Station_i, p)}
{\Interference_{\cA}(S - \{\Station_i\}, p) + \Noise}
~=~ \frac{\Power_i \cdot \dist{\Station_i, p}^{-\alpha}}{\sum_{j \neq i}
\Power_j
\cdot \dist{\Station_j, p}^{-\alpha} + \Noise} ~ .
\end{equation}
Observe that \(\SINR_{\cA}(\Station_i, p)\) is always positive since the
transmitting powers and the distances of the stations from \(p\) are always
positive and the background noise is non-negative.

In certain contexts, it is convenient to consider the reciprocal of the SINR function, namely, $\SINR^{-1}$ defined as
\begin{equation}
\label{eq:Def-SINR_reciprocal}
\SINR_{\cA}^{-1}(\Station_i, p)
~=~ \frac{\Interference_{\cA}(S - \{\Station_i\}, p)+\Noise}
{\Energy_{\cA}(\Station_i, p)}~.
\end{equation}

When the network \(\cA\) is clear from the context, we may omit it and write
simply \(\Energy(\Station_i, p)\), \(\Interference(\Station_j, p)\), \(\SINR(\Station_i, p)\) and \(\SINR^{-1}(\Station_i, p)\).

The fundamental rule of the SINR model is that the transmission of station
\(\Station_i\) is received correctly at point \( p \notin S \) if and only if
its SINR at \(p\) is not smaller than the reception threshold of the network,
i.e., \( \SINR_{\cA}(\Station_i, p) \geq \beta \).
If this is the case, then we say that \(\Station_i\) is \emph{heard} at \(p\).
We refer to the set of points that hear station \(\Station_i\) as the
\emph{reception zone} of \(\Station_i\), defined as
$$\ReceptionZone_{i}(\cA) ~=~
\{ p \in \Reals^d - S \mid \SINR_{\cA}(\Station_i, p) \geq \beta \}
\cup \{\Station_i\} ~.$$
This definition is necessary since \(\SINR(\Station_i, \cdot)\)
is undefined at points in \(S\) and in particular at \(\Station_i\) itself.
In the same manner we refer to the set of points that hear no station
\(\Station_i \in S\) (due to the background noise and interference) defined as
\commful
$ \ReceptionZone_{\emptyset}(\cA)=
\{ p \in \Reals^d - S \mid \SINR_{\cA}(\Station_i, p) < \beta,
~\forall \Station_i \in S\}.$
\commfulend
\commabs
$$\ReceptionZone_{\emptyset}(\cA) ~=~
\{ p \in \Reals^d - S \mid \SINR(\Station_i, p) < \beta,
~~\forall \Station_i \in S\}.$$
\commabsend
An SINR diagram $\ReceptionZone(\cA) = \{\ReceptionZone_{i}(\cA),~~
1 \leq i \leq n\} \cup \{\ReceptionZone_{\emptyset}(\cA)\}$
is a ``reception map'' characterizing the reception zones of the stations.
This map partitions the plane into $n+1$ zones; a zone for each station
\(\ReceptionZone_{i}(\cA)\), \( 1 \leq i \leq n \), and a zone
\(\ReceptionZone_{\emptyset}(\cA)\) where no successful reception exists
to any of the stations.

It is important to note that a reception zone, \(\ReceptionZone_{i}(\cA)\),
is not necessarily connected. A \emph{maximal connected component} within
a zone is referred to as a \emph{cell}. Let \(\ReceptionZone_{i,j}(\cA)\)
be the $j^{th}$ cell in \(\ReceptionZone_{i}(\cA)\).
\commabs
\par
\begin{figure}[htb]
\begin{center}
\includegraphics[scale=0.4]{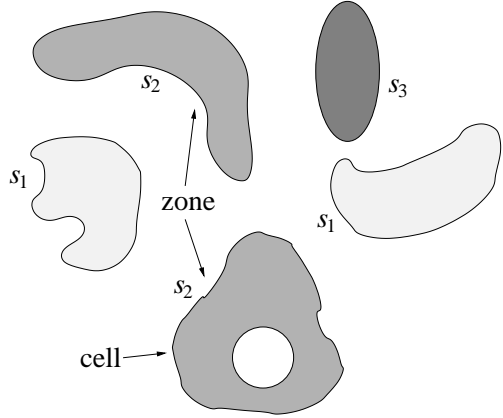}
\caption{ \label{figure:ReceptionMap}
\sf Schematic representation of a reception map, consisting of
reception zones, each composed of several connected components
referred to as \emph{cells}.
}
\end{center}
\end{figure}

\commabsend
Hereafter, the set of points where the transmissions of a given station
are successfully received is referred to as its {\em reception zone}, and
a {\em cell} is a maximal connected set or component in a given reception zone.
Hence the reception zone is a set of cells, given by
\(\ReceptionZone_{i}(\cA)=\{\ReceptionZone_{i,1}(\cA), \ldots
\ReceptionZone_{i,\NZones_{i}(\cA)}(\cA)\}\), where \(\NZones_{i}(\cA)\)
is the number of cells in \(\ReceptionZone_{i}(\cA)\).
Analogously, \(\ReceptionZone_{\emptyset}(\cA)\) is composed of
\(\NZones_{\emptyset}(\cA)\) connected cells,
\(\ReceptionZone_{\emptyset,j}(\cA)\).
Overall, the topology of a wireless network \(\cA\) is arranged
in three levels:
The \emph{reception map} is at the top of the hierarchy.
It is composed of $n+1$ reception zones, $\ReceptionZone_{i}(\cA)$,
$i \in \{1, \ldots n,\emptyset\}$.
Each zone \(\ReceptionZone_{i}(\cA)\) is composed of $\NZones_{i}(\cA)$
reception cells. For a pictorial description see Figure \ref{figure:ReceptionMap}.

The following definition is useful in our later arguments.
Let $F^{i}_{\cA}(p)$
\footnote{
When \(\cA\) is clear from context we may
omit it and simply write $\ReceptionZone_{i}, \NZones_{i}$ and $F_{i}(p)$.
When refereing to reception zones $\ReceptionZone_{i}(\cA_{d'})$ or
$\ReceptionZone_{i}(\cA_{\beta'})$ we may omit \(\cA\) and simply write
$\ReceptionZone_{i}(d')$ and $\ReceptionZone_{i}(\beta)$.}
, $p\in \R^{d}$ be the \emph{characteristic polynomial}
of $\ReceptionZone_i(\cA)$ given by
\commabs
\begin{equation}
\label{eq:reception_polynomial}
F^{i}_{\cA}(p) ~=~ \beta \left( \sum_{k \neq i}  \Power_{k} \prod_{l \neq k} \dist{\Station_{l},p}^{\alpha} +\Noise \cdot \prod_{k} \dist{\Station_{k} ,p}^{\alpha}\right)-\Power_{i}\prod_{k \neq i} \dist{\Station_{k},p}^{\alpha}~.
\end{equation}
\commabsend
Then $p \in \ReceptionZone_{i}(\cA)$ iff $F^{i}_{\cA}(p) \leq 0$.


Avin et al. \cite{Avin2009PODC} discuss the relationships between SINR diagram
on a set of stations $S$ with \emph{uniform} powers and the corresponding
{\em Voronoi diagram} on $S$. Specifically, it is shown that the $n$
reception zones \(\ReceptionZone_{i}(\cA)\) are strictly contained
in the corresponding
Voronoi cells $\vor_i$. SINR diagrams with \emph{non-uniform} powers
are related to the {\em weighted Voronoi diagram} of the stations
instead of to the {\em Voronoi diagram}.

In the weighted version of Voronoi diagram \cite{AurenhammerE84},
we consider a weighted system $\wvorsys=\langle S, \Vweights\rangle$,
where $S=\{\Station_1,...,\Station_{n}\}$ represents a set of $n$ points
in d-dimensional Euclidean space and
$\Vweights=\{\Vweight_1,...,\Vweight_{n}\}$ is an assignment of weights
$\Vweight_i \in \Reals_{> 0}$ to each point $\Station_i \in S$.
The {\em weighted voronoi diagram} of $\wvorsys=\langle S,\Vweights\rangle$
partitions the planes into $n$ zones, where
\commful
$\wvor_{i}(\wvorsys)=\left\{ p \in \R^{d} \mid
\frac{\Vweight_i}{\dist{\Station_i,p}} >
\frac{\Vweight_j}{\dist{\Station_j,p}},
\mbox{ for any } j\not=i \right\}$
\commfulend
\commabs
\begin{equation*}
\wvor_{i}(\wvorsys) ~=~
\left\{ p \in \R^{d} \mid \frac{\Vweight_i}{\dist{\Station_i,p}} ~>~
\frac{\Vweight_j}{\dist{\Station_j,p}}, \mbox{ for any } j\not=i \right\},
\end{equation*}
\commabsend
denotes the \emph{zones} (of influence) of a point $\Station_i$ in $S$,
for every $i\in \{1, \ldots, n\}$. The weighted Voronoi map denoted by
$\wvor(\wvorsys)$, is composed of {\em cells, edges} and {\em vertices}.
A cell corresponds to a \emph{maximal connected component} in
$\wvor_{i}(\wvorsys)$, $i \in \{1, \ldots ,n\}$. An \emph{edge} is
the relative interior of the intersection of two closed \emph{cells}.
Finally, a vertex is an endpoint of an edge. In the unweighted
Voronoi diagram each zone $\wvor_{i}(\wvorsys)$ corresponds to one
connected cell. On the contrary, a weighted Voronoi map is composed of
$O(n^{2})$ cells as was shown at \cite{AurenhammerE84}.
For a given wireless network
$\cA=\langle d, S,\Power, \Noise, \beta, \alpha \rangle$,
we define the corresponding weighted Voronoi system
$\wvorsys_{\cA}=\langle S^\cA, \Vweights^\cA\rangle$ in the following manner.
The set of points $S^{\cA}$ corresponds to $S$ positions and
$\Vweight^{\cA}_{i}=\Power_i^{1/\alpha}$, for every $1\leq i\leq n$.
\commful
We also consider the way the ``reception map'' $\ReceptionZone(\cA_{\alpha})$
of a given network $\cA_{\alpha}$ changes as $\alpha$ goes to infinity
while the other parameters (e.g., the set of stations, $\beta$,
the noise etc.) are fixed. The map $\ReceptionZone(\cA_{\alpha})$
converges to is denoted by
$\ReceptionZone(\cA_{\infty}) =
\lim_{\alpha \to \infty}\ReceptionZone(\cA_{\alpha}).$
In what follows we formally express the relations between
$\ReceptionZone(\cA)$ (respectively, $\ReceptionZone(\cA_{\infty})$)
and $\wvor(\wvorsys_{\cA})$ (resp., $\vor$).
(For lack of space, some proofs are moved to the appendix
and some are omitted.)
\begin{lemma}
\label{cl:wv_sinr}
(a) $\ReceptionZone_{i}(\cA) \subseteq \wvor_{i}(\wvorsys_{\cA})$,
for every $i\in\{1, \ldots ,n\}$ and $\beta \geq 1$.
\\
(b) $\ReceptionZone_{i}(\cA_{\infty}) \subseteq \vor_i$
for every $i\in\{1, \ldots ,n\}$.
\end{lemma}
\commfulend
\commabs
In what follows we formally express the relation between
$\ReceptionZone(\cA)$ and $\wvor(\wvorsys_{\cA})$.
\begin{lemma}
\label{cl:wv_sinr}
$\ReceptionZone_{i}(\cA) \subseteq \wvor_{i}(\wvorsys_{\cA})$,
for every $i\in\{1, \ldots ,n\}$ and $\beta \geq 1$.
\end{lemma}
\Proof
Let $d_{i}=\dist{\Station_{i},p}$. Let $p \in \R^{d}$ be such that
$p \in \ReceptionZone_{i}(\cA)$.
We prove that $p \in \wvor_{i}(\wvorsys_{\cA})$.
Since $p \in \ReceptionZone_{i}(\cA)$, by (\ref{eq:Def-SINR})
\begin{eqnarray*}
\frac{\Power_i}{d_{i}^{\alpha}} & \geq &
\beta \cdot \left(\sum_{j \neq i} \frac{\Power_j}{d_{j}^{\alpha}}+\Noise\right)
~\geq~ \frac{\Power_k}{d_{k}^{\alpha}}\left (1+ \sum_{j \neq k,i}
\frac{\Power_{j}/\Power_{k} }{\left(d_{j}/d_{k} \right)^{\alpha}}\right)
\end{eqnarray*}
where $\Power_k/d_{k}^{\alpha} =
\max_{j \neq i} \left(\Power_j/d_{j}^{\alpha}\right)$,
and hence
$$\frac{\Power_i}{d_{i}^{\alpha}} ~>~ \frac{\Power_k}{d_{k}^{\alpha}}
~~~~\mbox{and}~~~~
\frac{\Power_{i}^{1/\alpha}}{d_{i}} ~>~ \frac{\Power_{k}^{1/\alpha}}{d_{k}}.$$
The choice of $\Vweight_i$ implies that $p \in \wvor_{i}(\wvorsys_{\cA})$
and the claim holds.
\QED

Consider the way the ``reception map'' $\ReceptionZone(\cA_{\alpha})$
of a given network $\cA_{\alpha}$ changes as $\alpha$ goes to infinity
while the other parameters (e.g., the set of stations, $\beta$,
the noise etc.) are fixed. The map $\ReceptionZone(\cA_{\alpha})$
converges to is denoted by
$$\ReceptionZone(\cA_{\infty}) ~=~
\lim_{\alpha \to \infty}\ReceptionZone(\cA_{\alpha}).$$

\begin{lemma}
\label{cl:limit_wv_sinr}
$\ReceptionZone_{i}(\cA_{\infty}) \subseteq \vor_i$,
for every $i\in\{1, \ldots ,n\}$.
\end{lemma}
\Proof
By Lemma \ref{cl:wv_sinr},
$\ReceptionZone_{i}(\cA) \subseteq \wvor_{i}(\wvorsys_{\cA})$.
It follows that $\ReceptionZone_{i}(\cA_{\infty}) \subseteq \wvor_{i}(\wvorsys_{\cA_{\infty}})$ for $\Vweight_i=\lim_{\alpha \to \infty}\Power_i^{1/\alpha}=1$. But $\wvor_{i}(\wvorsys_{\cA_{\infty}})$ is simply $\vor_i$. This can also be seen by considering the SINR function: as $\alpha$  gets larger, the power of the station becomes negligible compared to distance between the station and the point $p$. In other words, it gets closer to the uniform Voronoi diagram.
\QED
\commabsend

\commful
We conclude this section by stating a technical lemma
(see \cite{Avin2009PODC}) to be used later on.
\commfulend
\commabs
We conclude this section by stating an important technical lemma
from \cite{Avin2009PODC} that will be useful in our later arguments.
\commabsend
\begin{lemma}
\label{lemma:Transformation}\cite{Avin2009PODC}
Let \( f : \Reals^d \rightarrow \Reals^d \)
be a mapping consisting of
rotation, translation, and scaling by a factor of \( \sigma > 0 \).
Consider some network \( \cA = \langle d, S, \Power, \Noise, \beta, \alpha \rangle \) and
let \( f(\cA) = \langle d, f(S), \Power, \Noise / \sigma^2, \beta, \alpha \rangle \),
where \( f(S) = \{ f(\Station_i) \mid \Station_i \in S \} \).
Then for every station \(\Station_i\) and for all points \( p \notin S \), we
have \( \SINR_{\cA}(\Station_i, p) ~ = ~ \SINR_{f(\cA)}(f(\Station_i), f(p))
\).
\end{lemma}

\commabs

\section{SINR diagrams of nonuniform networks: Basics}
\label{section:Basic}
\subsection{Disconnectivity of nonuniform power SINR maps}
\commabsend
\commful
\dnsparagraph{Disconnectivity of nonuniform power SINR maps:}
\commfulend
\commabs
The SINR diagram $\ReceptionZone(\cA)$ is a central concept to this paper.
We are interested in gaining some basic understanding of its topology.
Specifically, we aim toward finding some ``niceness'' properties of
reception zones and studying their usability in algorithmic applications.
In previous work \cite{Avin2009PODC}, Avin et al. consider the simplified case
where all stations transmit with the same power. For a \UPN,
the reception zone of each station is known to be connected and to exhibit
some desirable properties such as fatness and convexity.
In the current work we study the general (and common) case of
non-uniform transmission powers.
\commabsend
\commful
We study SINR diagrams with non-uniform transmission powers.
\commfulend
%
%
\commabs
\subsection{2-Station networks}
\label{subsection:2Stations}
This section provides a detailed characterization of the possible SINR
diagrams in a system with two stations.
Let $\cA = \langle d,\{\Station_1, \Station_2\}, (\Power_1,\Power_2),
\Noise, \beta, 2\alpha \rangle$ be a network consisting of two
stations \(\Station_1, \Station_2\)
embedded on the $x$-axis
with transmitting powers $\Power_1,\Power_2>0$ respectively,
with a threshold parameter $\beta \geq 1$ and path-loss parameter $2\alpha>0$.
For clarity of presentation, we first assume the simplified case where
there is no background noise (i.e., \( \Noise = 0 \)). This is represented
by the network $\cA_{\Noise=0}$. The case of $\cA_{\Noise>0}$, corresponding to
\( \Noise> 0 \), is discussed at the end of this section.

Assume without loss of generality that \(\Station_1\) is
located at the origin and \(\Station_2\) is located at $(a,0, \ldots, 0)$,
where $a>0$. Recall that for a 2-station network with no background noise,
the SINR formula takes the form
\begin{eqnarray*}
\SINR_{\cA_{\Noise=0}}(\Station_1,p)&=&
\frac{\Power_1\cdot\dist{\Station_1,p}^{-2\alpha}}
{\Power_2\cdot\dist{\Station_2,p}^{-2\alpha}}~=~
\frac{\Power_1}{\Power_2}\cdot
\left(\frac{\dist{\Station_2,p}^{2}}{\dist{\Station_1,p}^{2}}\right)^\alpha.
\end{eqnarray*}
Assuming that $p=(x_{1}, \ldots, x_{d})\in\Reals^{d}$, the formula takes the form
\begin{eqnarray}
\label{eq:SINR formula}
\SINR_{\cA_{\Noise=0}}(\Station_1,p)&=&
\frac{\Power_1}{\Power_2}\cdot\left(
\frac{(x_1-a)^2+\sum_{j=2}^{d}x_{j}^{2}}{x_1^2+\sum_{j=2}^{d}x_{j}^{2}}\right)^\alpha.
\end{eqnarray}
As may be expected, the parameter controlling the behavior of the system
is the ratio $\PowerRatio=\beta\Power_2/\Power_1$.
When $\PowerRatio \neq 1$, define
$\Center =
\left(\frac{a}{1-\sqrt[\alpha]{\PowerRatio}},0 \ldots, 0\right)\in\Reals^{d}$
and
$R = \left|\frac{a\cdot\sqrt[2\alpha]{\PowerRatio}}
{1-\sqrt[\alpha]{\PowerRatio}}\right|$.

\begin{figure}[htb]
\begin{center}
\includegraphics[scale=0.3]{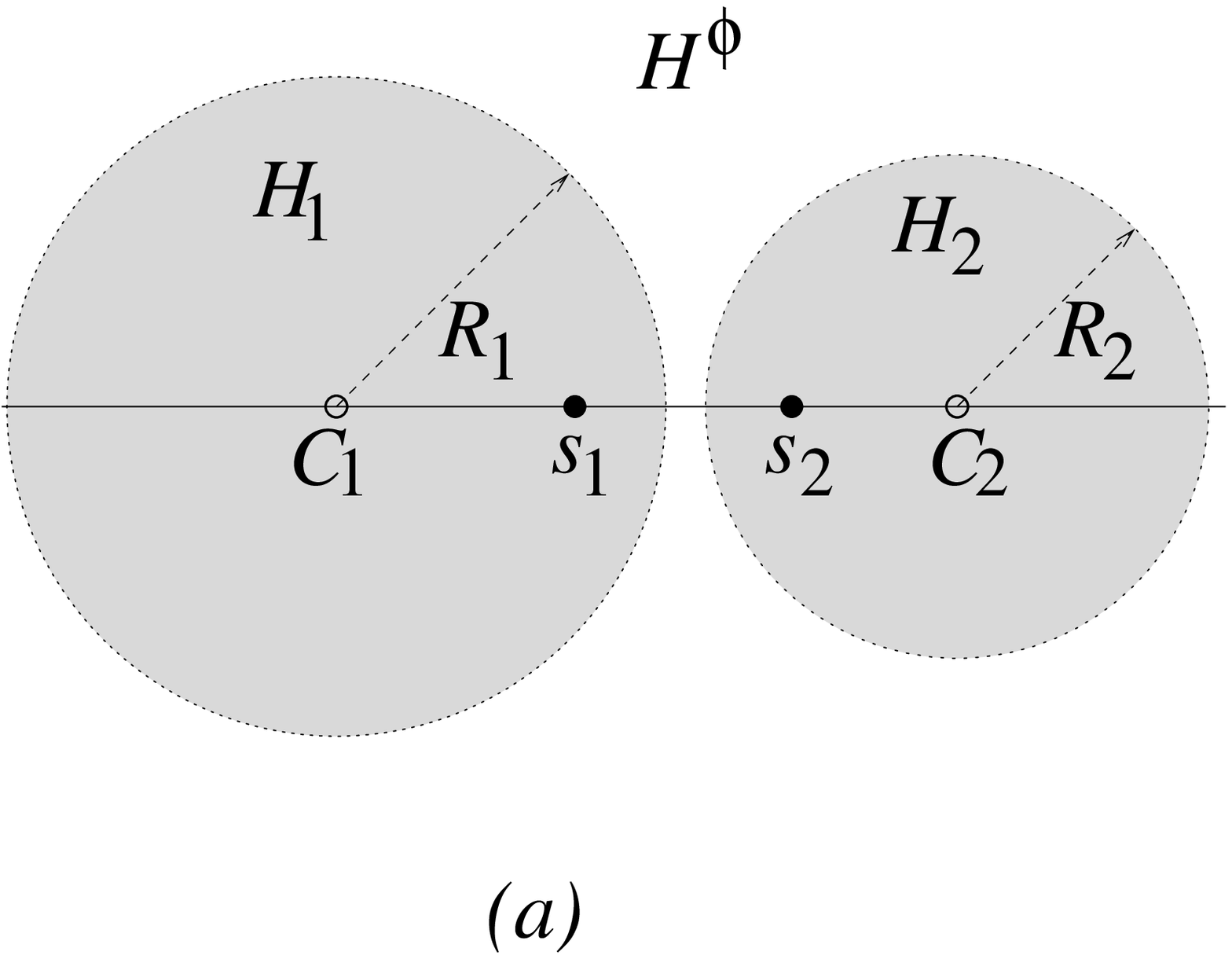}
\hfill
\includegraphics[scale=0.3]{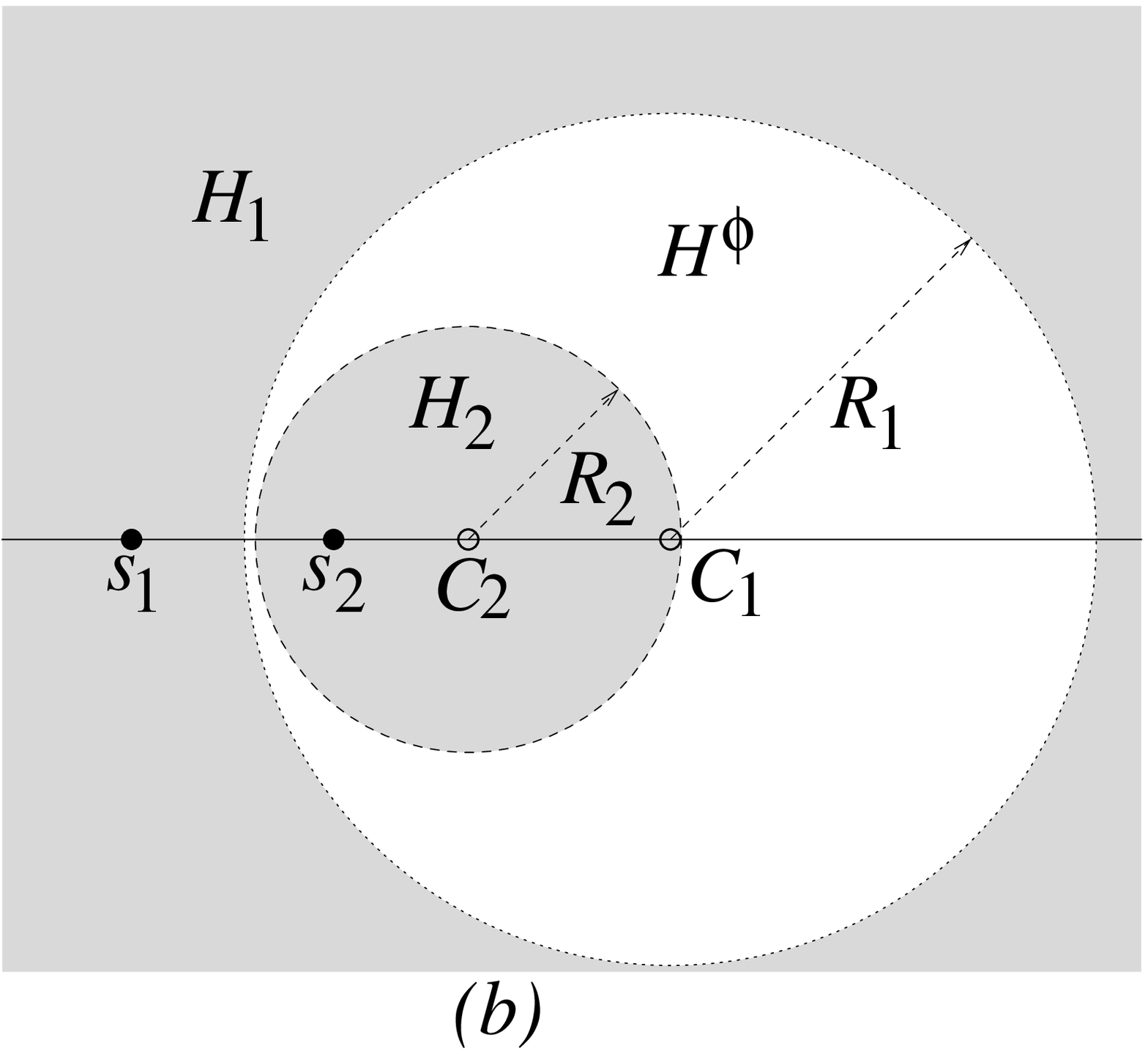}
\hfill
\includegraphics[scale=0.3]{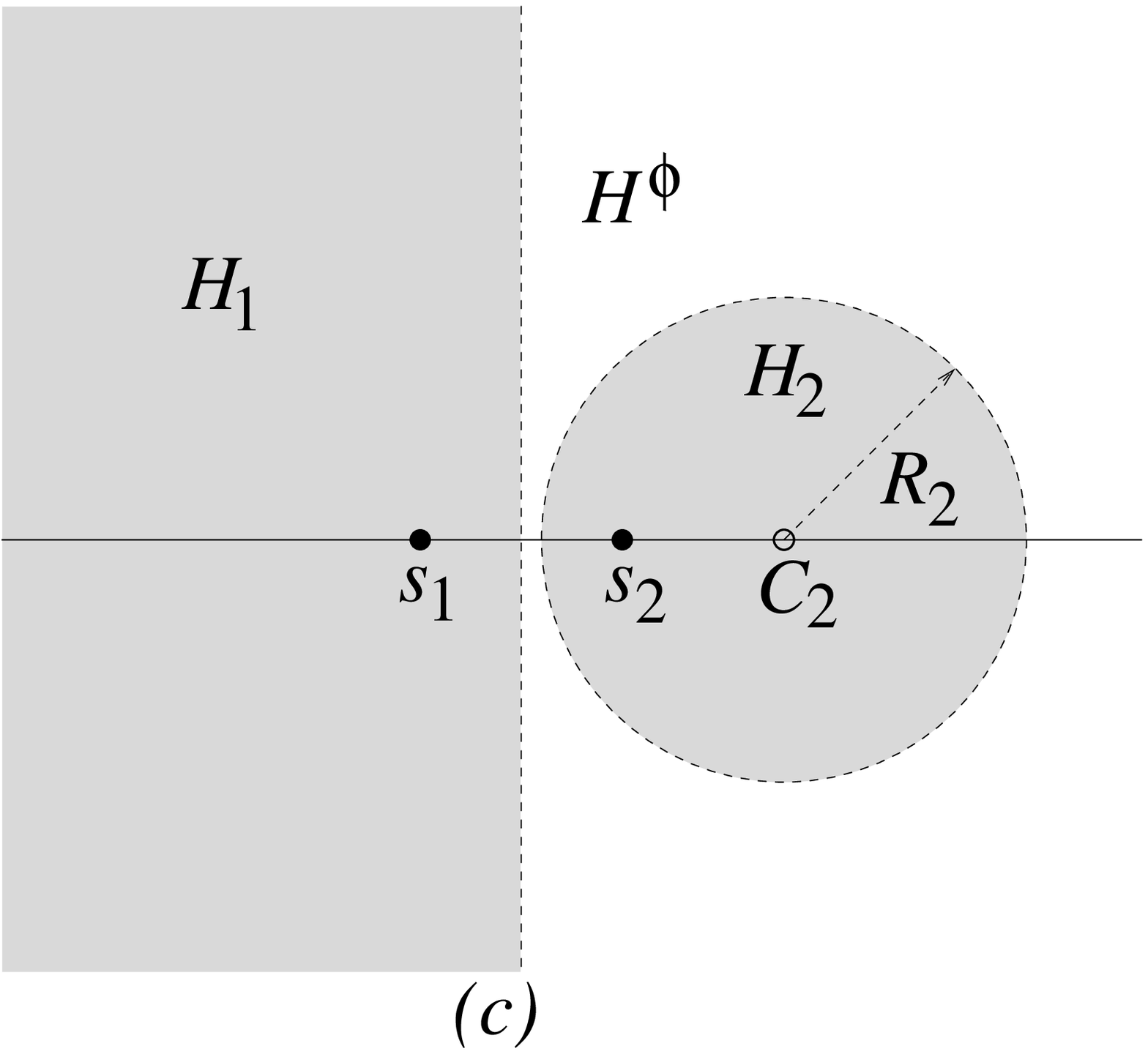}
\caption{ \label{figure:TwoStations}
\sf Possible configurations of a 2-stations network, where $\Station_1$
is located at the origin and $\Station_2$ is located at (0,2),
with the following energies and path loss parameter:
(a) $\Power_1=1.1$, $\Power_2=1$ and $\beta=2.1$:
$\PowerRatio>1$ and $\ReceptionZone_{1}(\cA_{\Noise=0})$
assumes configuration (C1),
(b) $\Power_1=2$, $\Power_2=1$ and $\beta=1.25$:
$\PowerRatio<1$ and $\ReceptionZone_{1}(\cA_{\Noise=0})$
assumes configuration (C2),
(c) $\Power_1=1.5$, $\Power_2=1$ and $\beta=1.5$:
$\PowerRatio=1$ and $\ReceptionZone_{1}(\cA_{\Noise=0})$
assumes configuration (C3).
}
\end{center}
\end{figure}

\begin{lemma}
\label{lemma: H1 structures}
The zone $\ReceptionZone_{1}(\cA_{\Noise=0})$ assumes one of the following
three possible configurations.
\begin{description}
\item{(C1)}
If $\PowerRatio>1$, then $\ReceptionZone_{1}(\cA_{\Noise=0})$ is
a $d$-dimensional disk, $\ReceptionZone_{1}(\cA_{\Noise=0})=\Ball^{d}(q,R)$.
\item{(C2)}
If $\PowerRatio<1$, then $\ReceptionZone_{1}(\cA_{\Noise=0})$ is
a complement of a $d$-dimensional disk,
$\ReceptionZone_{1}(\cA_{\Noise=0})=\Reals^d\setminus \Ball^{d}(q,R)$.
\item{(C3)}
If $\PowerRatio=1$, then $\ReceptionZone_{1}(\cA_{\Noise=0})$ is
a halfplane, $\ReceptionZone_{1}(\cA_{\Noise=0}) =
\left\{p=(x_1, \ldots, x_d)\in\Reals^{d} \mid x\leq a/2\right.\}$.
\end{description}
See Figure \ref{figure:TwoStations} for illustration assuming $d=2$.
\end{lemma}

\Proof
Eq. (\ref{eq:SINR formula}) implies that
$$\ReceptionZone_{1}(\cA_{\Noise=0}) ~=~
\{p=(x_1, \ldots, x_d)\in\Reals^{d} \mid \frac{\Power_1}{\Power_2}\cdot
\left(\frac{(x_1-a)^2+\sum_{j=2}^{d}x_{j}^{2}}
{x^2+\sum_{j=2}^{d}x_{j}^{2}}\right)^\alpha \geq \beta\}.$$
Letting  $A=\sqrt[\alpha]{\PowerRatio}$,
the condition on $p$ can be rewritten as
\begin{equation}
\label{eq:2stations_cond}
(x_1-a)^2+\sum_{j=2}^{d}x_{j}^{2} ~\geq~ A \cdot (x_1^2+\sum_{j=2}^{d}x_{j}^{2}).
\end{equation}
We begin with Claim (C3). If $A=1$, then condition (\ref{eq:2stations_cond})
can be written as
$(x_1-a)^2-x_1^2\geq 0$, implying $x_1\leq a/2$ and Claim (C3) follows.

Next, we prove (C1) and (C2). Assume that $A\not=1$.
We first rewrite condition (\ref{eq:2stations_cond}) in a circle form,
by rearranging it as $(1-A)(\sum_{j=1}^{d}x_{j}^{2})-2a x_{1}+a^2 \geq 0$, or
$$(1-A)\left(\left(x_1-\frac{a}{1-A}\right)^2+
\sum_{j=2}^{d}x_{j}^{2}\right)+a^2 - \frac{a^2}{1-A} ~\geq~ 0$$
$$\Leftrightarrow (1-A)\left(\left(x_1-\frac{a}{1-A}\right)^2+
\sum_{j=2}^{d}x_{j}^{2}\right) ~\geq~ \frac{a^2A}{1-A}~.$$
We consider two cases.

\noindent\textbf{Case 1:}
If $A=\sqrt[\alpha]{\PowerRatio}>1$, then
\begin{eqnarray*}
\ReceptionZone_{1}(\cA_{N=0}) &=&\left\{p=(x_1, \ldots, x_d)\in\Reals^d \mid
\left(x_1-\frac{a}{1-A}\right)^2+\sum_{j=2}^{d}x_{j}^{2} ~\leq~
\frac{a^2A}{(1-A)^2}\right\}
\\&=&
\left\{p=(x_1, \ldots, x_d)\in\Reals^d \mid \dist{\Center,p} ~\leq~
-\frac{a\cdot\sqrt[2\alpha]{\PowerRatio}}{1-\sqrt[\alpha]{\PowerRatio}}\right\}
~=~ \Ball^{d}(q,R).
\end{eqnarray*}
Hence the zone $\ReceptionZone_{1}(\cA_{\Noise=0})$ is composed of one cell
defined by a circle centered at $q$ of radius $R$. Claim (C1) follows.

\noindent\textbf{Case 2:}
If $A<1$, then
\begin{eqnarray*}
\ReceptionZone_{1}(\cA_{\Noise=0})&=&\left\{p=(x_1, \ldots, x_d)\in\Reals^{d} \mid
\left(x_1-\frac{a}{1-A}\right)^2+\sum_{j=2}^{d}x_{j}^{2} ~\geq~
\frac{a^2A}{(1-A)^2}\right\}
\\&=&
\left\{p=(x_1, \ldots, x_d)\in\Reals^{d} \mid \dist{\Center,p} ~\geq~
\frac{a\cdot\sqrt[2\alpha]{\PowerRatio}}{1-\sqrt[\alpha]{\PowerRatio}}\right\}
~=~ \Reals^{d}\setminus \Ball^{d}(q,R).
\end{eqnarray*}
Hence the zone $\ReceptionZone_{1}(\cA_{\Noise=0})$ is composed of
one cell defined by the complement of a circle centered at $q$ of radius $R$,
establishing (C2).
\QED

Finally, we turn to the case where $N > 0$.
It can be shown that $\ReceptionZone(\cA_{N >0})$ assumes three configurations
as well (see Lemma \ref{lemma: H1 structures}).
The key difference between $\ReceptionZone(\cA_{N >0})$ and
$\ReceptionZone(\cA_{N =0})$ is that the presence of noise induces only
bounded zones. Consequently, (C2) and (C3) where
$\ReceptionZone_{1}(\cA_{N =0})$ is unbounded, are no longer feasible.
These configurations are replaced by an equivalent ones where
$\ReceptionZone_{1}(\cA_{N =0})$ attains a bounded shape
(i.e., a large enclosing disc for (C2) and an elliptic shape for (C3)).
\commabsend

%
%

\begin{figure}[htb]
\begin{center}
\includegraphics[scale=0.6]{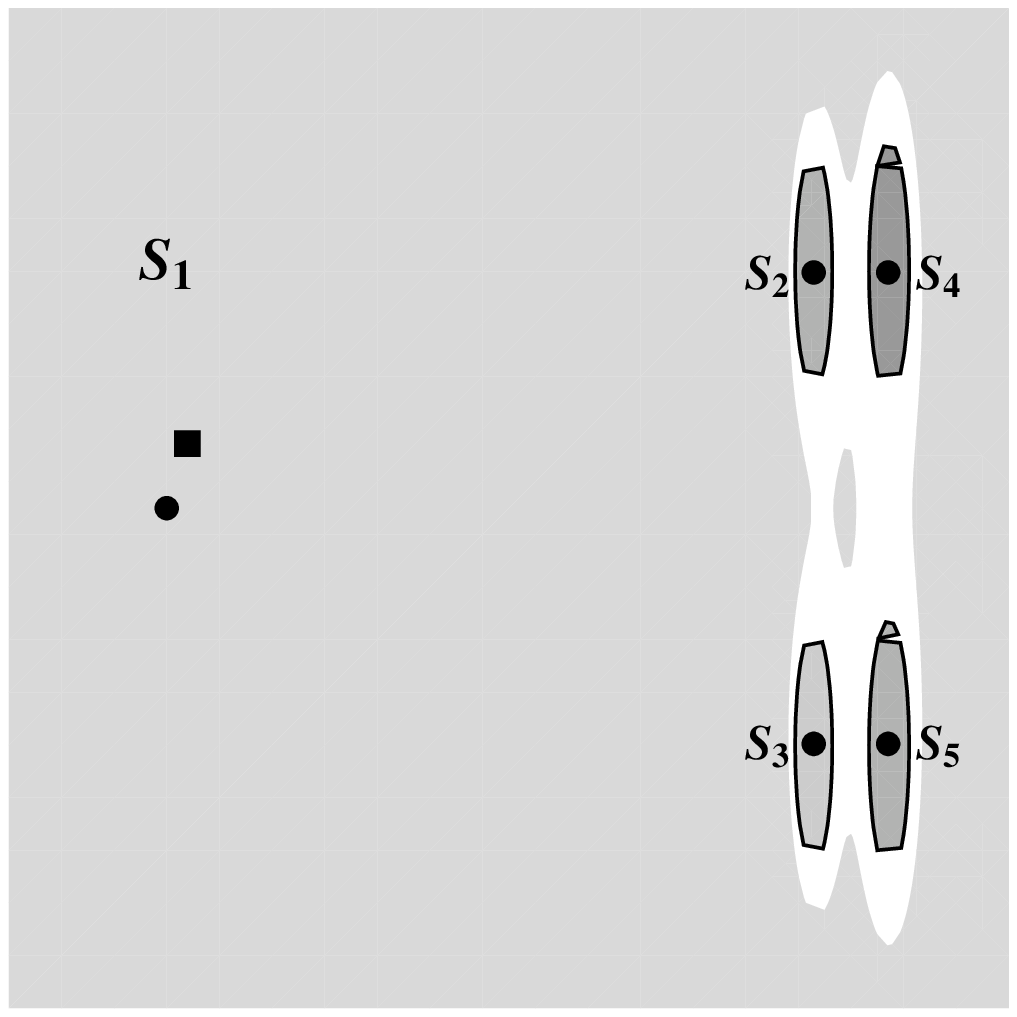}
\caption{ \label{figure:FiveStations}
\sf An instances of 5-station system with two connected cells
of $\ReceptionZone_{1}$.
}
\end{center}
\end{figure}

By considering a 2-station network with non-uniform power
\commful
(see Appendix \ref{subsection:2Stations}),
\commfulend
it is apparent that the reception zones of \NUPN{}s are not convex,
however connectivity is maintained.
Unfortunately, although this is true for 2-stations systems, it does not hold
in general. Connectivity might be broken even in networks with small number
of participants, as illustrated by the 5-station system
of Figure \ref{figure:FiveStations}, where the reception zone of $\Station_1$
is composed of two connected cells. This raises the immediate question of
bounding the maximal number of cells a given SINR diagram might have.

\commabs
\commabsend
A seemingly promising approach to studying this question is considering
the corresponding weighted Voronoi diagrams. Recall that by Lemma
\commful
\ref{cl:wv_sinr}(a),
\commfulend
\commabs
\ref{cl:wv_sinr},
\commabsend
$\ReceptionZone_{i}(\cA) \subseteq \wvor_{i}(\wvorsys_{\cA})$.
It therefore seems plausible that the number of weighted Voronoi cells
(bounded by $O(n^{2})$ \cite{AurenhammerE84}) might upper bound the number
of connected cells in the corresponding SINR diagram.
Unfortunately, this does not hold in general, since it might be the case
that a single weighted Voronoi cell corresponds to several connected
SINR cells. This phenomenon is formally stated in the following lemma.

\begin{lemma}
\label{cl:wv_sinr_ncell}
There exists a wireless network $\cA^{*}$ such that a given cell of
the corresponding weighted Voronoi diagram $\wvor(\wvorsys_{\cA^{*}})$
contains more than one cell of $\ReceptionZone(\cA^{*})$.
\end{lemma}
\commabs
\Proof
Let $\cA = \langle d, S, \Power, \Noise, \beta,\alpha \rangle$ be a wireless
network, where $S=\{\Station_1,...,\Station_{n}\}$ and $\ReceptionZone_{1}(\cA)$
is not connected, i.e., $\ReceptionZone_{1}(\cA)$ is composed of more than
one cell. Let \\
$\cA_{m}=\langle d, \cS_m, \WPower_m, \Noise, \beta,\alpha \rangle$,
where $\cS_m=\{\Station_1\}\cup \{\Station_2^1,...,\Station_2^m\} \cup...\cup\{\Station_{n}^1,...,\Station_{n}^m\}$, $\WPower_m=\{\wPower_1\}\cup\{\wPower_2^1,...,\wPower_2^m\}\cup...\cup\{\wPower_{n}^1,...,\wPower_{n}^m\}$, where $\wPower_1=\Power_1$ and $\wPower_i^l=\Power_i/m$, for every $i=2,...,n$ and $l=1,...,m$. To avoid cumbersome notation, let $\wvorsys_{m}=\wvorsys_{\cA_{m}}$ and let $\wvor(\wvorsys_{m})$ be the corresponding weighted Voronoi diagram of $\cA_{m}$.
In what follows, we show that for sufficiently large $m^{*}$, the network $\cA_{m^{*}}$ satisfies the conditions of the desired network $\cA^{*}$. Specifically, it is easy to verify that for large enough $m^{*}$, the weighted zone $\wvor_{1}(\wvorsys_{m^{*}})$ is connected. We next show that $\wvor_{1}(\wvorsys_{m^{*}})$ contains more than one connected cell of $\ReceptionZone_{1}(\cA_{m^{*}})$.  First, observe that $\ReceptionZone_{1}(\cA)= \ReceptionZone_{1}(\cA_{m^{*}})$, and therefore $\ReceptionZone_{1}(\cA_{m^{*}})$ is not connected as well. This follows by noting that $\Energy_\cA(\Station_1,p)=\Energy_{\cA_{m^{*}}}(\Station_1,p)$ and $\Interference_\cA(S \setminus \{\Station_1\},p)=\Interference_{\cA_{m^{*}}}(\cS_{m^{*}} \setminus \{\Station_1\},p)$.
Next, by the connectivity of $\wvor_{1}(\wvorsys_{m^{*}})$ and Lemma
\ref{cl:wv_sinr},
it follows that
$\ReceptionZone_{1}(\cA_{m^{*}}) \subseteq \wvor_{1}(\wvorsys_{m^{*}})$.
Since $\ReceptionZone_{1}(\cA_{m^{*}})$ is not connected, the lemma follows.
\QED
\commabsend

\commabs
This lemma illustrates that the structural complexity of the SINR diagram
cannot be fully captured by the weighted Voronoi diagram. Specifically,
it implies that the number of connected cells in a non-uniform SINR diagram
cannot be bounded by the number of weighted Voronoi cells,
hence a different approach is needed. This challenge is extensively discussed
in this paper, where we obtain bounds and provide extreme constructions
with respect to the the number of connected cells for a given station.
We conjecture that the obtained upper bounds are not tight, and our
constructions are close to the limit. Yet so far, no formal proof is available.
\commabsend

\section{The No-free-hole property}
\label{section:free-convex}
\begin{figure}[htb]
\begin{center}
\includegraphics[scale=0.3]{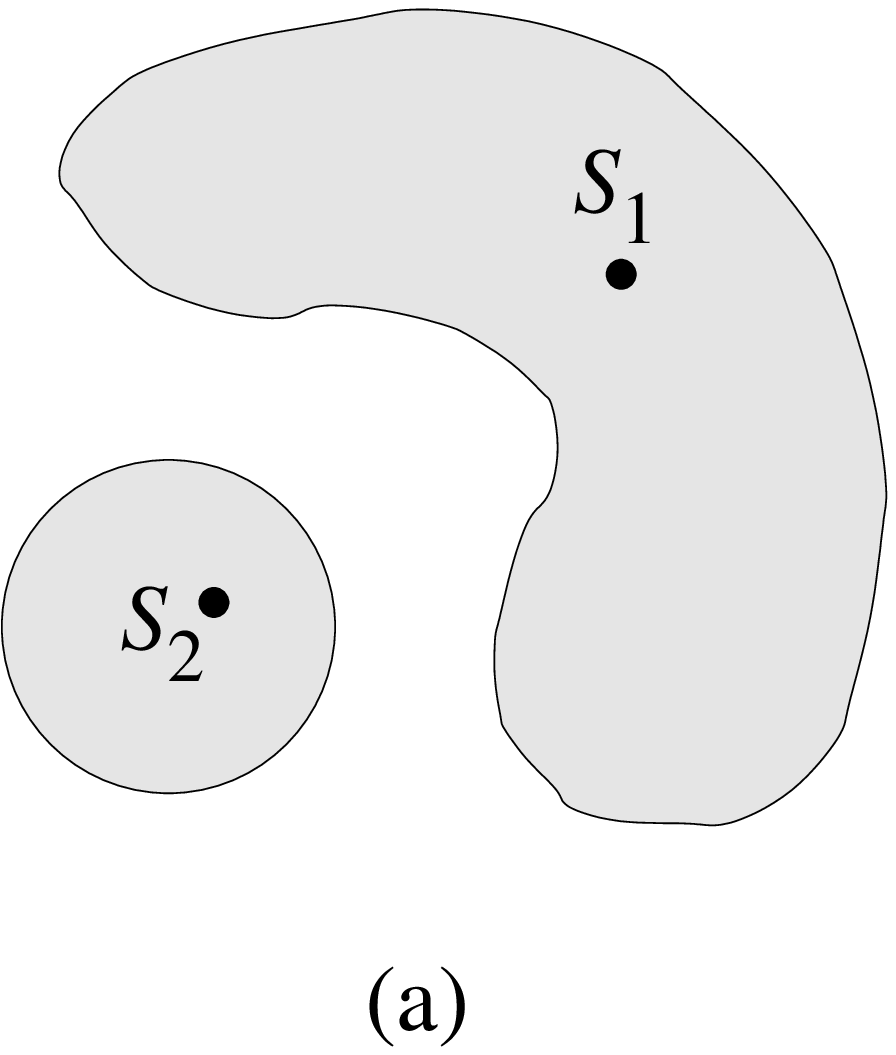}
\hfill
\includegraphics[scale=0.3]{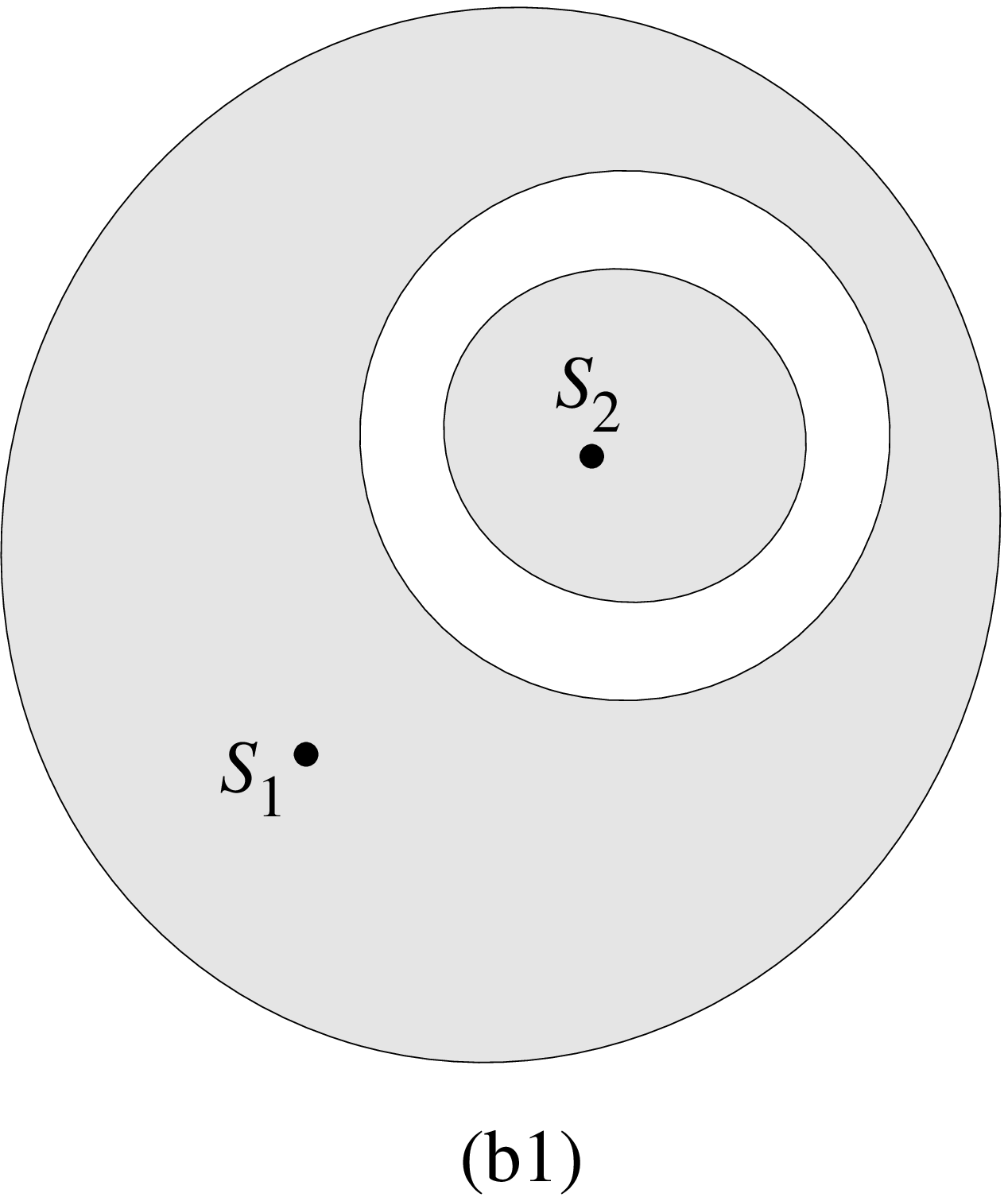}
\hfill
\includegraphics[scale=0.3]{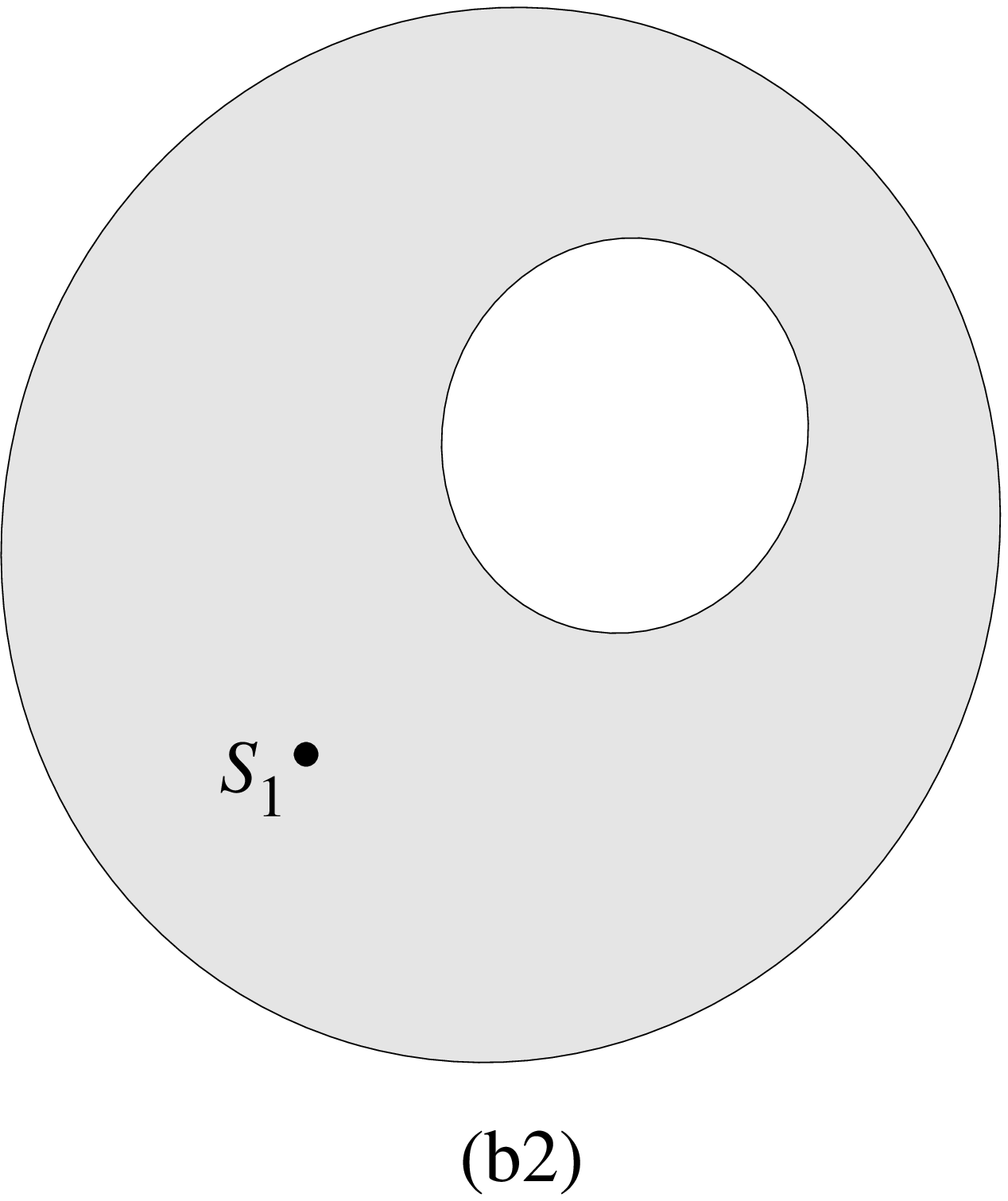}
\caption{ \label{figure:NonConvexshapes}
\sf Classification of non-convex cells.
(a) Non-convex contour;
(b1) Convex contour with a hole that occupied by some interfering station;
(b2) Convex contour with a hole that is free of stations.
}
\end{center}
\end{figure}

Convexity was shown in \cite{Avin2009PODC} to play a significant role in
showing that the reception zones of uniform SINR diagrams are connected.
Unfortunately, as discussed in the previous section, reception zones of
non-uniform SINR diagrams might be non-convex, even when the network is
composed of only two stations. Is there any form of weaker convexity that can
still be established?
Are there excluded configurations in non-uniform diagrams?
To address these questions, let's examine several examples of non-convex
shapes illustrated in Figure \ref{figure:NonConvexshapes}.
Non-convex shapes can be classified into two types:
(a) shapes with non-convex contour (Fig. \ref{figure:NonConvexshapes}a),
(b) shapes with a convex contour but with a hole.
Type (b) is further classified into two types;
(b1) the hole contains at least one interfering station
(Fig. \ref{figure:NonConvexshapes}b1) and
(b2) the hole is free of stations (Fig. \ref{figure:NonConvexshapes}b2).
Interestingly, though type (a) and (b1) are fairly common feasible
configurations of cells in non-uniform SINR diagrams, all our attempts to
generate a configuration of type (b2) have failed so far.
We conjecture that type (b2) is an excluded configuration of cells in
non-uniform SINR diagrams. In other words, we believe that every hole
in a reception cell must contain at least one interfering station.
This property (namely, that type (b2) is an excluded state) is hereafter
termed ``no-free-hole" or NFH for short, and is defined as follows.
A collection of closed shapes $\cC$ in $\R^{d}$ obeys the
{\em NFH property} with respect to a set $S$ of stations
if for every $C\in\cC$ that is free of stations, if all its border points are
reception points of $s_1$, then all points of $C$ are reception points as well.
Formally, if $C \cap S = \emptyset$ and
$\Boundary (C) \subseteq \ReceptionZone_1(\cA)$, then also
$C \subseteq \ReceptionZone_1(\cA)$.
This property turns out to be relevant for bounding the number of connected
cells in $\ReceptionZone(\cA)$. The next subsection is dedicated to proving
the conjecture in the 1-dimensional case.

\subsection{The one-dimensional case}
\label{subsec:freeconvexity_1d}
\commful
\dnsparagraph{The NFH property:}
We consider a closed one-dimensional shape given by a bounded segment $\segment=[a,b]$ such that $\Boundary (\segment)={a,b} \in \ReceptionZone_{1}(\cA)$ and $\segment \cap S = \emptyset$.
The purpose of this subsection is to show that $\segment$ follows the NFH property and use this fact to bound the number of connected cells in $\ReceptionZone(\cA)$. In what follows, we assume $\alpha=2$.
\commfulend
\commabs
The purpose of this subsection is to show that the NFH property
holds in the one dimensional case. This fact is later used in
Subsection \ref{subsection:NZones-1D} to bound the number of connected cells
in a one dimensional map. The analysis is organized as follows.
In Subsection \ref{subsubsec:1d_framework}, we introduce the framework
and establish some basic properties.
In Subsection \ref{subsubsec:1d_3stations}, we establish the NFH
property for 3-station network and no background noise.
In Subsection \ref{subsubsec:1d_nstations}, the NFH property
for $n$-station network either with or without background noise is established.

\subsubsection{Framework}
\label{subsubsec:1d_framework}

We consider a network of the form \( \cA = \langle d=1, S, \Power, \Noise, \beta \geq 1, \alpha=2 \rangle \). Let $x_{i}$ be the position of $\Station_i$. In the following we may abuse notion by confusing between a station and its geometric location. Without loss of generality, we focus on $\Station_1$.
\commabs
By adopting Eq. (\ref{eq:Def-SINR}) to the current setting we get that
\begin{equation}
\label{eq:1d_SINR}
\SINR_{\cA}(\Station_1,x) ~=~
\frac{\Power_1/(x-x_1)^2}{\sum\limits_{i=2}^n \Power_i/\left(x-x_i\right)^2+N}
\end{equation}
and $\Station_1$ is heard at $x$ iff $\SINR_{\cA}(\Station_1,x) \geq \beta$.
\commabsend
To establish NFH, we consider a segment
$\segment=[a,b] \subseteq \ReceptionZone_{1}(\cA)$ such that
$\Boundary (\segment)=\{a,b\}$ where $a,b\in \ReceptionZone_{1}(\cA)$ and
$\segment \cap S = \emptyset$, and show that under these conditions
$\segment \subseteq \ReceptionZone_{1}(\cA)$.
Continuity follows simply by the fact that no station is located on $\segment$.
We need to show that $p\in \ReceptionZone_{1}(\cA)$ for every $p \in [a,b]$.
\commabsend

\commabs
First, we provide some notation useful for this section.
Let $q_{i}(x)=\Power_i ((x-x_1)/(x-x_{i}))^{2}$ and let
$q_{\Noise}=\Noise (x-x_1)^{2}$. Then using Equation (\ref{eq:1d_SINR}),
the fundamental rule of the $\SINR$ model of can be expressed
by the 1-variate polynomial
\begin{equation}
\label{eq:1d_polynomial}
Q_{\cA}(x) ~=~ \sum\limits_{i=2}^n q_{i}(x) +q_{\Noise} -\Power_1/\beta~,
\end{equation}
such that $\Station_1$ is heard iff $Q_{\cA}(x) \leq 0$.
The first derivative of $Q_{\cA}(x)$ is given by
\begin{equation}
\label{eq:1d_poly_firstd}
\frac{\partial Q_{\cA}(x)}{\partial x} ~=~
\sum\limits_{i=2}^n \frac{\partial q_{i}(x)}{\partial x} +2\Noise (x-x_1) ~=~
2(x-x_1) \left( \Noise+ \sum\limits_{i=2}^n
\frac{x_1-x_{i}}{(x-x_{i})^{3}} \right)~.
\end{equation}
In the same manner, the second derivative of $Q_{\cA}(x)$ is given by
\begin{equation}
\label{eq:1d_poly_second_deriv}
\frac{\partial^{2} Q_{\cA}(x)}{\partial x^{2}} ~=~
\sum\limits_{i=2}^n \frac{\partial^{2} q_{i}(x)}{\partial x^{2}}+2 \Noise ~=~
2\sum\limits_{i=2}^n \frac{ (2 x - 3 x_1 + x_i) (x_i-x_1)} {(x-x_{i})^{4}}+2 \Noise~.
\end{equation}

Without loss of generality, let $a >x_{1}$. Our analysis relies on
the observation that a network \( \cA\) may assume one of the following
three configurations:\\
(C1) $x_{i} \geq x_{1}$, for every $\Station_{i} \in S$.\\
(C2) $\left(S \setminus \{\Station_1\}\right) \cap [x_{1},a] = \emptyset$.\\
(C3) $x_{j} < x_{1}$ for some $\Station_j \in S$ and
$\left(S \setminus \{\Station_1\}\right) \cap [x_{1},a] \neq \emptyset$.\\
The NFH property follows easily for networks in configuration (C1)
and (C2). In case the network $\cA$ assumes configuration (C3),
the proof is more involved.
We begin with configurations (C1) and (C2).
\begin{claim}
\label{claim:1d_polynomial_no_max_C1_C2}
If the network \( \cA = \langle d=1, S, \Power, \Noise\geq0, \beta, \alpha=2 \rangle \)
assumes configuration (C1) or (C2), then $Q_{\cA}(x)$ has no local maximum
in the interval $[a,b]$.
\end{claim}
\Proof
Without loss of generality, let $x_1=0$. Suppose first that \( \cA\) is in
configuration (C1). Since $x_{i} >0$ for every $i \in \{2, \ldots ,n\}$,
it follows by Eq. (\ref{eq:1d_poly_second_deriv}) that
$\partial^{2} Q_{\cA}(x)/\partial x^{2}>0$ for every $x>0$ and specifically
for every $p \in [a,b]$, which establishes the claim.
Next, assume that \( \cA\) is in configuration (C2). In this case,
the set of stations $S \setminus \{\Station_{1}\}$ can be partitioned into
two sets, namely, $S^{b^{+}}=\{\Station_i \mid i>1, x_{i} >b\}$ and
$S^{-}=\{\Station_i \mid i>1, x_{i} <0 \}$. Since $(x-x_{i})^{3}<0$
for every $\Station_i \in S^{b^{+}}$ and $x \in [a,b]$, we have that
$\partial q_{i}(x)/\partial x>0$ for every $x>0$ and $\Station_i \in S^{b^{+}}$.
In addition, one can verify that $\partial q_{i}(x)/\partial x>0$ for every
$x \in [a,b]$ and $\Station_i \in S^{-}$. In summary, we get that
$\partial Q_{\cA}(x)/\partial x > 0$ for every $p \in [a,b]$,
and the claim follows.
\QED

\begin{corollary}
\label{cor:1d_ClosedShape_NS_C1_C2}
If $\cA$ assumes configuration (C1) or (C2), then
$\segment \subseteq \ReceptionZone_{1}(\cA)$.
\end{corollary}
\Proof
By Eq. (\ref{eq:1d_polynomial}) it holds that $Q_{\cA}(a), Q_{\cA}(b) \leq 0$.
By Claim \ref{claim:1d_polynomial_no_max_C1_C2}, we have that $Q_{\cA}(x)$ has no local maximum in the interval $\segment=[a,b]$, hence for every $p \in \segment$ we have $Q_{\cA}(p) \leq \max \{Q_{\cA}(a),Q_{\cA}(b)\} \leq 0$, implying $p \in \ReceptionZone_{1}(\cA)$, and the claim follows.
\QED

\subsubsection{3-stations (no noise)}
\label{subsubsec:1d_3stations}
\begin{figure}[htb]
\begin{center}
\includegraphics[scale=0.45]{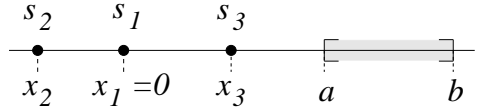}
\caption{ \label{fig:C3configuraion}
\sf An instances of a 3-station wireless network $\cA_3$ in configuration (C3).
}
\end{center}
\end{figure}

We now establish NFH for the special case of a \NUPN{}
with three stations and no background noise,
\( \cA_{3} = \langle d=1, S_{3}=\{\Station_1, \Station_2,\Station_3\}, \Power,
\Noise=0, \beta, \alpha=2 \rangle \).
Assume without loss of generality (by Lemma \ref{lemma:Transformation}) that
$x_{1}=0$, $x_{2} < x_{3}$ and $0<a<b$. Let
$S^{-}=\{\Station_{i} \in S_{3} \mid x_{i} < 0\}$,
$S^{b^{+}}=\{\Station_{i} \in S_{3} \mid x_{i} >b\}$ and
$S^{a^{-}}=\{\Station_{i} \in S_{3} \mid 0 < x_{i} < a\}$.
\begin{lemma}
\label{lem:1d_ClosedShape_3S}
Let $\segment=[a,b]$ be a segment such that
$a, b \in \ReceptionZone_{1}(\cA_{3})$ and $\segment \cap S =\emptyset$.
Then $\segment \subseteq \ReceptionZone_{1}(\cA_{3})$.
\end{lemma}
\Proof
Due to Corollary \ref{cor:1d_ClosedShape_NS_C1_C2}, it remains to consider
the case where $\cA_{3}$ assumes configuration (C3), i.e., $x_{2}<0$ and
$0 \leq x_{3} \leq a$, see Figure \ref{fig:C3configuraion}.
In this context, it is convenient to consider for $\ReceptionZone_{1}(\cA_{3})$
the following characterizing polynomial
$$P(x) ~=~
\Power_1(x-x_2)^2(x-x_3)^2- \Power_2x^2(x-x_3)^2 - \Power_3x^2(x-x_2)^2.$$
It follows that $P(x)\geq 0$ if and only if $x\in\ReceptionZone_{1}(\cA_{3})$.
Since $\deg(P(x))=4$, the polynomial $P(x)$ has at most 4 roots.
The claim follows by applying a counting argument on the number of roots
of $P(x)$. Clearly, a root of $P(x)$ is consumed whenever the polynomial
changes its sign. Since $P(x_2), P(x_3) >0$, $P(0)<0$ and $P(a),P(b) \leq 0$,
it follows that $P(x)$ has roots in each of the intervals $[x_{2}, 0]$,
$[0,x_{3}]$ and $[x_{3},a]$. Consequently, we are left with one undecided root.

We claim that every point $p \in \segment$ is a reception point of
$\Station_1$. This is proven by contradiction.
Assume to the contrary, that there is a non-reception point $q \in [a,b]$
such that $q \notin \ReceptionZone_{1}(\cA_{3})$ or $P(q)>0$. This would imply
the existence of two roots that correspond to the intervals $[a,q]$
and $[q,b]$. First, assume $P(a)<0$. Then the roots of each interval mentioned
do not overlap. We end with five roots which is infeasible by degree
consideration, a contradiction.
Else, assume $P(a)=0$ and that there is a non-reception point $q\in [a,b]$.
In this case there is one additional root (besides $a$) in the interval
$[a,b]$, and all roots of the polynomial are assigned.
As we assigned all four roots, it follows that $P(\infty)<0$
and $P(-\infty)>0$. By the fact that $P(\infty)<0$ it follows that
$\Power_1 \geq \Power_2+\Power_3$. But then it should also follow that
$P(-\infty)<0$ and we end with contradiction again. The claim follows.
\QED
\subsubsection{$n$-Stations}
\label{subsubsec:1d_nstations}

In this subsection, we extend Corollary \ref{cor:1d_ClosedShape_NS_C1_C2} and
Lemma \ref{lem:1d_ClosedShape_3S} and show that the NFH property
holds for any $n$-station network for $\Noise \geq 0$.
\commabsend

\commful
We next establish the following lemma, extending
Lemma \ref{lem:1d_ClosedShape_3S}.
\commfulend

\begin{figure}[htb]
\begin{center}
\includegraphics[width=0.5\textwidth]{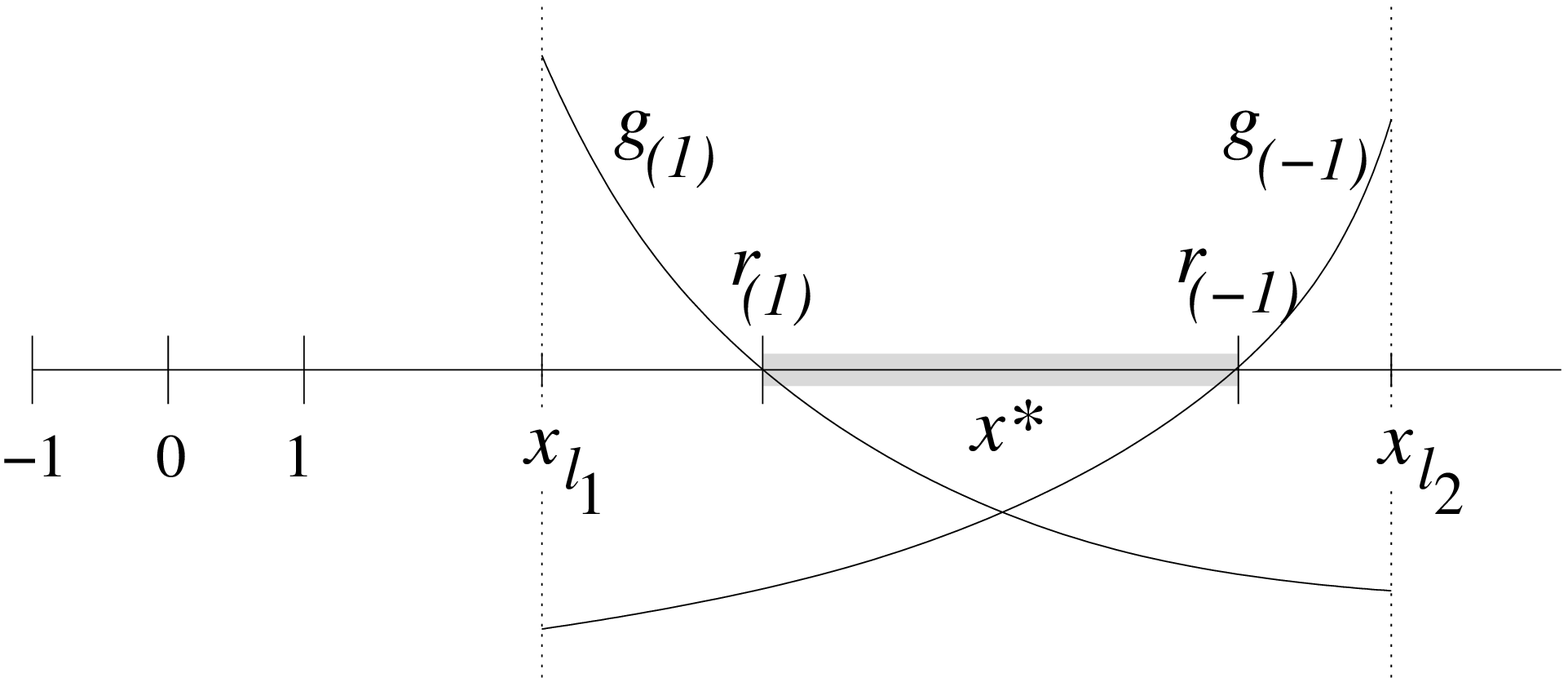}
\caption{ $\diffEnergyp(x)$ is monotonically decreasing and $\diffEnergym(x)$
is monotonically increasing in $[x_{l_1},x_{l_2}]$. In addition,
$\diffEnergyp(x_{l_1})>0$, $\diffEnergym(x_{l_1})<0$, $\diffEnergyp(x_{l_2})<0$ and
$\diffEnergym(x_{l_2})>0$. The case where $\rootp<\rootm$ implies that
$\diffEnergyp(x^*)<0$ and $\diffEnergym(x^*)<0$, for every
$x^*\in[\rootp,\rootm]$.
\label{figure:fisMonoton} }
\end{center}
\end{figure}

\begin{lemma}
\label{lem:1d_ClosedShape_nS}
Let $\segment=[a,b]$ be a segment, such that
$a, b \in \ReceptionZone_{1}(\cA)$ and $\segment \cap S =\emptyset$.
Then $\segment \subseteq \ReceptionZone_{1}(\cA)$.
\end{lemma}
\commabs
\Proof
Due to Corollary \ref{cor:1d_ClosedShape_NS_C1_C2}, it is sufficient to consider the case where $\cA$ assumes configuration (C3). For ease of computation, we assume without loss of generality that $a=-1$ and $b=1$ and the position of $\Station_1$ is arbitrary. We first note that since the SINR function is continuous on $\segment$, it is sufficient to consider the middle point of this section, $t=0$ and show that $t$ is in $\ReceptionZone_{1}(\cA)$, as the same argument can be re-applied to the segments $[a,t]$ and $[t,b]$ and so on. Lemma \ref{lem:1d_ClosedShape_nS} is proved by induction on the number of stations in the network, $n=|S|$. For the base of the induction, we consider the case $\cA$ assumes configuration (C1) or (C2) (indeed any 2-station network assumes one of these two configurations). The lemma clearly holds in this case due to Corollary \ref{cor:1d_ClosedShape_NS_C1_C2}. Next, we assume that the lemma holds for any $n-1$-station networks (in particular, networks in configuration (C3)). The induction step is more involved.
Let $\Station_{j},\Station_{k}$ be two stations positioned on the same side with respect to segment $\segment$ (since $\cA$ assumes (C3) such two stations are guaranteed to exist).
By Lemma \ref{lemma:Transformation} we may assume, without loss of generality, that
these stations are to the right of $b$. Note that since the position of $\Station_1$ is arbitrary, generality is indeed maintained. For clarity, let $\Station_{l_{1}}=\Station_{j}$ and $\Station_{l_{2}}=\Station_{k}$ where $x_{l_{1}} < x_{l_{2}}$. Informally, we prove the inductive step by showing that these two stations can be replaced by a single station $\Station^{*}$, resulting in a $(n-1)$-station network $\cA_{n-1}$ with set of stations $S_{n-1}$, that satisfies the following conditions:
\begin{enumerate}
\item
the interference experienced by a receiver located at $t=0$ is maintained, i.e., $\Interference_{\cA_{n-1}}(S_{n-1} \setminus \{\Station_1\},t)=\Interference_{\cA}(S \setminus \{\Station_1\},t)$; and
\item
the interference at segment endpoints does not increase, which guarantees that $a,b \in \ReceptionZone_{1}({\cA_{n-1}})$.
\end{enumerate}
Due to the inductive hypothesis for $(n-1)$-station networks, it then follows that $t \in \ReceptionZone_{1}({\cA_{n-1}})$, hence $t \in \ReceptionZone_{1}({\cA})$.
The next claim expresses this more formally.
\begin{claim}
\label{cl:1d_reduc_step_3S}
There exists a network
$$\cA^*_{n-1} ~=~
\langle d=1, S^*_{n-1} = (S \setminus\{\Station_{l_{1}},\Station_{l_{2}}\})
\cup \{\Station^{*}\}, (\Power \setminus\{\Power_{l_{1}},\Power_{l_{2}}\})
\cup \{\Power^{*}\},N,\beta \geq 1,\alpha=2 \rangle$$
such that:\\
(1)~$\Interference_{\cA^*_{n-1}}(S^*_{n-1} \setminus \{\Station_1\},0)=\Interference_{\cA}(S \setminus \{\Station_1\},0)$;\\
(2)~$\Interference_{\cA^*_{n-1}}(S^*_{n-1} \setminus \{\Station_1\},q) \leq \Interference_{\cA}(S \setminus \{\Station_1\} ,q)$ for $q \in \{a,b\}.$
\end{claim}
\Proof
Recall that $a=-1$ and $b=1$. Let
\begin{equation}
\label{eq:1d_power_induc}
\Power(x) ~=~ x^{2} \cdot
\Energy_{\cA}(\{\Station_{l_{1}},\Station_{l_{2}}\},0),
~\mbox{for}~ x \in [x_{l_{1}},x_{l_{2}}].
\end{equation}
Consider a station $\Station$ with position $x$ and transmission energy $\Power(x)$. By replacing the stations $\{\Station_{l_{1}},\Station_{l_{2}}\}$ by $s$, we get the $(n-1)$-station network
\begin{equation*}
\cA_{n-1}(x) ~=~ \langle d=1,S_{n-1}=(S \setminus\{\Station_{l_{1}},\Station_{l_{2}}\})
\cup \{\Station\}, (\Power \setminus\{\Power_{l_{1}},\Power_{l_{2}}\})
\cup \{\Power(x)\},N,\beta \geq 1,\alpha=2 \rangle.
\end{equation*}
It is easy to verify that $\Interference_{\cA_{n-1}(x)}(S_{n-1} \setminus \{\Station_1\},0) =\Interference_{\cA}(S \setminus \{\Station_1\},0)$ for every $x \in [x_{l_{1}},x_{l_{2}}]$, which establishes condition (1). Consider condition (2). We show that there exists $x^{*} \in [x_{l_{1}},x_{l_{2}}]$ such that $\cA_{n-1}(x^{*})$ satisfies condition (2) (in addition to Condition (1) that is satisfied for any $x \in [x_{l_{1}},x_{l_{2}}]$).
Let
\begin{eqnarray*}
\diffEnergyp(x)&=&\Interference_{\cA_{n-1}(x)}(S_{n-1} \setminus \{\Station_1\},1)-\Interference_{\cA}(S \setminus \{\Station_1\},1)
~~\mbox{and}\\
\diffEnergym(x)&=&\Interference_{\cA_{n-1}(x)}(S_{n-1} \setminus \{\Station_1\},-1)-\Interference_{\cA}(S \setminus \{\Station_1\},-1),
\end{eqnarray*}
for every $x \in [x_{l_{1}},x_{l_{2}}]$. We next show that there exists $x^{*} \in [x_{l_{1}},x_{l_{2}}]$ such that $\diffEnergyp(x^{*}) \leq 0$ and $\diffEnergym(x^{*}) \leq 0$.\\
Rearranging, we get that,
\begin{eqnarray}
\diffEnergyp(x)&=&\Interference_{\cA_{n-1}(x)}(S_{n-1} \setminus \{\Station_1\},1)-\Interference_{\cA}(S \setminus \{\Station_1\},1)\nonumber\\
&=&
\Energy_{\cA_{n-1}(x)}(\{\Station\},1)-\Energy_{\cA}(\{\Station_{l_{1}},\Station_{l_{2}}\},1) \nonumber
\\&=&
\left(\frac{x}{x-1} \right)^{2} \cdot \Energy_{\cA}(\{\Station_{l_{1}},
\Station_{l_{2}}\},0)-\Energy_{\cA}(\{\Station_{l_{1}},\Station_{l_{2}}\},1)~.
\label{eqn:1d_diffEnergyp}
\end{eqnarray}
In the same manner,
\begin{eqnarray}
\diffEnergym(x)&=&
\left(\frac{x}{x+1} \right)^{2} \cdot \Energy_{\cA}(\{\Station_{l_{1}},
\Station_{l_{2}}\},0)-\Energy_{\cA}(\{\Station_{l_{1}},\Station_{l_{2}}\},-1)~.
\label{eqn:1d_diffEnergym}
\end{eqnarray}
Note that $(x/x-1)^{2}$ is monotonically decreasing in the segment $(1,\infty)$, and $(x/x+1)^{2}$ is monotonically increasing in $(1,\infty)$. Correspondingly, $\diffEnergyp(x)$ is monotonically decreasing and $\diffEnergym(x)$ is monotonically increasing in the range of their definition, $[x_{l_{1}},x_{l_{2}}]\subseteq(1,\infty)$, as illustrated in Figure \ref{figure:fisMonoton}.

\begin{figure}[htb]
\begin{center}
\includegraphics[width=0.5\textwidth]{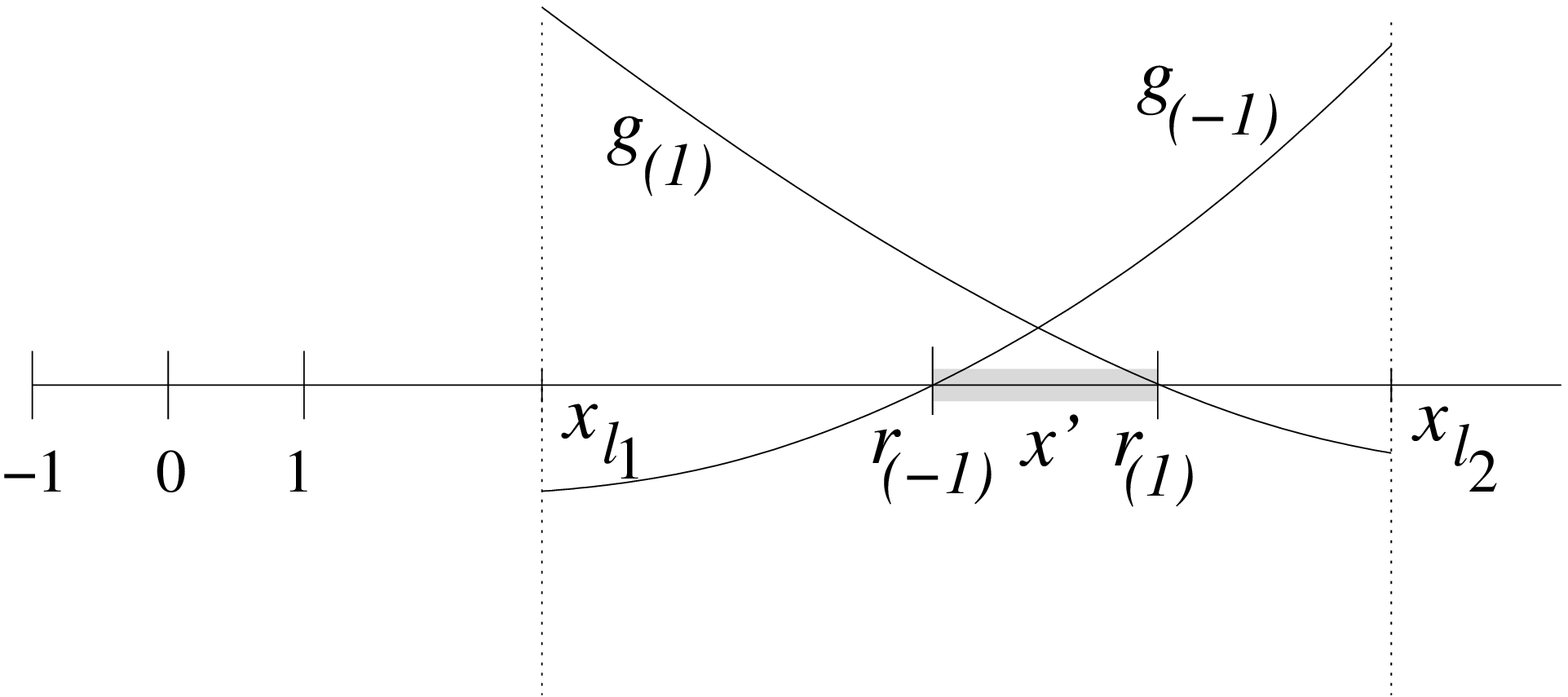}
\caption{ $\rootm<\rootp$ implying that $\diffEnergyp(x')>0$ and
$\diffEnergym(x')>0$, for every $x'\in(\rootm,\rootp)$.
\label{figure:ContrfisMonoton}}
\end{center}
\end{figure}

\begin{figure}[htb]
\begin{center}
\includegraphics[width=0.5\textwidth]{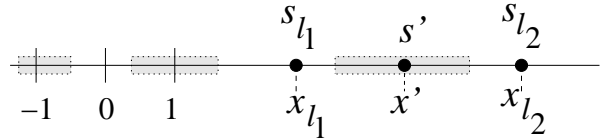}
\caption{ A wireless system $\cA_{3}^{\epsilon}$, where $1<x_{l_{1}} < x' <x_{l_{2}}$.
\label{figure:ContrWireless}}
\end{center}
\end{figure}

Recall that we aim to show that there exists some $x^{*}$ such that
$\diffEnergyp(x^{*}) \leq 0$ and $\diffEnergym(x^{*}) \leq 0$.
We proceed by showing that both functions $\diffEnergyp(x)$ and
$\diffEnergym(x)$ have a single root in $[x_{l_{1}},x_{l_{2}}]$.
For this, we use the following inequalities
(proved later on, in Claim \ref{cl:append_diffEnergypm}).
\begin{equation}
\diffEnergyp(x_{l_{1}})>0 ~~~~\mbox{ and }~~~~ \diffEnergyp(x_{l_{2}})<0,
\label{ineq: diffEnergyp(l1)>0 diffEnergyp(l2)<0}
\end{equation}
\begin{equation}
\diffEnergym(x_{l_{1}})<0 ~~~~\mbox{ and }~~~~ \diffEnergym(x_{l_{2}})>0.
\label{ineq: diffEnergym(l1)<0 diffEnergym(l2)>0}
\end{equation}
By inequality (\ref{ineq: diffEnergyp(l1)>0 diffEnergyp(l2)<0}) and the fact
that $\diffEnergyp(x)$ is monotonically decreasing in $[x_{l_{1}},x_{l_{2}}]$,
it follows that there exists $\rootp\in [x_{l_{1}},x_{l_{2}}]$
such that $\diffEnergyp(\rootp)=0$.
In the same manner, by inequity
(\ref{ineq: diffEnergym(l1)<0 diffEnergym(l2)>0}) and the fact that
$\diffEnergym(x)$ is monotonically increasing in $[x_{l_{1}},x_{l_{2}}]$,
it follows that there exists $\rootm\in [x_{l_{1}},x_{l_{2}}]$
such that $\diffEnergym(\rootm)=0$.
%
%
%
%
Due to the monotonicity of $\diffEnergyp(x)$ and $\diffEnergym(x)$
it turns out that
\begin{eqnarray}
\diffEnergyp(x) &>& 0,~~ \mbox{ for every } x\in[x_{l_{1}},\rootp)
~~\mbox{ and }~~
\label{ineq: diffEnergyp(s)><0 : 1}
\\
\diffEnergyp(x) &\leq& 0,~~ \mbox{ for every } x\in[\rootp,x_{l_{2}}),
\label{ineq: diffEnergyp(s)><0 : 2}
\end{eqnarray}
and similarly,
\begin{eqnarray}
\diffEnergym(x) &\leq& 0,~~ \mbox{ for every } x\in[x_{l_{1}},\rootm]
~~\mbox{ and }~~
\label{ineq: diffEnergym(s)><0 : 1}
\\
\diffEnergym(x) &>& 0,~~ \mbox{ for every } x\in(\rootm,x_{l_{2}}),
\label{ineq: diffEnergym(s)><0 : 2}
\end{eqnarray}
as illustrated in Figure \ref{figure:fisMonoton}.
By inequalities (\ref{ineq: diffEnergyp(s)><0 : 2}) and
(\ref{ineq: diffEnergym(s)><0 : 1}), it turns that taking $x^{*}$
to be in the range $[\rootm,\rootp]$ achieves the desire, as
$\diffEnergyp(x) \leq 0$ and $\diffEnergym(x) \leq 0$ for every $x \in [\rootm,\rootp]$.
Finally, it remains to prove the range $[\rootm,\rootp]$ is not empty, or that
\begin{eqnarray*}
\rootp ~\leq~ \rootm.
\label{ineq: rootp geq rootm}
\end{eqnarray*}
Assume, by the way of contradiction that $\rootp > \rootm$ as illustrated in Figure \ref{figure:ContrfisMonoton}. Consider a station $\Station'$ positioned at $x'\in(\rootm , \rootp)$ with transmitting power of $\Power(x')$.
By inequality (\ref{ineq: diffEnergyp(s)><0 : 1}) it then turns out that
$\diffEnergyp(x')>0$, and similarly, by inequality
(\ref{ineq: diffEnergym(s)><0 : 2}), $\diffEnergym(x')>0$.
Let \(\cA_{3}\) denote a three station network,
$$\cA_{3} ~=~ \langle d=1,\{\Station',\Station_{l_{1}},\Station_{l_{2}}\},
\{\Power(x'),\Power(x_{l_{1}}),\Power(x_{l_{2}})\}, N=0,\beta,\alpha=2 \rangle.$$
By the above, it holds that
\begin{equation*}
\Energy_{\cA_{3}}(\Station',q) ~>~
\Energy_{\cA_{3}}(\{\Station_{l_{1}},\Station_{l_{2}}\},q),
\end{equation*}
for $q \in \{-1,1\}$. Whereas by the energy function $\Power(x)$, Eq. (\ref{eq:1d_power_induc}), it follows that
\begin{equation*}
\Energy_{\cA_{3}}(\Station',0) ~=~
\Energy_{\cA_{3}}(\{\Station_{l_{1}},\Station_{l_{2}}\},0)~.
\end{equation*}
Since $N=0$, we end with a situation where $\SINR_{\cA_{3}}(\Station_1,q) > 1$ for $q \in \{-1,1\}$ and $\SINR_{\cA_{3}}(\Station_1,0) = 1$. Consider a network $\cA_{3}^{\epsilon}=(\{\Station',\Station_{l_{1}},\Station_{l_{2}}\},\{\Power(x'),\Power(x_{l_{1}}),\Power(x_{l_{2}})\}, 0,\beta+\epsilon, \alpha=2 \rangle$, where
$$\epsilon ~=~ \min\left\{\SINR_{\cA_{3}}(\Station',1)~,~
\SINR_{\cA_{3}}(\Station',-1)\right\}-1,$$
see Figure \ref{figure:ContrWireless}.
Hence, we end with the following situation, in which $\SINR_{\cA_{3}^{\epsilon}}(\Station_1,q) \geq 1+\epsilon$, for $q\in\{-1,1\}$, and
$\epsilon>0$, implies that $\SINR_{\cA_{3}^{\epsilon}}(\Station_1,0) < 1+\epsilon$.
Thus,
$\{a,b\}\in \ReceptionZone_{1}(\cA_{3}^{\epsilon})$ and
$p=0 \notin \ReceptionZone_{1}(\cA_{3}^{\epsilon})$,
contradiction by Lemma \ref{lem:1d_ClosedShape_3S}. The claim follows.
\QED

The induction step of Lemma \ref{lem:1d_ClosedShape_nS} is established,
the lemma follows.
\QED

Finally, we establish inequalities
(\ref{ineq: diffEnergyp(l1)>0 diffEnergyp(l2)<0}) and
(\ref{ineq: diffEnergym(l1)<0 diffEnergym(l2)>0}),
required for Lemma \ref{lem:1d_ClosedShape_nS}.
\begin{claim}
The functions $\diffEnergyp$ and $\diffEnergym$,
defined in Equations (\ref{eqn:1d_diffEnergyp}) and (\ref{eqn:1d_diffEnergym}),
satisfy Inequalities
(\ref{ineq: diffEnergyp(l1)>0 diffEnergyp(l2)<0})
and
(\ref{ineq: diffEnergym(l1)<0 diffEnergym(l2)>0}).
\label{cl:append_diffEnergypm}
\end{claim}
\Proof
Let $p \in \{1, \ldots, n\}$ and $q \in \R_{>1}$ then define
$$c_{p}^{q} ~=~ \Energy(\Station_{p},q) / \Energy(\Station_{p},0).$$
It then follows that
\begin{eqnarray*}
c_{l_{1}}^{1} &=& \left(\frac{x_{l_{1}}}{x_{l_{1}}-1} \right)^{2} ~\mbox{and}~
c_{l_{2}}^{1} ~=~ \left(\frac{x_{l_{2}}}{x_{l_{2}}-1}\right)^{2} \\
c_{l_{1}}^{-1}&=& \left(\frac{x_{l_{1}}}{x_{l_{1}}+1}\right)^{2} ~\mbox{and}~
c_{l_{2}}^{-1}~=~ \left(\frac{x_{l_{2}}}{x_{l_{2}}+1}\right)^{2}~.
\end{eqnarray*}
Recall that $x_{l_{1}} < x_{l_{2}}$. Then by the decreasing (respc. increasing) monotonicity of $(x/x-1)^{2}$ (respc. $(x/x+1)^{2}$), we have that
\begin{eqnarray}
c_{l_{1}}^{1} & > & c_{l_{2}}^{1} ~\mbox{and}~
\label{eqn:append_diffEnergyp_ineqc}
c_{l_{1}}^{-1} ~<~ c_{l_{2}}^{-1}~.
\label{eqn:append_diffEnergym_ineqc}
\end{eqnarray}
Using these inequalities, we get that
\begin{eqnarray}
\Energy_{\cA}(\{\Station_{l_{1}},\Station_{l_{2}}\},1)&=&\Energy_{\cA}(\{\Station_{l_{1}}\},1)+\Energy_{\cA}(\{\Station_{l_{2}}\},1)
~=~ c_{l_{1}}^{1}\cdot\Energy_{\cA}(\{\Station_{l_{1}}\},0)+c_{l_{2}}^{1}\cdot\Energy_{\cA}(\{\Station_{l_{1}}\},0) \nonumber
\\& < &
c_{l_{1}}^{1}\cdot \Energy_{\cA}(\{\Station_{l_{1}},\Station_{l_{2}}\},0)~.
\label{eqn:append_diffEnergyp_ineq_aux1}
\end{eqnarray}
In the same manner,
\begin{eqnarray}
\Energy_{\cA}(\{\Station_{l_{1}},\Station_{l_{2}}\},1)&>&
c_{l_{2}}^{1}\cdot \Energy_{\cA}(\{\Station_{l_{1}},\Station_{l_{2}}\},0)~,
\label{eqn:append_diffEnergyp_ineq_aux2}
\\
\Energy_{\cA}(\{\Station_{l_{1}},\Station_{l_{2}}\},-1)&>&
c_{l_{1}}^{-1}\cdot \Energy_{\cA}(\{\Station_{l_{1}},\Station_{l_{2}}\},0)~,
\mbox{and}~
\label{eqn:append_diffEnergym_ineq_aux1}
\\
\Energy_{\cA}(\{\Station_{l_{1}},\Station_{l_{2}}\},-1)&<&
c_{l_{2}}^{-1}\cdot \Energy_{\cA}(\{\Station_{l_{1}},\Station_{l_{2}}\},0)~.
\label{eqn:append_diffEnergym_ineq_aux2}
\end{eqnarray}

Finally, the left and right Inequalities of
Ineq. (\ref{ineq: diffEnergyp(l1)>0 diffEnergyp(l2)<0})
and (\ref{ineq: diffEnergym(l1)<0 diffEnergym(l2)>0})
follow, respectively, by Ineq. (\ref{eqn:append_diffEnergyp_ineq_aux1}),
(\ref{eqn:append_diffEnergyp_ineq_aux2}),
(\ref{eqn:append_diffEnergym_ineq_aux1}) and
(\ref{eqn:append_diffEnergym_ineq_aux2})).
\QED
\commabsend
\subsection{Beyond 1-d}
Our conjecture states that zones of the $d$-dimensional map are free convex for every $d \geq 1$. Currently, no proof is available. However, two positive results are presented in this context. In Section \ref{section:Connectivity}, we consider the case where stations are embedded in $\R^{d}$, but study the topological properties of their reception-zones in $\R^{d+1}$, where niceness properties emerge. We show that zones in $\R^{d+1}$ obey a stronger property, namely, \emph{hyperbolic convexity}, which consequently implies that the zones enjoy the NFH property (see Corollary \ref{corollary:hyperbolic_ClosedShape}). In addition, in Subsection \ref{subsec:maximum_principle}, we study the interference function and show that it satisfies the maximum principle. Whether the SINR function (whose denominator is the interference function) follows the maximal principle it is not yet known, if so - the NFH property of $d-$dimensional zones follows.

\section{The number of connected cells in non-uniform SINR diagram}
\commabs
In this section, we aim to achieve bounds for the number of connected cells in non-uniform diagrams. The scheme is as follows. In Section \ref{subsection:NZones-1D}, we consider the 1-dimensional case,
where stations are embedded in $\R^{1}$. We use the NFH property to derive a tight bound on the number of cells. In Subsection \ref{subsec:Rd_numofcells}, we consider the general case of $\R^{d}$ and provide upper bounds for the number of cells. In Subsection  and \ref{subsec:NConstruct},  we present an extreme construction that maximizes the number of cells of a single transmitter. In Section \ref{section:Connectivity}, we study networks of stations embedded in $\R^d$ but draw the map in $\R^{d+1}$. It is shown that the $d+1$ dimensional zones are connected.

\commful
To analyze the number of connected cells in the zones of $\ReceptionZone(\cA)$,
it is instructive to consider a useful property termed {\em No Free-Hole}, NFH for short.
A collection of closed shapes $\cC$ in $\R^{d}$ obeys this property
with respect to a set $S$ of stations
if for every $C\in\cC$ that is free of stations, if all its border points are
reception points of $s_1$, then all points of $C$ are reception points as well.
(Formally, if $C$ is convex, $C \cap S = \emptyset$ and
$\partial C \subseteq \ReceptionZone_1(\cA)$, then also
$C \subseteq \ReceptionZone_1(\cA)$.)

In Section \ref{section:Connectivity}, we consider the case where stations
are embedded in $\R^{d}$, but study the topological properties of their
reception-zones in $\R^{d+1}$, where niceness properties emerge.
We show that zones in $\R^{d+1}$ obey a stronger property, namely,
\emph{hyperbolic convexity}, which consequently implies that they also enjoy
NFH (see Corollary \ref{corollary:hyperbolic_ClosedShape}).
\commfulend
\subsection{The one-dimensional case}
\label{subsection:NZones-1D}
We follow the notations of Subsection \ref{subsec:freeconvexity_1d}.
Recall that it is assumed that $\beta \geq 1$, otherwise the number of cells is bounded by $\Omega(n^{2})$.
\commabsend
\commabs
To see this, consider the case for infinitesimally small $\beta = \epsilon$. In this case, there exists a point $p\in\ReceptionZone_{i}(\cA)$ in
$(-\infty,x_{j^i_1})$ and $(x_{j^i_{n-1}},\infty)$ and in any segment
$(x_{j^i_k},x_{j^i_{k+1}})$, for every $k=1,....,n-2$,
where $j^i_k$ is the index of the $k$'th element in $\{x_j\mid j\in\{1,...,n\}\setminus\{i\}\}$, i.e., $x_{j^i_1}<x_{j^i_{2}}<...<x_{j^i_{n-1}}$.
Since $\Station_{j} \notin \ReceptionZone_{i}$ for any $\Station_{j} \in S\setminus\{\Station_{i}\}$, it follows that any station has $n$ reception cells (separated by the positions of $n-1$ stations), hence the overall number of cells is $n^{2}$.
\commabsend
\par In one dimension, a connected cell $\ReceptionZone_{1,j}(\cA) \subseteq \ReceptionZone_{1}(\cA)$, $j \in [1, \NZones_{1}(\cA)]$, is represented by a segment $\segment=[a,b]$, where $a,b \in \R$.
\commabs
Recall that $\segment$ is a cell of $\ReceptionZone_{1}(\cA)$ iff it satisfies two properties: (a) Continuity with respect to successful reception of $\Station_{1}$, i.e., $\SINR_{\cA}(\Station_1,x) \geq \beta ~~~~\mbox{for every} ~~~~ a\leq x \leq b$. (b) Maximality with respect to property (a), i.e., for any $a'<a$ and $b'>b$, neither of the segments $[a',b]$ nor $[a,b']$ enjoys property (a).
\commabsend
\par We now use the NFH property established in
Lemma \ref{lem:1d_ClosedShape_nS} for the one dimensional case to show that the number of connected cells in 1-dimensional SINR diagrams is linear, establishing the following theorem.
\begin{theorem}
\label{thm:1d_nzones_bound}
In a 1-dimensional network \( \cA = \langle d=1, S, \Power, \Noise, \beta,
\alpha=2 \rangle \), the number of connected cells satisfies
$\sum_{\Station_i \in S} \NZones_{i}(\cA) \leq 2n-1$.
\end{theorem}
\Proof
We begin by observing that the reception zone
of the weakest station is connected.
\begin{lemma}
\label{lem:lowest_energy}
Consider a network \( \cA = \langle d=1, S, \Power, \Noise, \beta \geq 1,
\alpha=2 \rangle \) where station $\Station_{1} \in S$ has the lowest transmission power, i.e., $\Power_{1}=\min\{\Power_{i} \mid i=1, \ldots ,n\}$. Then $\ReceptionZone_{1}(\cA)$ is connected.
\end{lemma}
\commabs
\Proof
Assume, towards contradiction, that there exist (at least) two non-empty (disconnected) cells $\ReceptionZone_{1,1}(\cA)$ and $\ReceptionZone_{1,2}(\cA)$, corresponding to the segments $\segment_1=[a_{1},b_{1}]$ and $\segment_2=[a_{2},b_{2}]$ respectively.
Without loss of generality, assume that $b_{1} < a_{2}$ and $\Station_1 \in \segment_1$. By the definition of a cell, $p \notin  \ReceptionZone_{1}(\cA)$ for any $p \in (b_{1},a_{2})$. Consider two cases. First, assume that there exists some $j>1$, such that $\Station_{j} \in (b_{1},a_{2})$. We show that in this case $\segment_{2} \cap \wvor_{i}(\wvorsys_{\cA})=\emptyset$. Recall that $\Vweight^{\cA}_{i}=\Power_i^{1/\alpha}$ for every $\Station_i \in S$. Since $\Power_{j} \geq \Power_{1}$, it follows that $\Vweight^{\cA}_{j} \geq \Vweight^{\cA}_{1}$. In addition, $\dist{\Station_{j},p} < \dist{\Station_{1},p}$ for any $p \in \segment_{2}$. Thus $\segment_{2} \cap \wvor_{i}(\wvorsys_{\cA})=\emptyset$, leading to contradiction by Lemma
\ref{cl:wv_sinr}.
Next, consider the complementary case, where $[b_{1},a_{2}] \cap S=\emptyset$. Since $b_{1},a_{2} \in \ReceptionZone_{1}(\cA)$, by the NFH property (Lemma \ref{lem:1d_ClosedShape_nS}), $p \in \ReceptionZone_{1}(\cA)$ for every $p \in (b_{1},a_{2})$, contradiction. The Lemma is established.
\QED
\commabsend

We proceed by bounding the total number of cells in $\ReceptionZone(\cA)$.
Let the stations of $S=\{\Station_{1}, \ldots ,\Station_{n}\}$ be ordered in non-increasing order of transmission energies, i.e.,
$\Power_1\geq\Power_2\geq...\geq\Power_n$.
Consider a process in which the stations are added to the system sequentially, placing $\Station_{t}$ at position $x_{t}$ in step $t$ for any $t=1, \ldots ,n$.

Let $S_{t}=\{\Station_{1}, \ldots ,\Station_{t}\}$ be the set of stations already in place on the end of the $t^{th}$ iteration. Let \( \cA_{t} = \langle d=1, S_{t}, \Power, \Noise, \beta,
\alpha=2 \rangle \) denote the wireless network at this stage, and let $\mu_{t}$ denote the number of connected cells in \( \cA_{t}\). To analyze the increase in $\mu_{t}$ on the $t^{th}$ iteration, in which the station $\Station_{t}$ was added to \( \cA_{t-1}\) at point $x_{t}$, we distinguish between two cases:
\begin{description}
\item{(a)} The point $x_{t}$ could not receive correctly any of the stations in $S_{t-1}$, i.e., $\SINR_{\cA_{t-1}}(\Station_{k},x_{t}) < \beta$ for every $\Station_{k}\in S_{t-1}$.
\item{(b)} The point $x_{t}$ was a successful reception point for some of $\Station_k\in S_{t-1}$ on the end of iteration $t-1$, that is $\SINR_{\cA_{t-1}}(\Station_{k},x_{t}) \geq \beta$.
\end{description}
We state the following two claims (one for each case).
\commful
\begin{lemma}
\label{cl:1d_induct_step}
(a) If $\SINR_{\cA_{t-1}}(\Station_{k},x_{t}) < \beta$ for every
$\Station_{k}\in S_{t-1}$, then $\mu_{t} \leq \mu_{t-1}+1$.
\\
(b) If $\SINR_{\cA_{t-1}}(\Station_k,x_{t}) \geq \beta$
for some $\Station_k\in S_{t-1}$, then $\mu_{t} \leq \mu_{t-1}+2$.
\end{lemma}
\commfulend
\commabs
\begin{claim}
\label{cl:1d_induct_step}
If $\SINR_{\cA_{t-1}}(\Station_{k},x_{t}) < \beta$ for every $\Station_{k}\in S_{t-1}$, then $\mu_{t} \leq \mu_{t-1}+1$.
\end{claim}
\Proof
We prove a slightly stronger property, namely, that if the cell $\ReceptionZone_{k,\ell}(\cA_{t-1}) \subseteq \ReceptionZone_{k}({\cA_{t-1}}),$ for $\Station_k \in S_{t-1}$, $\ell \in [1, \NZones_{k}(\cA_{t-1})]$, corresponds to the segment $\segment=[a,b]$, then adding to \(\cA_{t-1}\) any station $\Station_j$ for $j \geq t$, even out of order, at step $t$ to the point $x_j\notin [a,b]$ cannot split $\segment$ into two or more reception cells, which implies that $\mu_{t} \leq \mu_{t-1}+1$.
(Note that, $\segment$ may disappear in the manner that any station cannot be heard in $\segment$.)
This property is proven by contradiction. Assume to the contrary, that $\Station_j$ does split $\segment$ into $\segment_{l}=[c,d]$ and $\segment_{r}=[e,f]$ where $a\leq c < d < e < f\leq b$. This implies the existence of a point $q$, $d < q <e$, such that $q$ cannot receive $\Station_{k}$ correctly under simultaneous transmission of the stations of $S_{t}=S_{t-1} \cup \{\Station_j\}$, whereas the points $d$ and $e$ do receive $\Station_{k}$'s transmission correctly. As $\segment=[a,b] \subseteq \ReceptionZone_{k}(\cA_{t-1})$, the subsegment $[d,e]$ is free of $S_{t-1}$ stations (except maybe $\Station_{k}$), i.e., $x_{f} \notin [d,e]$ for any $\Station_f \in S_{t-1} \setminus \{\Station_{k}\}$. In addition $x_j\notin [a,b]$, hence also $x_j\notin [d,e]$, implying that $[d,e] \cap \left(S_{t}\setminus \{\Station_{k}\}\right) =\emptyset$. It follows by the NFH property (Lemma \ref{lem:1d_ClosedShape_nS}) that $\Station_{k}$ is received at every point in $[d,e]$, including $q$, leading to contradiction and the claim follows.
\QED
\commabsend

\commabs
\begin{claim}
\label{cl:1d_induct_stepB}
If $\SINR_{\cA_{t-1}}(\Station_k,x_{t}) \geq \beta$ for some $\Station_k\in S_{t-1}$, then $\mu_{t} \leq \mu_{t-1}+2$.
\end{claim}
\Proof
As $\Station_{t}$ is the weakest station in \(\cA_{t}\), by Lemma \ref{lem:lowest_energy}, $\ReceptionZone_{t}(\cA_{t})$ is composed of a single reception cell, containing the point $x_{t}$. Let $\segment^{t}=\ReceptionZone_{t}(\cA_{t})$ and let $\ReceptionZone_{k,\ell}(\cA_{t-1})$ correspond to the segment $\segment^{k}=[a,b]$ such that $x_{t} \in [a,b]$, for $\ell \in [1,\NZones_{k}(\cA_{t-1})]$, therefore $\segment^{k} \cap \segment^{t}\neq \emptyset$. We begin by showing that $\Station_{t}$ cannot split $\segment^{k}$ into more than two parts, namely, to the left and to the right of $x^{t}$. Assume, to the contrary, that adding $\Station_{t}$ creates more than two additional cells of $\Station_k$. Without loss of generality, assume that there are at least two cells of $\Station_k$ to the left of $x^{t}$, denoted by $\segment^{k}_{l}=[a_{l},b_{l}]$ and $\segment^{k}_{l'}=[a_{l'},b_{l'}]$, such that $b_{l}<a_{l'}$. Since $[b_{l},a_{l'}] \cap S_{t}=\emptyset$, we end with a contradiction to Lemma \ref{lem:1d_ClosedShape_nS}. Moreover, by the stronger property proved in Lemma \ref{cl:1d_induct_step}, it follows that none of the cells of $\left (\cup_{\Station_{f} \in S_{t-1}}\ReceptionZone_{f}(\cA_{t-1})\right) \setminus \{\segment_k\}$ is divided at step $t$, which completes our argument. Overall, due to step $t$, we have at most two fragments of a previous existing reception cell and one addition of new reception cell, namely, $\segment^{t}$. The claim follows.
\QED
\commabsend
Combining Claims \ref{cl:1d_induct_step} and \ref{cl:1d_induct_stepB},
it follows that after $n$ steps, $\mu_{n} \leq 2n-1$,
establishing Theorem \ref{thm:1d_nzones_bound}.
\QED

\subsection{The \emph{\large d}-dimensional case}
\label{subsec:Rd_numofcells}

We now consider the general case of a network of the form
\( \cA = \langle d, S, \Power, \Noise, \beta, \alpha=2 \rangle \),
and establish upper and lower bounds on the number of connected cells.
\commabs
To obtain an upper bound on the number of connected cells we apply the following theorem due to Milnor \cite{Milnor64} and Thom \cite{Thom65}.
\begin{theorem}(Milnor (1964), Thom(1965))
\label{thm:milnor}
Let $f_{1} \ldots, f_{m}$ be polynomials in $\R^{d}$ with $\deg(f_{i}) < K$. Then $V=\{x=(x_1, \ldots, x_d) \mid f_{i}(x)=0 ~\mbox{for every}~ i \in \{1, \ldots m \}\}$ has at most $K(2K-1)^{d-1}$ connected cells.
\end{theorem}
The following is a direct consequence of Theorem \ref{thm:milnor}.
\begin{lemma}
\label{cor:nzones}
$\sum_{i=1}^{n} \NZones_{i} = O(n^{d+1})$.
\end{lemma}
\Proof
Consider $F^{i}_{\cA}(p)$, the \emph{characteristic polynomial} of $\ReceptionZone_{i}(\cA)$ given in Eq. (\ref{eq:reception_polynomial}). As $\deg(F^{i}_{\cA}(p)) \leq 2 \cdot n$, Theorem \ref{thm:milnor} implies that $\NZones_{i}(\cA)= O(n^{d})$. Summing over all $n$ stations yields the claim.
\QED
In the same manner we can also bound the number of connected cells in
$\ReceptionZone_{\emptyset}(\cA)$, where no station is received correctly.

\begin{corollary}
\label{cor:nzones_noise}
$\NZones_{\emptyset}(\cA) = O(n^{2d})$.
\end{corollary}
\Proof
We first show that for $\beta \geq 1$, the characteristic polynomial of $\ReceptionZone_{\emptyset}(\cA)$ (also known as the \emph{noise polynomial}) is
\begin{equation}
\label{eq:func_noise}
F^{\emptyset}_{\cA}(p) ~=~ -\prod_{i=1}^{n} F^{i}_{\cA}(p)~.
\end{equation}
It is required to show that $p \in \ReceptionZone_{\emptyset}(\cA)$ iff $F^{\emptyset}_{\cA}(p) <0$. The first direction is trivial, as if $p \in \ReceptionZone_{\emptyset}(\cA)$ then $F^{i}_{\cA}(p)>0$ for every $i \in \{1, \ldots ,n\}$ and hence $F^{\emptyset}_{\cA}(p) <0$. For the opposite direction, observe that if $p \notin \ReceptionZone_{\emptyset}(\cA)$, then there exists exactly one station $\Station_j$ such that $p \in \ReceptionZone_{j}(\cA)$ and $F^{j}(\cA,p)\leq 0$. This is due to the fact that when $\beta \geq 1$, reception zones for different stations do not overlap. Hence $F^{i}_{\cA}(p) >0$ for any $i \neq j$, and therefore $F^{\emptyset}_{\cA}(p) \geq 0$ as required.
\par Consequently, the degree of the noise polynomial $F^{\emptyset}_{\cA}(p)$ is bounded by $\deg(F^{\emptyset}_{\cA}(p)) \leq 2 \cdot n^{2}$. By Theorem \ref{thm:milnor} it then follows that $\NZones_{\emptyset}(d) <O(n^{2d}).$
\QED

Throughout the reminder of this section we consider the case where $\alpha=2$ and the 2-dimensional Euclidean plane, i.e., $d=2$. We focus on the station $\Station_{1}$ with transmission power $\Power_{1}$ and devise two construction schemes that aim to maximize the number of connected cells $\NZones_1$. These constructions achieve $\NZones_1=\Omega(n)$ and $\NZones_1=\Omega(\log \Power_{1})$ respectively. We believe the first construction is close to the maximum possible, i.e., we suspect that $\sum_{i=1}^{n}\NZones_{i} = \Theta(n),$ yet no proof is currently available.
\commabsend
\commful
The upper bound is a direct consequence of the Theorem by
Milnor \cite{Milnor64} and Thom \cite{Thom65},
and the lower bound is established by direct construction,
see Appendix and illustration in Figure \ref{fig:Construction}.

\begin{lemma}
\label{cl:1d_construct}
(a) $\sum_{i=1}^{n} \NZones_{i} = O(n^{d+1})$.
\\
(b) There exists network $\cA$ such that $\NZones_{1}= \Omega(n)$.
\end{lemma}
\commfulend


\commabs
\subsection{Construction of $\Omega(n)$ connected cells for a single station}
\label{subsec:NConstruct}

The goal of the construction is as follows. Given $n \geq 1$, find a placement of $4n+1$ stations $S=\{\ZStation_{0}, \ldots ,\ZStation_{4n}\}$ and a power assignment $\Power$ such that $\NZones_{0}=n+1$, that is, $\ZStationC$ is correctly received in $n+1$ different connected cells.
\par Let us partition $S \setminus \{\ZStation_{0}\}$ into $n$ quadruples $
\ZStationSet_{i}=\{\ZStation_{4i+1}, \ldots ,\ZStation_{4i+4}\}$, $0 \leq i \leq n-1$, each corresponding to the vertices of an axis-aligned square. We assume the SINR parameters $\alpha = 2$, $\Noise = 1$, $\beta = 1$ and $\Power_{i} = 1$ for $i >0$. The value of $\Power_0$ and the positions of $S$ will be determined later on. The resulting network is \( \cA = \langle d=2, S, \Power, \Noise=1, \beta=1,\alpha=2 \rangle \). We next present the construction and then analyze the resulting structure.
\subsubsection{Proposed scheme for station locations}
\begin{figure}[htb]
\epsfxsize=8cm \center{\leavevmode
\epsfbox{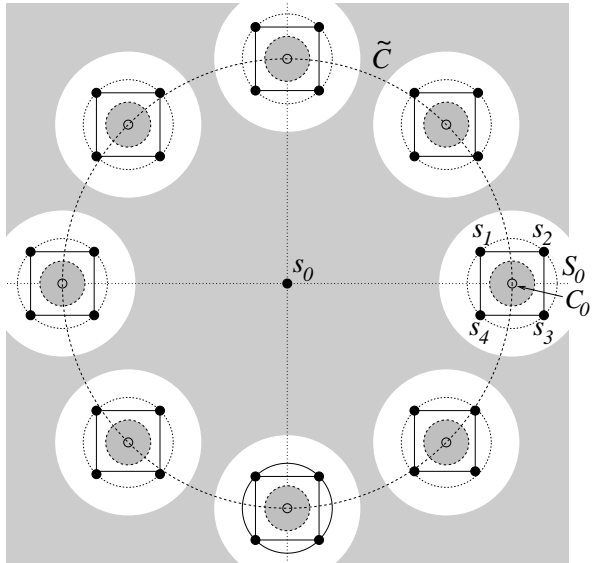}
\caption{Geometric view of the construction }
\label{fig:Construction} }
\end{figure}

Locate station $\ZStationC$ at the origin $(0,0)$ and draw a circle $\tilde{C}$ of radius $R$ around it. Place $n$ points $C_{0}, \ldots, C_{n-1}$ at equidistant locations on $\tilde{C}$, with
$C_{i} =\left \langle{R\cos \left (\frac{2\pi}{n}i \right),
~ R\sin \left (\frac{2\pi}{n}i \right)}\right \rangle$ for $0 \leq i \leq n-1$.
Around each point $C_{i}$ draw a unit circle. Locate the stations
of $\ZStationSet_{i}$ on the vertices of the axis-aligned
$\sqrt{2}\times\sqrt{2}$ square enclosed by $i^{th}$ unit-circle.
Let $\sqS{i}$ be the square defined by its four vertices $S_i$.
See Figure ~\ref{fig:Construction}.

We make use of the following
equalities.
\begin{fact}
\label{fc:NConstruct_DistCenters}
(a)~~$\dist{C_{0},C_{i}}= 2R\sin\left (\frac{\pi}{n}i \right)$.\\
(b)~~$\displaystyle\sum_{i=1}^{n-1}{\frac{1}{\sin^2
\left (\frac{\pi}{n}i \right)}} ~=~ \frac{n^2-1}{3}~, \cite{MATH}.$
\end{fact}

\begin{corollary}
\label{cor:NConstruct_DistCenters}
$\displaystyle\sum_{i=1}^{n-1}{\frac{1}{\dist{C_{0},C_{i}}^2}} ~=~
\frac{n^2-1}{12R^2}~.$
\end{corollary}

\begin{figure}[htb]
\epsfxsize=8cm \center{\leavevmode
\epsfbox{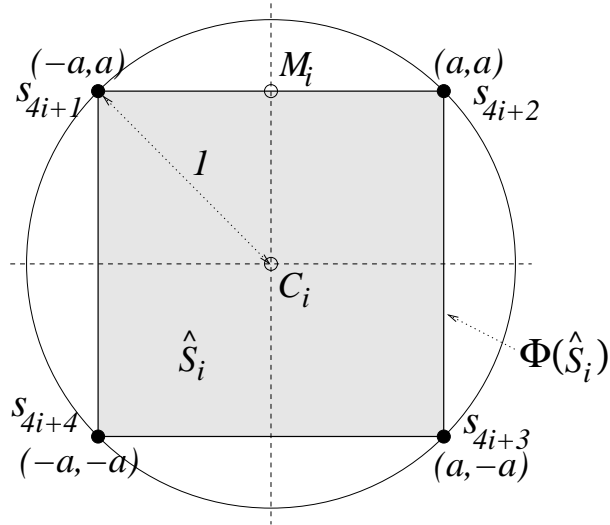}
\caption{Zoom into on the unit circle of $S_i$ and on the square $\sqS{i}$.}
\label{fig:UnitCircle} }
\end{figure}

\begin{lemma}
\label{lem:construct-1}
For every $0 \leq i \leq n-1$:\\
(a) For the center point $C_{i}$, $\Interference(\ZStationSet_{i}, C_{i})= 4.$ \\
(b) For any point $p \in \Boundary (\sqS{i})$, $\Interference(\ZStationSet_{i}, p) \geq 4\frac{4}{5}$.
\end{lemma}
\Proof
For convenience, let us translate the square $\ZStationSet_{i}$ to the
origin, i.e., map $C_{i}$ to $(0,0)$. Let
$\ZStationSet_{i}=\{(-a,a),(a,a),(a,-a),(-a,-a)\}$ be the vertices
of the resulting $2a \times 2a$ square, where $a=1/\sqrt{2}$
(see Figure ~\ref{fig:UnitCircle}).

The interference of $\ZStationSet_{i}$ on the center point $C_{i}=(0,0)$ is given by $\Interference(\ZStationSet_{i},C_{i}) = \Interference(\ZStationSet_{i},(0,0))=4 \cdot \left(1/\left(\sqrt{2}a \right)^{2} \right) = 4,$ implying part (a) of the lemma.

\begin{figure}[htb]
\epsfxsize=8cm \center{\leavevmode
\epsfbox{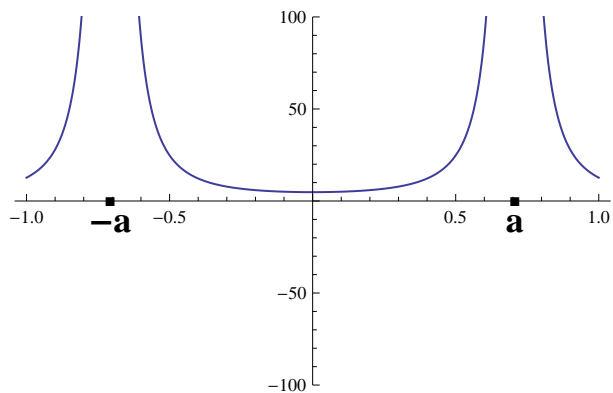}
\caption{Interference function along one square edge}
\label{fig:IntFunc} }
\end{figure}

We next prove part (b).
Due to symmetry, we may restrict attention to a single square edge, say, the upper edge
$e= \{p=(x,y) \mid -a \leq x \leq a\ , y = a\}$. The interference of the four stations of $\ZStationSet_{i}$ on a point $p=(x,y)\in e$ is given by
\begin{equation}
\label{eq:construct-6}
\Interference(\ZStationSet_{i},(x,a)) ~=~
\frac{1}{(x-a)^{2}} + \frac{1}{(x+a)^{2}} + \frac{1}{(x-a)^{2}+4a^{2}} +
\frac{1}{(x+a)^{2} + 4a^{2}}~.
\end{equation}
Let $M_{i}=(0,a)$ be the middle point on edge $e$.
The point
$M_{i}$ is the only local optimum of $\Interference(\ZStationSet_{i},(x,a))$ in the range $x\in (-a,a)$, as
\begin{eqnarray*}
\frac{\partial\Interference(\ZStationSet_{i},(x,a))}{\partial x}
 &=&
-\frac{2}{\left(x-a \right)^{3}}-\frac{2}{\left(x+a \right)^{3}}-
\frac{2 \left (x-a \right)}{\left (4a^{2}+\left (x-a \right)^{2} \right)^{2}}-
\frac{2\left(x+a \right)}{\left(4a^2+\left (x+a\right)^2 \right)2} ~=~ 0
\end{eqnarray*}
implies $x=0$. Since the second derivative $\partial ^{2} \Interference(\ZStationSet_{i},(x=0,a))/\partial x^{2}= 1496/(125a^{2}) > 0$, we conclude that $M_{i}$ is indeed a local minimum (see Figure ~\ref{fig:IntFunc}).
In particular, we get that $\Interference(\ZStationSet_{i},M_{i}) =\Interference(\ZStationSet_{i},(0,a))=12/(5a^{2}) = 24/5,$ and $\Interference(\ZStationSet_{i},M_{i})\leq\Interference(\ZStationSet_{i},p)$
for any $p=(x,y)\in \Boundary (\sqS{i})$,
that is not an edge midpoint,
establishing part (b) of the lemma.
\QED

\subsubsection{Construction strategy}
A desired construction should impose two requirements for each $0 \leq i \leq n-1$:
\begin{description}
\item{(R1)}
\(\SINR(\ZStationC,C_{i}) \geq1 \),
\item{(R2)}
\(\SINR(\ZStationC,p) < 1 \) for every $p \in \Boundary (\sqS{i})$.
\end{description}
Requirement (R1) guarantees that $\ZStationC$ is correctly received at $n$ \emph{regions}, namely, the immediate $\varepsilon$-neighborhoods of the points $C_{i}$ for sufficiently small $\epsilon >0$, whereas requirement (R2) implies also that $\Station_0$ is not received on any point on the perimeters of the $n$ squares, and hence guarantees the $n$ reception regions to be disconnected cells.

Having fixed the station locations up to the choice of $R$, and the
transmission powers of all stations except $\ZStationC$, it remains to select
values for $R$ and $\Power_0$ that will ensure (R1) and (R2).
We employ the following strategy. For each $p \in \Boundary (\sqS{i})$,
we establish an \emph{overestimate} for the energy received at $M_{i}$
from $\ZStationC$, and an \emph{underestimate} for the interference caused
by $S \setminus \{\ZStationC\}$.
For each $C_{i}$, we establish an \emph{underestimate} for the energy received
at $C_{i}$ from $\ZStationC$, and an \emph{overestimate} for the interference
caused by $S \setminus \{\ZStationC\}$.
We then select $\Power_0$ and $R$ that satisfy Requirements ($R1$) and ($R2$)
under these stricter conditions.


\subsubsection{Satisfying (R1) at the center points $C_{i}$}
\begin{lemma}
\label{cl:NConstruct_R1}
If $R \geq \sin^{-1} (\pi/n)$ and $\Power_0 \geq 5R^{2}+4(n^{2}-1)/3,$ then requirement (R1) holds, namely, \(\SINR(\ZStationC,C_{i}) \geq1 \) for every~$0\leq i \leq n-1$.
\end{lemma}
\Proof
Let $\widehat{\ZStation}_{j}=\ZStation_{4j+k}$ for some $k \in \{1, \ldots, 4\}$ be the closest station to $C_{i}$ in $\ZStationSet_{j}$, i.e., such that $\dist{C_{i},\widehat{\ZStation}_{j}} =~ \min_{1\leq l \leq 4} ~\{\dist{C_{i},\ZStation_{4j+l}}\}$. To overestimate the interference of $\ZStationSet_{j}$ on $C_{i}$ we eliminate the other three stations of $\ZStationSet_{j}$, and assign $\widehat{\ZStation}_{j}$ transmission power $\widehat{\Power}_{j}=4$. By the triangle inequality, $\dist{\widehat{\ZStation}_{j}, C_{i}} > \dist{C_{i},C_{j}}-1,$ and therefore
\begin{eqnarray*}
\Interference(\ZStationSet \setminus \left( \ZStationSet_{i} \cup
\{\ZStationC\} \right),C_{i}) &=&
\sum_{j \neq i }{\Interference(\ZStationSet_{j},C_{i}) } ~<~
\sum_{j \neq i }{\Interference(\widehat{\ZStation}_{j},C_{i}) } ~<~
\sum_{j \neq i }{\frac{4}{\left (\dist{C_{i},C_{j}}-1 \right)^{2}}}~.
\end{eqnarray*}
By Fact \ref{fc:NConstruct_DistCenters} (a),
\begin{eqnarray*}
\Interference(\ZStationSet \setminus \left( \ZStationSet_{i} \cup
\{\ZStationC\} \right),C_{i}) &=&
\sum_{i \neq 0 }{\frac{4}{ \left(2R\cdot\sin \left (\frac{\pi}{n}i \right)-1
\right)^{2}}} ~\leq~
\sum_{i \neq 0 }{\frac{4}{ (R \cdot \sin(\frac{\pi}{n}i))^{2}}}~,
\end{eqnarray*}
where the last inequality follows by the fact that $R \cdot \sin(\frac{\pi}{n}i) \geq 1$ for every $i$ (by the first assumption of the lemma). By Corollary \ref{cor:NConstruct_DistCenters},
\begin{equation*}
\Interference(\ZStationSet \setminus \left(\ZStationSet_{i} \cup
\{\ZStationC\}\right),C_{i}) ~<~ \frac{4(n^{2}-1)}{3R^{2}}~.
\end{equation*}
By Lemma \ref{lem:construct-1} (a) it follows that $\Interference(\ZStationSet \setminus \left(\ZStationSet_{i} \cup \{\ZStationC\}\right),C_{i}) < 4+ 4(n^{2}-1)/(3R^{2})$. Finally, by plugging this into Equation (\ref{eq:Def-SINR}), recalling that $\Noise =1$, we get that
\begin{eqnarray*}
\SINR(\ZStationC,C_{i}) &\geq&
\frac{\Power_{0} \cdot R^{-2}}{5 + 4(n^{2}-1)/(3R^{2})} ~>~ 1,
\end{eqnarray*}
where the last inequality follows by the second assumption of the lemma. Hence requirement (R1) holds.
\QED

\subsubsection{Satisfying (R2) at the boundary points $\Boundary (\sqS{i})$}
We now turn to selecting $R$ and $\Power_0$ ensuring Requirement $(R2)$ at every point $p \in \Boundary (\sqS{i})$. By construction, $R-1$ is the minimal distance from the origin to any point on a unit circle centered at $C_{i}$. Hence we have the following.
\begin{observation}
\label{obz:NConstruct_R2_min_dist}
$\Energy(\ZStationC,p) < \Power_{0}/(R-1)^{2},~\mbox{for every}~ p \in \Boundary (\sqS{i})$.
\end{observation}

\begin{lemma}
\label{cl:NConstruct_R2}
If $R \geq \sin^{-1}(\pi/n)$ and $\Power_{0} < (5 \frac{4}{5} + \frac{n^{2}-1}{27R^{2}})\cdot(R-1)^{2}$, then requirement (R2) holds, namely, \(\SINR(\ZStationC,p) < 1 \) for every  $p \in \Boundary (\sqS{i}), ~0 \leq i \leq n-1$.
\end{lemma}
\Proof
We underestimate $\Interference(\ZStationSet_{j},p), ~p \in \Boundary (\sqS{i})$ by considering only the station $\widehat{\ZStation}_{j}=\ZStation_{4j+k}$ (for some $k \in \{1,\ldots, 4\}$) closest to $p$ in $\ZStationSet_{j}$. The distance $\dist{\widehat{\ZStation}_{j},p}$ can be overestimated by the distance between $p$ and center $C_{j}$.
Formally, we have:
\begin{eqnarray}
\label{eqn:inter_middel_ineq}
\Interference(\ZStationSet_{j},p) &>& \Interference(\widehat{\ZStation}_{j},p)
~>~ \Interference(C_{j},p) >\frac{1}{(\dist{C_{j},C_{i}}+1)^{2}}
~>~ \frac{4}{(3\cdot \dist{C_{j},C_{i}})^{2}}~.
\end{eqnarray}
To see the last inequality, note that since $R \geq  \sin^{-1}(\pi/n)$ by the first assumption of the lemma, Fact \ref{fc:NConstruct_DistCenters}(a) guarantees that $\dist{C_{0},C_{1}} \geq 2$. As $\dist{C_{0},C_{1}}=\min_{i \neq j} \dist{C_{i},C_{j}}$ it follows that also $\dist{C_{i},C_{j}} \geq 2$ for every $i$ and $j$.
We therefore have, by Inequality (\ref{eqn:inter_middel_ineq}) and by Fact \ref{fc:NConstruct_DistCenters} (b), that
\begin{eqnarray*}
\Interference(\ZStationSet \setminus \left(\ZStationSet_{i} \cup
\{ \ZStationC\}\right),p) &=&
\displaystyle\sum_{\substack{j \neq i }}{\Interference(\ZStationSet_{j},p)} ~>~
\frac{4}{9}\cdot \sum_{j \neq i }\dist{C_{j},C_{i}}^{-2} ~=~
\frac{n^{2}-1}{27R^{2}}.
\end{eqnarray*}
Next, by combining Observation \ref{obz:NConstruct_R2_min_dist} and
Equation (\ref{eq:Def-SINR}), we have that
\begin{eqnarray*}
\SINR(\ZStationC,p) &\leq& \frac{\Power_{0} \cdot
(R-1)^{-2}}{5 \frac{4}{5}+(n^{2}-1)/(27R^{2})} ~<~ 1,
\end{eqnarray*}
where last inequality follows by the second assumption of the lemma. The lemma follows.
\QED

\subsubsection{Putting it all together}
Finally, we combine the conditions developed in the previous subsections for of requirements (R1) and (R2) (Lemmas \ref{cl:NConstruct_R1} and \ref{cl:NConstruct_R2}) and show that there exists a feasible solution, namely, a choice of $R$ and $\Power_0$ such that both requirements hold. A summary of the conditions is provided in Table \ref{tab:NConstruct_Sum_RQ}.

\begin{table}[htbp]
 \centering
  \begin{tabular}[t]{|c|c|c|c|c|}

   \hline
      Point ($q$) & $\Energy(\ZStationC,q)~~$ &  $\Interference(\ZStationSet_{i},q)+ ~\displaystyle\sum_{\substack{j\neq i}}{\Interference(\ZStationSet_{j},q)}~~$ &  $R$ & $\Power_0~~$ \\
   \hline

    &&&&\\

     $C_{i}$ &   $\Power_{0}/R^{2}$ & $<4~+~\frac{4(n^{2}-1)}{3R^{2}}$ &   $\geq \sin^{-1} (\pi/n)$ &   $\geq 5R^{2}+4(n^{2}-1)/3$ \\
     &&&&\\

     $p \in \Boundary (\sqS{i})$ &   $<\Power_{0}/(R-1)^{2}$ & $>4\frac{4}{5}+\frac{(n^{2}-1)}{27R^{2}}$ &   $\geq \sin^{-1}(\pi/n)$ &   $< (5 \frac{4}{5} + \frac{n^{2}-1}{27R^{2}})\cdot(R-1)^{2}$ \\
     &&&&\\

   \hline
 \end{tabular}
 \caption{Summary of construction requirements }
 \label{tab:NConstruct_Sum_RQ}
\end{table}
Clearly, $R$ should be greater than $\sin^{-1} (\pi/n)$. Let $U=\left(5 \frac{4}{5}+\frac{n^{2}-1}{27R^{2}}\right)\cdot {\left(R-1 \right)^{2}}$ and $L=\left(5+\frac{4 \left(n^{2}-1 \right)}{3R^{2}} \right)\cdot {R^{2}}$. Then by Lemmas \ref{cl:NConstruct_R1} and \ref{cl:NConstruct_R2}, $\Power_0$ should be chosen to satisfy $\Power_0 < U$ and $\Power_0 \geq L$. It is left to verify that for every $n$ there exists a choice of $R>\sin^{-1} (\pi/n)$ such that $U > L$. If this holds, then any choice of $\Power_0$ in the range $U > \Power_0 \geq L$ satisfies the requirements. Letting
\begin{eqnarray}
\label{eq:NConstruct_Sum}
\Delta &=& U -L ~=~ \left(5 \frac{4}{5}+\frac{n^{2}-1}{27R^{2}}\right)\cdot
{\left(R-1 \right)^{2}}-\left(5+\frac{4 \left(n^{2}-1 \right)}{3R^{2}} \right)
\cdot {R^{2}}~,
\end{eqnarray}
it suffices to show that $\Delta>0$ for sufficiently large $R$.
This is done by developing Equation (\ref{eq:NConstruct_Sum}) taking into account leading factors. For ease of analysis, let $n^{*}=n^{2}-1$. Then by Equation (\ref{eq:NConstruct_Sum}) we need $R$ to satisfy $
R^{2}\cdot \left(4n^{*}/\left(3R^{2}\right)+5 \right) < (R-1)^{2}\cdot \left(n^{*}/\left (27R^{2} \right)+29/5 \right)$. Multiplying by $R^{2}$ and rearranging, the requirement becomes
\begin{eqnarray*}
\frac{4}{5}R^{4}-\frac{58}{5}R^{3}+\frac{29}{5}R^{2} &>&
\left(\frac{35}{27}R^{2}-\frac{2}{27}R+\frac{1}{27} \right)n^{*}~.
\end{eqnarray*}
For sufficiently large $R$, the left hand side expression is greater than $3/5 R^{4}$ and the right hand side expression is smaller than $12/5 R^{2} \cdot n^{*}$, so it suffices to require that $3/5 R^{4} > 12/5 R^{2}\cdot n^{*}$, or after simplification, that $R > 2n$.
We therefore established the following.
\begin{theorem}
\label{thm:construction}
There exists a network \(\cA\) such that $\NZones_{1}=\Omega(n)$.
\end{theorem}

\section{Connectivity of reception zones in $\R^{d+1}$}
\label{section:Connectivity}

\commabs
\subsection{Hyperbolic convexity in SINR diagrams ($\alpha \geq 2$)}
\label{section:HyperbolicConvexity}
\commabsend
\commful
\paragraph{Hyperbolic convexity in SINR diagrams:}
\commfulend
Let \( S = \{ \Station_1, \ldots, \Station_{n} \} \) be a set of
stations embedded in the $d$-dimensional space $\R^{d}$. We consider on the network \( \cA = \langle d, S, \Power, \Noise, \beta,2 \alpha \geq 2 \rangle \) in $\R^{d}$ and the \emph{reception map} $\ReceptionZone(\cA_{d+1})$ created for it in $\R^{d+1}$.
We assume without loss of generality that the stations are embedded
in the hyperplane $x_{d+1}=0$ in $\R^{d+1}$, with positions
$(x_1^{\Station_{i}},...,x_d^{\Station_{i}},0)$.
Throughout this section we slightly abuse notation by occasionally considering
a point $p=(x_1^p,...,x_d^p)$ in $\R^d$ as a point in $\R^{d+1}$, namely,
$(x_1^p,...,x_d^p,0)$.
This section concerns what happens when we go one dimension higher, and consider the SINR diagram in dimension $d+1$ for $S$. Recall that %
\commful
$ \ReceptionZone_{i}(\cA_{d+1}) =
\{ p \in \Reals^{d+1} \setminus \{S\} \mid \SINR_{\cA_{d+1}}(\Station_i, p) \geq \beta \}
\cup \{\Station_i\} .$
\commfulend
\commabs
$$ \ReceptionZone_{i}(\cA_{d+1}) ~=~
\{ p \in \Reals^{d+1} \setminus \{S\} \mid \SINR(\Station_i, p) \geq \beta \}
\cup \{\Station_i\} ~.$$
\commabsend
The following theorem shows that the situation improves dramatically in this setting.
\begin{theorem}
\label{thm:d+1_Connected}
Given a network $\cA = \langle d,S, \Power, N,\beta,2\alpha \geq 2 \rangle$, $\ReceptionZone_{i}(\cA_{d+1})$ is connected for every $i \in \{1, \ldots ,n\}$.
\end{theorem}

In what follows, we concentrate on
$\Station_{1}$ and show that $\ReceptionZone_{1}(\cA_{d+1})$ is connected.
\commabs
Let $p=(x_1^p,...,x_d^p,x_{d+1}^p) \in \R^{d+1}$ be any point that correctly
receives the transmission of station $\Station_{1}=(x_1^{s_1},...,x_d^{s_1},0)$.
To prove that $\ReceptionZone_{1}(\cA_{d+1})$ is connected, we show that
there exists a continuous curve connecting $\Station_{1}$ and $p \in \R^{d+1}$
such that $\Station_{1}$ is correctly received at any point along this curve.
In fact, we establish a stronger property, namely, that for any
\commabsend
\commful
We establish the following property.
\commfulend
\commabs
two points $p_{1}=(x_1^{p_{1}},...,x_d^{p_{1}},x_{d+1}^{p_{1}})$ and
$p_{2}=(x_1^{p_{2}},...,x_d^{p_{2}},x_{d+1}^{p_{2}})$ in
$\ReceptionZone_{1}(\cA_{d+1})$, residing on the same side of the hyperplane
$x_{d+1}=0$, i.e., satisfying
\commabsend
\commful
Let $p_{1},p_{2} \in \ReceptionZone_{1}(d+1)$ satisfying
\commfulend
\begin{equation}
\label{eq:hyperbolic-condition}
sign(x_{d+1}^{p_{1}}) \cdot sign(x_{d+1}^{p_{2}}) ~\geq~ 0~,
\end{equation}
there exists a continuous curve connecting $p_{1}$ and $p_{2}$ in $\R^{d+1}$
such that $\Station_{1}$ is correctly received at any point along this curve.
In particular, this curve corresponds to the {\em hyperbolic geodesic}
of $p_{1}$ and $p_{2}$ denoted by $\hypergeodesic{p_{1},p_{2}}$. Note that this indeed guarantees the connectivity
of $\ReceptionZone_{1}(\cA_{d+1})$ by taking $p_{1}=\Station_{1}$.

We begin by recalling some facts about hyperbolic geometry, see \cite{THU97}
for details. Specifically, we consider a standard model of hyperbolic planes,
namely, the upper half-plane model. Under this model, the geodesic of two
points $p_{1},p_{2} \in \R^{d}$  is  either a vertical line or an arc,
as will be formulated later. A point set $P$ is {\em hyperbolic star-shaped}
with respect to point $p_{1} \in P$ if the hyperbolic geodesic of $p_{1}$
and every point $p_{2} \in P$,  is contained in the point-set $P$ as well, e.g., $\hypergeodesic{p_{1},p_{2}} \subseteq P$ (where $p_{1}$ and $p_{2}$ satisfy Ineq. (\ref{eq:hyperbolic-condition})).
A point set $P$ is {\em hyperbolic convex} if it is star-shaped with respect
to any point $p_{1} \in P$. In other words, for any two points
$p_{1},p_{2} \in \textit{P}$ obeying (\ref{eq:hyperbolic-condition}),
$\hypergeodesic{p_{1},p_{2}} \subseteq P$ as well.
In this section we show that the reception zone $\ReceptionZone_{1}(\cA_{d+1})$
is hyperbolic convex and therefore connected.

\commabs
\begin{figure}[htb]
\begin{center}
\includegraphics[scale=0.7]{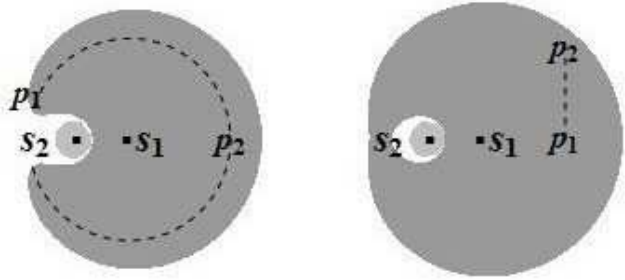}
\caption{ \label{figure:Hyper2}
\sf Hyperbolic convexity in $\R^{2}$. The stations $s_{1}$ and $s_{2}$
are embedded in $\R^{1}$. (a) Convexity on a straight vertical line in $\R^{2}$,  $\Segment{p_{1}}{p_{2}}$. (b) Hyperbolic convexity on a circular arc in $\R^{2}$, $\arc{p_{1}}{p_{2}}$.
}
\end{center}
\end{figure}
\commabsend


We proceed by considering two cases, one for each type of hyperbolic geodesics.
\par
\begin{figure}[htb]
\begin{center}
\includegraphics[scale=0.5]{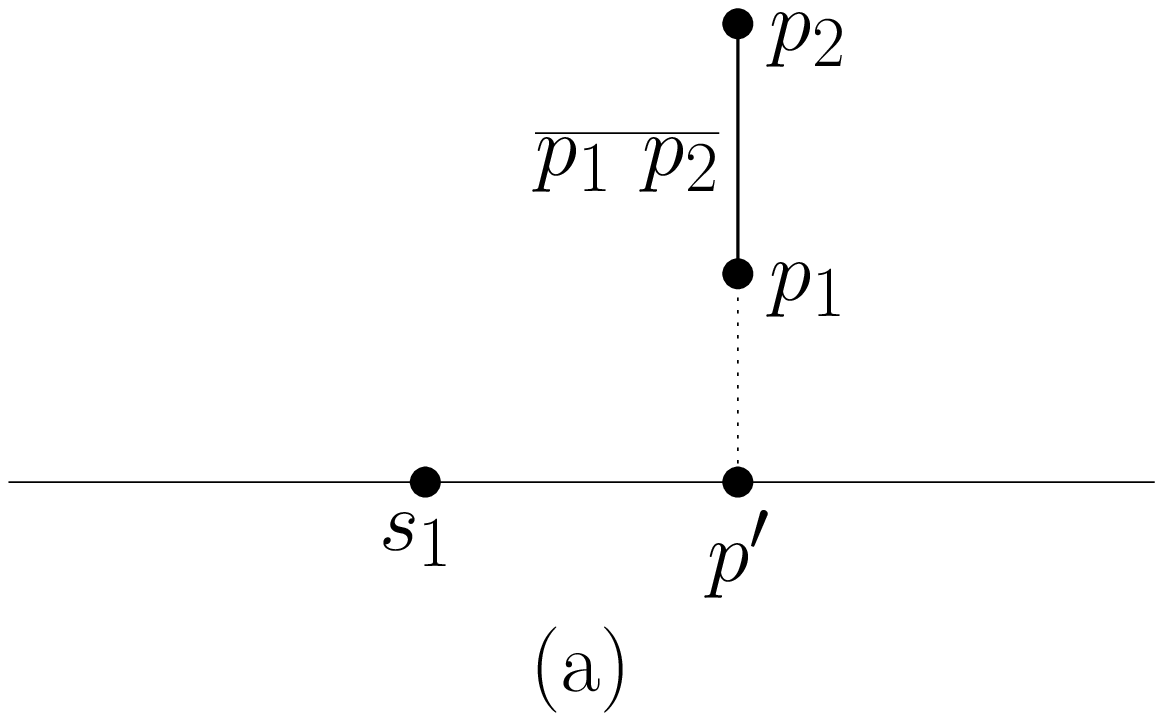}
\hfill
\includegraphics[scale=0.5]{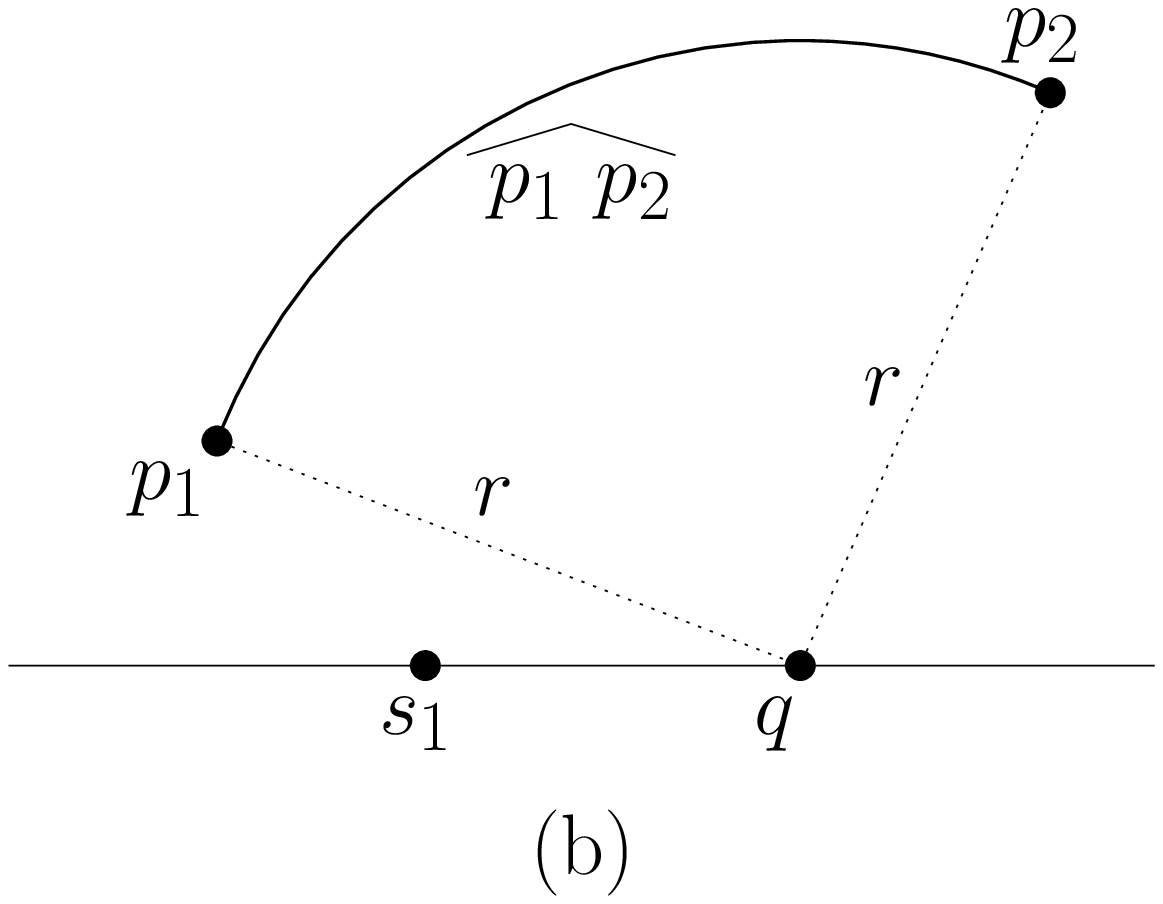}
\caption{ \label{fig:2d-hypercon}
\sf The hyperbolic geodesic of points $p_1$ and $p_2$ corresponds to either
(a) a vertical line (case HC1), or
(b) a hyperbolic arc (case HC2).
}
\end{center}
\end{figure}

\noindent{\bf Case HC1:}
$x_i^{p_{1}}=x_i^{p_{2}}$ for $i \in \{1,\ldots ,d\}$;
$\hypergeodesic{p_{1},p_{2}}$  corresponds to a vertical line
denoted by $\Segment{p_{1}}{p_{2}}$, see points $p_{1}$ and $p_{2}$ of
Figure ~\ref{fig:2d-hypercon}(a).
\par\noindent{\bf Case HC2:}
There exists some $i \in \{1,\ldots ,d\}$ such that $x_i^{p_{1}} \neq x_i^{p_{2}}$;
$\hypergeodesic{p_{1},p_{2}}$ corresponds to an arc,
denoted by $\arc{p_{1}}{p_{2}}$, see points $p_{2}$ and $p_{3}$ of
Figure ~\ref{fig:2d-hypercon}(b).

In Subsection \ref{subsec:HC1}, we consider Case HC1 and show that if
$p_{1}$ and $p_{2}$ are in $\ReceptionZone_{1}(\cA_{d+1})$, then so is any point
on the segment $\Segment{p_{1}}{p_{2}}$.
In Subsection \ref{subsec:HC2}, we refer to Case HC2 and show that if
$p_{1}$ and $p_{2}$ are in $\ReceptionZone_{1}(\cA_{d+1})$, then there exists
an arc $\arc{p_{1}}{p_{2}}$ fully contained in $\ReceptionZone_{1}(\cA_{d+1})$.
In particular, for $p_{1}=\Station_{1}$, there exists an arc
$\arc{\Station_{1}}{p_{2}}$, for every reception point
$p_{2} \in \ReceptionZone_{1}(\cA_{d+1})$, such that
$\arc{\Station_{1}}{p_{2}} \subseteq \ReceptionZone_{1}(\cA_{d+1})$, i.e.,
the zone is hyperbolic star-shaped with respect to $\Station_1$,
hence it is connected.

\subsubsection{Analysis of Case HC1}
For Case HC1, we state the following lemma.
\label{subsec:HC1}
\begin{lemma}
\label{lem:HCl:hyperbolic_segment}
Let $p_{1}, p_{2} \in \ReceptionZone_{1}(\cA_{d+1})$ be points obeying Inequality  (\ref{eq:hyperbolic-condition}) such that $x_j^{p_{1}}=x_j^{p_{2}}$ for $j \in \{1,\ldots ,d\}$. Then
$\Segment{p_{1}}{p_{2}} \in \ReceptionZone_{1}(\cA_{d+1})$.
\end{lemma}
\commabs
\Proof
Assume without loss of generality that $x_{d+1}^{p_1}<x_{d+1}^{p_2}$.
Consider an internal point
$p=(x_1^{p_{1}},...,x_d^{p_{1}},x^p_{d+1})\in \Segment{p_{1}}{p_{2}}$, i.e.,
$x_{d+1}^{p_{1}}<x_{d+1}^{p}<x_{d+1}^{p_{2}}$.
Due to symmetry we may restrict attention to $x_{d+1}^{p_{1}} \geq 0$. (To simplify notations, when it is clear from the context, we may omit $p$ from $x^p_{d+1}$ and write $x_{d+1}$.)
Let $p'=(x_1^{p_{1}},...,x_d^{p_{1}},0)$.
For ease of notation, let $a_{i}=\dist{s_i,p'}^2$, $b_{i}(x_{d+1})=a_{i}+x^2_{d+1}$.
Note that $\dist{s_i,p}^{2\alpha}=b_{i}^{\alpha}(x_{d+1})$, for
every $i\in\{1,...,n\}$.  Thus, the SINR function of $\Station_{1}$ restricted to such point $p$ is given by
\begin{equation*}
\label{eq:hyperbolic-SINR_segment}
\SINR(\Station_1, p) ~=~
\frac{\frac{\Power_{1}}{b_{1}(x_{d+1})^{\alpha}}}{\displaystyle\sum_{\substack{i=2}}^{n}
\frac{\Power_{i}}{b_{i}(x_{d+1})^{\alpha}}+N}~.
\end{equation*}
Let $l_{i}(x_{d+1})=b_{1}(x_{d+1})/b_{i}(x_{d+1})$  and $m_{i}(x_{d+1})=\left(a_{i}-a_{1} \right)/b_{i}(x_{d+1})^{2}$.
In this context, it may be convenient to consider the reciprocal of the SINR function (Eq. (\ref{eq:Def-SINR_reciprocal})),
\begin{equation}
\label{eq:hyperbolic-SINR-1_segment}
\SINR^{-1}(\Station_1, p) ~=~ \displaystyle\sum_{\substack{i=2}}^{n}
\frac{\Power_{i}}{\Power_{1}} \cdot l_{i}^{\alpha}(x_{d+1})+\frac{N \cdot
b_{1}^{\alpha}(x_{d+1})}{\Power_{1}}.
\end{equation}
We first show that this function is twice differentiable in $x_{d+1}$ on $\Segment{p_{1}}{ p_{2}}$.
In particular, it is sufficient to show that it is continuous.
Assume the contrary. Since the function is undefined only at stations positions, discontinuity implies that there might exist some station $\Station_i \in \Segment{p_{1}}{p_{2}}$, where $2\leq i \leq n$. Since $x_{d+1}^{\Station_i}=0$, only $p_{1}$ might correspond to such $\Station_i$. But $p_{1}$ is a reception point of $\Station_{1}$, contradiction.
To characterize the optimum points, we next consider the first and second derivatives of the function $\SINR^{-1}$ on $\Segment{p_1}{p_2}$.

Note that, $\frac{\partial l_{i}(x_{d+1})}{\partial x_{d+1}}=2x_{d+1}\cdot m_{i}(x_{d+1})$~,~$\frac{\partial m_{i}(x_{d+1})}{\partial x_{d+1}}=-4x_{d+1}m_i(x_{d+1})/b_i(x_{d+1})$ and $\frac{\partial b_{i}(x_{d+1})}{\partial x_{d+1}}=2x_{d+1}$.
Thus,
\begin{eqnarray}
\frac{\partial \SINR^{-1}(\Station_{1},p)}{\partial x_{d+1}} &=&
2 \alpha \cdot  x_{d+1}\left( \displaystyle\sum_{\substack{i=2}}^{n}
\frac{\Power_{i}}{\Power_{1}} \cdot l_{i}^{\alpha-1}(x_{d+1}) \cdot m_{i}(x_{d+1})+
\frac{N \cdot b_{1}^{\alpha-1}(x_{d+1})}{\Power_{1}}\right),
\label{eq:hyperbolic-dSINR-1_segment}
\end{eqnarray}
and
\begin{eqnarray}
\frac{\partial^{2} \SINR^{-1}(\Station_{1},p)}{\partial x_{d+1}^{2}} &=&
2 \alpha \left( \displaystyle\sum_{\substack{i=2}}^{n} \frac{\Power_{i}}{\Power_{1}}
\cdot l_{i}^{\alpha-1}(x_{d+1}) \cdot m_{i}(x_{d+1})+\frac{N \cdot  b_{1}^{\alpha-1}(x_{d+1})}{\Power_{1}}\right)
\\&+&
4\alpha \cdot (\alpha-1) \cdot x_{d+1}^{2} \cdot \left( \displaystyle\sum_{\substack{i=2}}^{n} \frac{\Power_{i}}{\Power_{1}}l_{i}^{\alpha-2}(x_{d+1}) \cdot m_{i}^{2}(x_{d+1})\nonumber
+
\frac{N \cdot b_{1}^{\alpha-2}(x_{d+1})}{\Power_{1}} \right)\nonumber
\\&-&
8 \alpha x_{d+1}^{2}\displaystyle\sum_{\substack{i=2}}^{n}
\frac{\Power_{i}}{\Power_{1}} \cdot  l_{i}^{\alpha-1}(x_{d+1}) \cdot \frac{m_{i}(x_{d+1})}{b_{i}(x_{d+1})} ~.\nonumber
\label{eq:hyperbolic-d2SINR-1_segment}
\end{eqnarray}
Let $\cJ_{pos}=\{i\in\{2, \ldots ,n\} \mid a_{i} \geq a_{1}\}$ and $\cJ_{neg}=\{i\in\{2, \ldots ,n\} \mid a_{i}<a_{1}\}$.

We distinguish between two cases.\\
{\bf Case 1:} $\cJ_{neg}=\emptyset$.
In this case, $m_{i}(x_{d+1})\geq 0$, therefore, by Eq. (\ref{eq:hyperbolic-dSINR-1_segment}) we get that
$\frac{\partial \SINR^{-1}(\Station_{1},p)}{\partial x_{d+1}}\geq 0$, for every $p\in\Segment{p_1}{p_2}$.
This implies that $\SINR^{-1}(s_1,p)\leq\SINR^{-1}(s_1,p_2)$, thus
$\SINR(\Station_{1},p)\geq\SINR(\Station_{1},p_{2})\geq \beta$ as required.
\\\noindent {\bf Case 2:} $\cJ_{neg}\not=\emptyset$.
There exists some $2 \leq i \leq n$ such that $a_{i}<a_{1}$. This implies the possible existence of other optimum points.
Consider an optimum point of the form $p_{opt}=(x_1^{p_{1}},...,x_d^{p_{1}},x_{d+1}^{opt})$, where $x_{d+1}^{opt}\neq0$.
Thus by Eq. (\ref{eq:hyperbolic-dSINR-1_segment}) we have that
\begin{equation}
\label{eq:hyperbolic-segment_ext}
\frac{\partial \SINR^{-1}(\Station_{1},p_{opt})}{\partial x_{d+1}^{opt}} ~=~
2 \alpha \cdot  x_{d+1}^{opt}\left( \displaystyle\sum_{\substack{i=2}}^{n}
\left(\frac{\Power_{i}}{\Power_{1}} \cdot l_{i}^{\alpha-1}(x_{d+1}^{opt}) \cdot m_{i}(x_{d+1}^{opt}) \right)+
\frac{N \cdot b_{1}^{\alpha-1}(x_{d+1}^{opt})}{\Power_{1}}\right)~=~ 0~.
\end{equation}
In turn, this implies that
\begin{equation}
\label{eq:hyperbolic-segment_ext_notzero}
\displaystyle\sum_{\substack{i=2}}^{n}
\frac{\Power_{i}}{\Power_{1}}\left( l_{i}^{\alpha-1}(x_{d+1}^{opt}) \cdot m_{i}(x_{d+1}^{opt}) \right)+\frac{N \cdot b_{1}^{\alpha-1}(x_{d+1}^{opt})}{\Power_1} ~=~ 0.
\end{equation}
Plugging this equality into Equation (\ref{eq:hyperbolic-d2SINR-1_segment}), the second derivative of $\SINR^{-1}$ at $p_{opt}$ becomes
\begin{eqnarray}
\label{eqn:hyperbolic_HC1_2derivative}
\frac{\partial^{2} \SINR^{-1}(\Station_{1},p_{opt})}{\partial (x_{d+1}^{opt})^{2}} &=&
4\alpha(\alpha-1) (x_{d+1}^{opt})^{2} \left(\displaystyle\sum_{\substack{i=2}}^{n} \frac{\Power_{i}}{\Power_{1}} l_{i}^{\alpha-2}(x_{d+1}^{opt}) \cdot m_{i}^{2}(x_{d+1}^{opt})+
\frac{N \cdot b_{1}^{\alpha-1}(x_{d+1}^{opt})}{\Power_{1}} \right)
\\&-&
8 \alpha (x_{d+1}^{opt})^{2}\displaystyle\sum_{\substack{i=2}}^{n}
\frac{\Power_{i}}{\Power_{1}}\cdot  l_{i}^{\alpha-1}(x_{d+1}^{opt}) \cdot \frac{m_{i}(x_{d+1}^{opt})}{b_{i}(x_{d+1}^{opt})} \nonumber~.
\end{eqnarray}
%
To prove the lemma, we wish to show that the SINR function has no-local minimum on the vertical line segment, $\Segment{p_{1}}{ p_{2}}$, or that the second derivative of $\SINR^{-1}$ restricted to extreme internal points in the segment is non negative.  Define
\begin{eqnarray}
\label{eqn:hyperbolic_HC1_2derivative_partial}
\wp(x_{d+1}^{opt}) &=&
-8 \alpha (x_{d+1}^{opt})^{2}\displaystyle\sum_{\substack{i=2}}^{n}
\frac{\Power_{i}}{\Power_{1}} \cdot  l_{i}^{\alpha-1}(x_{d+1}^{opt}) \cdot \frac{m_{i}(x_{d+1})}{b_{i}(x_{d+1})}~.
\end{eqnarray}
Since $\alpha \geq 1$, $l_i(x_{d+1}^{opt})\geq 0$ and $b_1(x_{d+1}^{opt})\geq 0$, thus by Eq. (\ref{eqn:hyperbolic_HC1_2derivative}), it is sufficient to show that
$\wp(x_{d+1}^{opt}) \geq 0$.
Note that, $\cJ_{pos}$ and $\cJ_{neg}$ separate Eq. (\ref{eqn:hyperbolic_HC1_2derivative_partial}) into its positive and negative terms. Let
$$S_{pos} ~=~
\displaystyle\sum_{i \in \cJ_{pos} } \frac{\Power_{i}}{\Power_{1}} \cdot  l_{i}^{\alpha-1}(x_{d+1}^{opt}) \cdot \frac{m_{i}(x_{d+1}^{opt})}{b_{i}(x_{d+1}^{opt})}
~~~~ \mbox{and} ~~~~
S_{neg} ~=~ \displaystyle\sum_{i \in \cJ_{neg}} \frac{\Power_{i}}{\Power_{1}} \cdot  l_{i}^{\alpha-1}(x_{d+1}^{opt}) \cdot \frac{m_{i}(x_{d+1}^{opt})}{b_{i}(x_{d+1}^{opt})}~.$$
Then $\wp(x_{d+1}^{opt})= -8(x_{d+1}^{opt})^{2} \left(S_{pos}+S_{neg} \right)$.
Recall that by the definition of $b_{1}(x_{d+1}^{opt})$, it follows that
$$b_{1}(x_{d+1}^{opt}) ~<~ b_{i}(x_{d+1}^{opt})~~~~ \mbox{iff}~~~~
a_{i} ~>~ a_{1}~,$$
for any $1\leq i\leq n$.
Since $b_{i}(x_{d+1}^{opt})>0$ for every $i$, we have that
\begin{equation*}
S_{pos} ~\leq~ \displaystyle\sum_{i \in \cJ_{pos}} \frac{\Power_{i}}{\Power_{1}}
\cdot  l_{i}^{\alpha-1}(x_{d+1}^{opt}) \cdot \frac{m_{i}(x_{d+1}^{opt})}{b_{1}(x_{d+1}^{opt})}
~~~~ \mbox{and} ~~~~
S_{neg} ~<~ \displaystyle\sum_{i \in \cJ_{neg}} \frac{\Power_{i}}{\Power_{1}}
\cdot  l_{i}^{\alpha-1}(x_{d+1}^{opt}) \cdot \frac{m_{i}(x_{d+1}^{opt})}{b_{1}(x_{d+1}^{opt})}~,
\end{equation*}
%
This implies that
$$S_{pos}+S_{neg} ~<~ \frac{1}{b_{1}(x_{d+1}^{opt})}\cdot
\left(\displaystyle\sum_{\substack{i=2}}^{n} \frac{\Power_{i}}{\Power_{1}} \cdot
l_{i}^{\alpha-1}(x_{d+1}^{opt}) \cdot m_{i}(x_{d+1}^{opt}) \right) =\frac{1}{b_{1}(x_{d+1}^{opt})}\cdot\frac{-N \cdot b_{1}^{\alpha-1}(x_{d+1}^{opt})}{\Power_{1}} \leq 0, $$
where the right equality hold by Eq. (\ref{eq:hyperbolic-segment_ext_notzero}).  Thus,  $\wp(x_{d+1}^{opt}) \geq 0$ and also $\partial^{2} \SINR^{-1}(\Station_{1},p_{opt})/\partial (x_{d+1}^{opt})^{2} \geq 0$. I.e., any local optimum point other than
$p'=(x_1^{p_{1}},...,x_d^{p_{1}},0)$ is a local minimum.
Since Equation (\ref{eq:hyperbolic-SINR-1_segment}) is continuous and twice differentiable in $\Segment{p_{1}}{p_{2}}$,
this case corresponds to three local optimum points: two local minima, namely, $(x_1^{p_{1}},...,x_d^{p_{1}},x_{d+1}^{opt})$ and $(x_1^{p_{1}},...,x_d^{p_{1}},-x_{d+1}^{opt})$, and one local maximum point, $p'=(x_1^{p_{1}},...,x_d^{p_{1}},0)$ in between.
In sum, there is no local maxima inside $\Segment{p_{1}}{p_{2}}$, which implies that
$\SINR^{-1}(\Station_{1},p)\leq \max \{(\SINR^{-1}(\Station_{1},p_{1})),\SINR^{-1}(\Station_{1},p_{2})\}$ and this
implying that
$$\SINR(\Station_{1},p)\geq \min \{\SINR(\Station_{1},p_{1}),\SINR(\Station_{1},p_{2})\}\geq\beta.$$ The lemma holds.
\QED
\commabsend

\commabs
For a pictorial description of Lemma \ref{lem:HCl:hyperbolic_segment},
see Figure \ref{figure:Hyper2}a.
\par
\commabsend
\commabs
A direct consequences of this claim is the following.
\begin{corollary}
\label{cor:hyperbolic_segment_intersection}
Let $p_{1}, p_{2} \in \ReceptionZone_{1}(\cA_{d+1})$ be points satisfying
Equation (\ref{eq:hyperbolic-condition}) s.t $x_j^{p_{1}}=x_j^{p_{2}}$
for $j \in \{1,\ldots ,d\}$.
Then the line $L$ extrapolated by the segment $\Segment{p_{1}}{p_{2}}$
intersects $\ReceptionZone_{1}(\cA_{d+1})$ at most 4 times.
\end{corollary}
\Proof
The proof follows immediately by the fact that $L$ has at most three extremum points. Note that in general, the number of intersections is bounded by $O(n)$, due to the degree of the SINR function.
\QED
\commabsend

\subsubsection{Analysis of Case HC2}
\label{subsec:HC2}
We next deal with the complementary case CH2.
Let $p_{1},p_{2} \in \R^{d+1}$ be two points of interest such that
$x_j^{p_{1}}\neq x_j^{p_{2}}$ for some $j \in \{1,\ldots ,d\}$
(recall that $p_{1}$ and $p_{2}$ obey Inequality
(\ref{eq:hyperbolic-condition})). The hyperbolic geodesic
of $p_{1}$ and $p_{2}$, $\arc{p_{1}}{p_{2}}$, is defined as follows.
Let $p_{1}^{d}=(x_1^{p_{1}},...,x_d^{p_{1}},0)$ and $p_{2}^{d}=(x_1^{p_{2}},...,x_d^{p_{2}},0)$ be the projection of the points $p_{1}$ and $p_{2}$ to the hyperplane $x_{d+1}=0$, respectively. Consider a point $q \in \R^{d}\times\{0\}$ equidistant from $p_{1}$ and $p_{2}$ and positioned on the line defined by the points $p_{1}^{d}$ and $p_{2}^{d}$. Let $r=\dist{p_{1},q}=\dist{p_{2},q}$.
The hyperbolic geodesic, $\arc{p_{1}}{p_{2}}$, corresponds to the shorter arc
connecting $p_{1}$ and $p_{2}$ on the circumference
$\Boundary (\Ball^{d+1}(q,r))$.

\begin{lemma}
\label{cl:hyperbolic_arc}
Let $p_{1}, p_{2} \in \ReceptionZone_{1}(\cA_{d+1})$ obeying
(\ref{eq:hyperbolic-condition}). Then
$\arc{p_{1}}{p_{2}} \subseteq \ReceptionZone_{1}(\cA_{d+1})$.
\end{lemma}
\commabs
\Proof
By Lemma \ref{lemma:Transformation}, we may assume without loss of generality that $x_j^{p_{1}}=x_j^{p_{2}}$ for $j \in \{2,\ldots, d\}$ and by $q$ definition it follows that $x_j^{q}=x_j^{p_{1}}$ for $j \in \{2,\ldots, d\}$.
Due to symmetry, we may restrict attention to the case where
$x_{d+1}^{p_{1}} \geq 0$ and $x_{d+1}^{p_{2}} \geq 0$. We begin by showing that
the SINR function has no local minimum on $\arc{p_{1}}{p_{2}}$.
Recall that $r=\dist{q,p_{1}}$.
The circumference $\Boundary (\Ball^{d+1}(q, r))$ is defined by the equation
\begin{equation}
\label{eq:hyperbolic-ball}
\sum_{j=1}^{d}(x_j-x_j^{q})^2+x_{d+1}^2 ~=~ r^2~.
\end{equation}
Equivalently, the $x_{d+1}$ coordinate of points on the circumference
can be expressed as
$x_{d+1}=\pm \sqrt{r^{2}-\sum_{j=1}^{d}(x_j-x_j^{q})^2}~$.
Let $g(x_1, \ldots ,x_{d})=\sqrt{r^{2}-\sum_{j=1}^{d}(x_j-x_j^{q})^2}~$,
for every $(x_1,...,x_d)\in\mathbb{R}^d$.
We consider the function $\SINR^{-1}(\Station_{1},p)$ of Eq. (\ref{eq:Def-SINR_reciprocal}), restricted to a point $p=(x_1,...,x_d,g(x_1, \ldots ,x_{d}))$ on $\Boundary (\Ball^{d+1}(q, r))$. For ease of notation, let $a_{i}=(x_1^{\Station_i}-x_1^{q})$ and $b_{i}=\sum_{j=1}^{d} \left((x_j^{\Station_i})^{2}-(x_j^{q})^2 \right)-2\sum_{j=2}^{d}(x_j^{\Station_i}-x_j^{q})x_j+r^2$.
We then have that
\begin{equation}
\label{eq:hyperbolic-dist_ball}
\dist{\Station_i,p}^{2} ~=~
\sum_{j=1}^{d+1}\left( x_{j}^{\Station_{i}}-x_{j}\right)^{2} ~=~
b_{i}-2a_{i}x_1.
\end{equation}
Let $l_{i}(x_1)=\dist{\Station_1,p}^{2}/\dist{\Station_i,p}^{2}$.
By plugging Equation (\ref{eq:hyperbolic-dist_ball}) into the $\SINR^{-1}$ function (Eq. (\ref{eq:Def-SINR_reciprocal})), we get that
\begin{equation}
\label{eq:hyperbolic-SINR-1_ball}
\SINR^{-1}(\Station_{1},p) ~=~ \displaystyle\sum_{\substack{i=2}}^{n}
\frac{\Power_{i}}{\Power_{1}} \cdot l_i^{\alpha}(x_1) +
\frac{N(b_{1}-2a_{1}x_1)^{\alpha}}{\Power_{1}}~.
\end{equation}
Note that since $x_{j}^{p_{1}}=x_{j}^{p_{2}}$ for $j \in \{2,\ldots,d\}$, it follows by the definition of $\widehat{p_{1} ~ p_{2}}$, that $x_{j}^{p}=x_{j}^{p_{1}}$ for $j \in \{2,\ldots,d\}$ for every $p \in \widehat{p_{1} ~ p_{2}}$. To characterize the optimum points, it is sufficient, therefore, to consider the derivatives of Equation (\ref{eq:hyperbolic-SINR-1_ball}) with respect to $x_1$ only (i.e., treating $x_j$ for $j \in \{2,\ldots,d\}$ as constants). It is important to note that Equation (\ref{eq:hyperbolic-SINR-1_ball}) is twice differentiable on $\widehat{p_{1} ~ p_{2}}$. To see this, it in enough to argue that it is continuous or that no station $\Station_i$ other than $\Station_{1}$ belongs to $\widehat{p_{1} ~ p_{2}}$ (this is indeed a sufficient condition for continuity). Assume, to the contrary, that there might be some station $\Station_i \in \widehat{p_{1} ~ p_{2}}$ for $2 \leq i \leq n$. Since the arc endpoints, $p_{1}$ and $p_{2}$, are in $\ReceptionZone_{1}(\cA_{d+1})$, neither of them correspond to $\Station_{i}$, for $i>1$. It follows that $\Station_i$ occurs at some internal point on the arc. Since $p_{1}$ and $p_{2}$ satisfy Inequality (\ref{eq:hyperbolic-condition}), it follows that $x_{d+1}^{p} > 0$ for every $p\in \widehat{p_{1} ~ p_{2}}\setminus\{p_{1},p_2\}$~. Yet, $x_{d+1}^{\Station_i}=0$, for every $\Station_i \in S$, yielding a contradiction.
Define $m_{i}(x_1)=2\left(a_{i}b_{1}-a_{1}b_{i}\right)/\left(b_{1}-2a_{1}x_1\right)^2$, e.g., $m_{i}(x_1)={\partial l_{i}(x_{1})}/{\partial x_{1}}$.
Note that, $\partial m_{i}(x_{1})/\partial x_{1}=\frac{4 a_i\cdot m_i(x_1)}{b_{1}-2a_{1}x_1}=\frac{-4 a_i\cdot m_i(x_1)}{\dist{s_i,p}^2}$.

Consider an optimum point $p_{opt}\in \widehat{p_{1} ~ p_{2}}\setminus\{p_{1},p_2\}$. This optimum point satisfy
\begin{equation}
\label{eq:hyperbolic-condition0}
\frac{\partial \SINR^{-1}(\Station_{1},p_{opt})}{\partial x_1^{opt}} ~=~
\alpha \left(\displaystyle\sum_{\substack{i=2}}^{n} \frac{\Power_{i}}{\Power_{1}}
\cdot l_{i}(x_1^{opt})^{\alpha-1} \cdot m_{i}(x_1^{opt})- \frac{2a_{1} \cdot N \cdot \left(b_{1}-2a_{1}x_1^{opt}\right)^{\alpha-1} } {\Power_{1}}\right)=0~.
\end{equation}
The second derivative with respect to $x_1^{opt}$ is given by
\begin{eqnarray*}
\frac{\partial^{2} \SINR^{-1}(\Station_{1},p_{opt})}{\partial (x_1^{opt})^{2}} &=&
\alpha  (\alpha-1)\left( \displaystyle\sum_{\substack{i=2}}^{n} \frac{\Power_{i}}{\Power_{1}} \cdot l_{i}^{\alpha-2}(x_1^{opt}) \cdot m_{i}^{2}(x_1^{opt})+
4 a_{1}^{2} \cdot \frac{N \left(b_{1}-2a_{1}x_1^{opt}\right)^{\alpha-2}}{\Power_1}\right)
\\&+&
4 \alpha \displaystyle\sum_{\substack{i=2}}^{n} \frac{\Power_{i}}{\Power_{1}} \cdot l_{i}^{\alpha-1}(x_1^{opt}) \cdot \frac{ a_{i} \cdot m_{i}(x_1^{opt})}{\dist{\Station_i,p_{opt}}^{2}}~.
\end{eqnarray*}
Define
\begin{equation}
\label{eq:hyperbolic-condition2}
\wp(p_{opt}) ~=~
4 \alpha \displaystyle\sum_{\substack{i=2}}^{n} \frac{\Power_{i}}{\Power_{1}} \cdot l_{i}^{\alpha-1}(x_1^{opt}) \cdot \frac{ a_{i} \cdot m_{i}(x_1^{opt})}{\dist{\Station_i,p_{opt}}^{2}}~. \nonumber
\end{equation}
Since $\alpha \geq 1$, it is sufficient to show that $\wp(p_{opt}) \geq 0$.
We 
separate the summation of Equation (\ref{eq:hyperbolic-condition2}) into two parts, i.e., $S_{pos}$ and $S_{neg}$, the summation of elements that correspond to positive (respectively, negative) elements in the left term of Equation (\ref{eq:hyperbolic-condition0}). Formally, letting $\cJ_{pos}=\{i\in \{2, \ldots ,n\}\mid a_{i}b_{1}\geq a_{1}b_{i} \}$ and $\cJ_{neg}=\{i \in \{2, \ldots ,n\} \mid a_{i}b_{1}<a_{1}b_{i} \}$, we have that $\wp(p_{opt}) = 4\alpha(S_{pos}+S_{neg}),$ where
$$S_{pos} ~=~ \displaystyle\sum_{i \in \cJ_{pos}}
\frac{\Power_{i}}{\Power_{1}} \cdot l_{i}(x_1^{opt})^{\alpha-1} \cdot \frac{ a_{i} \cdot m_{i}(x_1^{opt})}{\dist{\Station_i,p_{opt}}^{2}}
~~~~\mbox{and}~~~~
S_{neg} ~=~ \displaystyle\sum_{i \in \cJ_{neg}} \frac{\Power_{i}}{\Power_{1}}
\cdot l_{i}(x_1^{opt})^{\alpha-1} \cdot \frac{ a_{i} \cdot m_{i}(x_1^{opt})}{\dist{\Station_i,p_{opt}}^{2}}~.$$
Let $c_{i}(x)=a_{i}/\dist{\Station_i,p_{opt}}^{2}$. Then,  $sign(c_{i}(x_{1}^{p}))=sign(a_{i})$ for any $p \in \Boundary (\Ball^{d+1}(q, r)) \setminus \{\Station_{1}\}$.
Therefore it follows that
$c_{1}(x_1^{opt}) \leq c_{i}(x_1^{opt})$ if $a_{i}b_{1}\geq a_{1}b_{i}$ (i.e., $i \in \cJ_{pos}$), and that $c_{1}(x_1^{opt})>c_{i}(x_1^{opt})$ if $a_{i}b_{1}<a_{1}b_{i}$ (i.e., $i \in \cJ_{neg}$), implying that
\begin{eqnarray*}
S_{pos}\geq \displaystyle\sum_{i \in \cJ_{pos}} \frac{\Power_{i}}{\Power_{1}} \cdot l_{i}^{\alpha-1}(x_1^{opt}) \cdot c_{1}(x_1^{opt}) \cdot  m_{i}(x_1^{opt})~~ \mbox{and}~~
S_{neg}>\displaystyle\sum_{i \in \cJ_{neg}} \frac{\Power_{i}}{\Power_{1}} \cdot l_{i}^{\alpha-1}(x_1^{opt}) \cdot c_{1}(x_1^{opt}) \cdot m_{i}(x_1^{opt})~.
\end{eqnarray*}
Therefore,
\begin{eqnarray*}
\wp(p_{opt})&\geq& 4 \alpha \cdot c_{1}(x_1^{opt})\displaystyle\sum_{i=2}^{n} \frac{\Power_{i}}{\Power_{1}} \cdot l_{i}^{\alpha-1}(x_1^{opt}) \cdot m_{i}(x_1^{opt})
\\&=&
8 \alpha \cdot c_{1}(x_1^{opt})\cdot \frac{a_1 \cdot N \cdot \left(b_{1}-2a_{1}x_1\right)^{\alpha-1} } {\Power_{1}}
\\&=&
8 \alpha \cdot \frac{a_1^{2} \cdot N}{\Power_1 \cdot \dist{\Station_1,p_{opt}}^{4-2\alpha}}\geq 0,
\end{eqnarray*}
where the second equality follows by Eq. (\ref{eq:hyperbolic-condition0}). It therefore holds that $\partial^{2} \SINR^{-1}(\Station_{1},p_{opt})/\partial (x_1^{opt})^{2} \geq 0$ as required.
%
%
We showed that there is no local maximum point of $\SINR^{-1}(\Station_1,p)$ on $\widehat{p_{1} ~ p_{2}}$.
Thus, there is no local minimum point of $\SINR(\Station_1,p)$ on $\widehat{p_{1} ~ p_{2}}$.
Hence, $\SINR(\Station_1,p)\geq \min(\SINR(\Station_1,p_{1}),\SINR(\Station_1,p_{2})))\geq\beta$ for every point $p \in \widehat{p_{1} ~ p_{2}}$, as required.
\QED
For a pictorial description of Lemma \ref{cl:hyperbolic_arc},
see Figure \ref{figure:Hyper2}b.
\commabsend

Finally, we turn to complete the proof for Thm. \ref{thm:d+1_Connected}.
By Lemma \ref{cl:hyperbolic_arc}, $\ReceptionZone_{1}(\cA_{d+1})$ is hyperbolic
convex. It follows that $\ReceptionZone_{1}(\cA_{d+1})$ is hyperbolic
star-shaped with respect to $\Station_{1}$ and is therefore connected.
\QED

\subsection{Application to testing reception conditions}
We now describe a direct implication of the hyperbolic convexity property of $\ReceptionZone_{i}(\cA_{d+1})$.
Let $C \in \R^{d+1}$ be a closed shape (not necessarily convex) that does not contain any station, $C \cap S= \emptyset$, contained in the positive (or negative) half-plane $x_{d+1}>0$ (resp. $x_{d+1}<0$), i.e., Inequality (\ref{eq:hyperbolic-condition}) is satisfied for every two points $p_{1},p_{2} \in C$. The following corollary uses the hyperbolic convexity of $\ReceptionZone_{i}(\cA_{d+1})$ to show that if $\Boundary (C)$ receive the transmission by $\Station_i$ successfully, so is any internal point $p \in C$. In addition, if no point on the boundary, $\Boundary (C)$, is able to receive the transmission by $\Station_i$ successfully, then $\SINR(\Station_i,p) <\beta$ for any internal point $p \in C$. In other words, for any closed shape $C$ such that $\Boundary(C) \cap \Boundary(\ReceptionZone_{i}(\cA_{d+1}))=\emptyset$, by testing merely the boundary $\Boundary(C)$ for reception of $\Station_{i}$, one can deduce about the reception of an internal point $p \in C$.
\begin{corollary} \label{corollary:hyperbolic_ClosedShape}
(a) if $\Boundary (C) \subseteq \ReceptionZone_{i}(\cA_{d+1})$, then $C \subseteq \ReceptionZone_{i}(\cA_{d+1})$.
(b) if $\Boundary (C) \cap\ReceptionZone_{i}(\cA_{d+1})=\emptyset$, then $C \cap\ReceptionZone_{i}(\cA_{d+1})=\emptyset$.
\end{corollary}
\commabs
\Proof
Property (a) follows by Lemma \ref{lem:HCl:hyperbolic_segment}.
To prove property (b) assume, by way of contradiction, that there exists a point $p \in C$ such that $\SINR_{\cA}(\Station_i, p) \geq \beta$. By Thm. \ref{thm:d+1_Connected}, $\ReceptionZone_{i}(\cA_{d+1})$ is connected and is hyperbolic star-shaped with respect to $\Station_i$. This implies that there exists an arc
$\arc{p}{\Station_i}$ such that
$\arc{p}{\Station_{i}} \subseteq \ReceptionZone_{i}(\cA_{d+1})$.
Since $p$ is an internal point and $\Station_i \notin C$, the arc
$\arc{p}{\Station_i}$ must intersect $\Boundary(C)$, implying that
there exists some point $q \in \Boundary(C)$ such that
$\SINR_{\cA}(\Station_i, q) \geq \beta$, contradiction.
\QED
\commabsend

\commabs
\section{Systems of infinitely many weak stations (wires)}
\label{section:wires_formulation}

\subsection{Wire Stations}
\label{subsection:Wire Stations}

The next construction we present achieves $\NZones_1=\log \Power_{1}$ zones.
It is obtained by using ``wires" composed of infinitely many weak stations
as described next. We assume $\R^{2}$ and $\alpha=2$.

Let $\Boundary (\Ball(q, r))$ be the circumference of a ball of radius $r>0$
centered at $q \in \R^{2}$.
To avoid cumbersome notation, without loss of generality, let $q=(0,0)$.
Let  $p(r,\theta)=(r \cos \theta, r \sin \theta)$ be a point
on $\Boundary (\Ball(q, r))$.
Consider a positive integer $\chi\geq 1$. Let $\Delta_\chi=2\pi/\chi$ and let
$\thetaci=i\cdot\Delta_\chi$.
Denote by $W_{\chi}$ the ``discrete wire" composed of $\chi$ equally spaced
stations positioned on $\Boundary (\Ball(q, r))$ with total energy $\Power$.
That is, the stations of the wire are positioned at the points
$\{p(r,\thetaci) \mid 0 \leq i \leq \chi-1\}$,
and the power of each such station is fixed to $\Power/\chi$.

In what follows we extend and slightly abuse the notion of a station in a wire
and its geometric location. We define a \emph{continuous wire} (or just
a \emph{wire}) $W(q,r, \Power)$ as the limit of the discrete wire
$W_{\chi}(q, r, \Power)$ as $\Delta_\chi$ gets infinitesimally large.
That is,
\begin{equation}
\label{eq:cont_wire}
W(q,r, \Power) ~=~ \lim_{\chi \to \infty} W_{\chi}(q,r, \Power).
\end{equation}
Let $\Interference(W, k)$ denote the interference experienced at point $k$ due to the wire $W(q,r, \Power)$. In what follows we derive a formulation for $\Interference(W, k)$ and describe its geometric interpretation.

Finally, we present two tasks for which this formulation is found to be useful.

\begin{claim}
\label{cl:wire_inter}
Let $W=(q,r,\Power)$ be a continuous wire.
Then $\Interference(W,k)=\Power/| r^{2}- \dist{q,k}^{2}|$
for $k \notin \Boundary(\Ball(q,r))$.
\end{claim}
\Proof
Without loss of generality, let $q=(0,0)$ and $k=(-x,0)$.
Note that generality is maintained since the interference caused by the wire
is equal in all directions (\emph{omnidirectional}).
We begin with the case where $x > r$. The interference experienced at point
$k=(-x,0)$ due to the discrete wire $W_{\chi}$ is given by
\begin{eqnarray}
\label{eqn:disc_wire_inter}
\Interference(W_{\chi}, k) &=& \sum_{i=0}^{\chi -1}
\Interference(p(r,\thetaci ), k) ~=~
\frac{\Power}{\chi} \cdot \sum_{i=0}^{\chi -1}
\frac{1}{\dist{p(r,\thetaci), k}^{2}} \nonumber
\\&=&
\frac{\Power}{2\pi} \cdot \sum_{i=0}^{\chi -1} \frac{\Delta_\chi}
{\dist{p(r,\thetaci ), k}^{2}}~.
\end{eqnarray}
By Equation (\ref{eq:cont_wire}) it follows that for the continuous wire $W$,
\begin{eqnarray}
\label{eqn:cont_wire_inter}
\Interference(W, k) &=&
\lim_{\chi\to \infty}\Interference(W_{\chi},k) ~=~
\frac{\Power}{2\pi} \cdot \lim_{\chi\to \infty} \sum_{i=0}^{\chi -1}
\frac{\Delta_\chi}{\dist {p(r,\thetaci), k}^{2}} \nonumber
\\&=&
\frac{\Power}{2\pi} \cdot \int_0^{2\pi} \frac{d\theta}{\dist{p(r,\theta),k}^2}~,
\end{eqnarray}
where the last equality holds by definition of the integral of Riemann , where $\Delta_\chi=\theta^\chi_{i+1}-\thetaci$, for every $0\leq i\leq \chi-1$.
Recalling that $p(r,\theta)=(r \cos \theta, r \sin \theta)$, we get that
\begin{eqnarray*}
\dist{p(r,\theta), k}^{2} &=& x^{2}+2 r \cdot x \cdot \cos\theta+r^{2} ~=~
2 r \cdot x \cdot \left (\frac{x^{2}+r^{2}}{2 r \cdot x}+\cos \theta \right).
\end{eqnarray*}
For simplicity of notation, let $a=2 r \cdot x$ and let $b=(x^{2}+r^{2})/a$. Then $\dist{p(r,\theta), k}^{2}=a \cdot(b+\cos \theta)$. Plugging this in Eq. (\ref{eqn:cont_wire_inter}) we get that
\begin{eqnarray}
\label{eqn:cont_wire_inter_final}
\Interference(W, k) &=& \frac{\Power}{2\pi} \cdot \int_{0}^{2\pi} \frac{1}{a}
\cdot \frac{1}{b+\cos \theta} ~=~ \frac{\Power}{2\pi} \cdot \frac{1}{a}
\cdot \frac {2 \pi}{\sqrt{b^{2}-1}} \nonumber
\\&=&
\frac {\Power}{\sqrt{a^{2}\cdot b^{2}-a^{2}}} ~=~
\frac {\Power}{\sqrt{ (x^{2}+r^{2})^{2}-4 r^{2} \cdot x^{2}}} ~=~
\frac {\Power}{x^{2}-r^{2}}~,
\end{eqnarray}
as required. The complementary case, $x < r$, is analogous,
details are omitted.
\QED

\begin{figure}[htb]
\begin{center}
\includegraphics[scale=0.3]{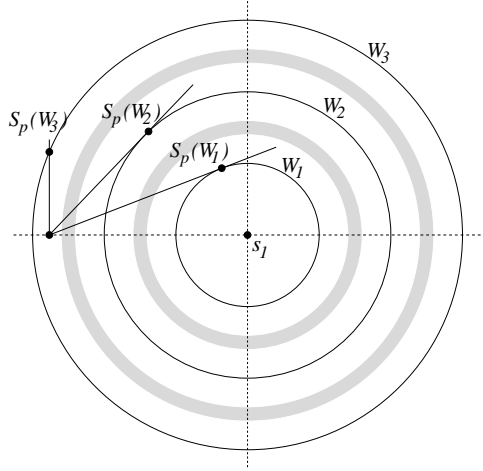}
\caption{ \label{figure:WireGeo}
\sf Schematic Representation of the Wires Construction.
Each wire is composed of infinitely many weak stations positioned on
the circumference of a ball. The shadowed area corresponds to reception cells
of $\Station_1$. $S_{p}(W_{i})$, $i \in \{1,2,3\},$ is a dummy station
whose interference on $p$ is equivalent to that of $W_{i}$
when transmitting with power $\Power_i$.
}
\end{center}
\end{figure}

It is noteworthy that this formulation has a nice geometric interpretation (see Figure \ref{figure:WireGeo}). Given a point $k$, we say that $W(q,r,\Power)$ is \emph{inner} if $k \notin \Ball(q,r)$ and \emph{outer} otherwise.
Without loss of generality, let $q=(0,0)$ and $k=(-x,0)$. The interference of $W(q,r,\Power)$ on point $k$ can be represented by placing a single station $S_{k}(W)$ with transmitting power $\Power$ as follows. When wire $W(q,r,\Power)$ is \emph{outer} with respect to $k$, the coordinates of $S_{k}(W)$ are given by $(-x,r^{2}-x^{2})$. Similarly, when $W(r,\Power)$ is \emph{inner} with respect to $p$, the coordinates of $S_{p}(W_{i})$ are given by $(-r^{2}/x,r\sqrt{x^{2}-r^{2}}/x)$ (i.e., $S_{k}(W)$ corresponds to the touching point of the straight line going through $k$ and tangent to $\Boundary (\Ball(q,r))$).

In the remainder of this section we present two applications
of this formulation.

\subsection{Construction of $\Omega(\log \Power_{1})$ reception zones
 for a single station}

Let $\Station_{1}$ be a station positioned at the origin $q=(0,0)$, with power $\Power_{1}$.
In this section we show that one can induce $\rho+1$ cells of $\Station_{1}$ by using $\rho$ wires, where $\rho=\Omega(\log \Power_{1})$. Consider a collection of wires of increasing radii $r_{i}$ around $\Station_1$ given by the sequence $\overline{W}=\{W_{1}(q,r_{1},1), \ldots ,W_{\rho}(q,r_{\rho},1)\}$, where $r_{i}=4^{i}$. The network is given by \( \cA = \langle d=2, \{\Station_{1},\overline{W}\}, \{\Power_1,\overline{1}\}, N, \beta, \alpha=2 \rangle \).
In what follows, we show that this setting induces $\rho+1$ reception cells for $\Station_{1}$. For a pictorial description of the generated cells see Figure \ref{figure:WireGeo}.
Let $p_{i}=(x_{p_{i}},0)$ be such that $x_{p_{i}} \in [r_{i-1},r_{i}]$.
\begin{claim}
There exists a sequence $\{p_{i}\}$ such that $p_{i} \in \ReceptionZone_{1,i}$ for $i \in [1, \rho]$.
\end{claim}
\Proof
Let $p_{1}=(0,0)$ and let $x_{p_{i}}= (2r_{i-1}+r_{i})/3,$ for $i \in [2, \rho+1]$. Note that by taking $r_{i}=4^{i}$ it follows that $p_{i}=2r_{i-1}= r_{i}/2$.
We now verify that $\Station_{1}$ is correctly received at each $p_{i}$. Note that any two points $p_{i},p_{j}$ are separated by at least one impenetrable wire, in the sense that the weak stations of the wire cannot hear any other transmitter. Consequently, any two points $p_{i},p_{j}$ indeed correspond to two disconnected cells of $\Station_{1}$. The case of $p_{1}$ is trivial as it corresponds to the station itself. Next, consider $p_{i}$ for some $i \in \{2, \ldots ,\rho+1\}$. The interference by the wires experienced at $p_{i}$ can be divided into two terms
\begin{equation}
\label{eq:small_wire_inter}
\Interference_{1}(\overline{W},p_{i}) ~=~
\sum_{j=1}^{i-1} \Interference(W_{j},p_{i}),
\end{equation}
i.e., interference caused by wires $W_{j}$ where $r_{j} < x_{p_{i}}$, and in addition,
\begin{equation}
\label{eq:large_wire_inter}
\Interference_{2}(\overline{W},p_{i}) ~=~
\sum_{j=i}^{\rho} \Interference(W_{j},p_{i}),
\end{equation}
corresponding to interference caused by wires $W_{j}$ where $r_{j} > x_{p_{i}}$. Clearly, $\Interference(\overline{W},p_{i})=\Interference_{1}(\overline{W},p_{i})+\Interference_{2}(\overline{W},p_{i})$. By Claim \ref{cl:wire_inter}, Equation (\ref{eq:small_wire_inter}) can be rewritten as
\begin{eqnarray}
I_{1}(\overline{W},p_{i}) &=& \sum_{j=1}^{i-1} \frac{1}{x_{p_i}^2-r_j^2} ~<~
\sum_{j=1}^{i-1} \frac{1}{3r_{j}^{2}} ~=~
\sum_{j=1}^{i-1} \frac{1}{3 \cdot 4^{2j}} ~<~ 1,
\end{eqnarray}
where the first inequality follows by the fact that
$x_{p_{i}} \geq 2 \cdot r_{j}$, $j \in [1,i-1]$.
In the same manner, Eq. (\ref{eq:large_wire_inter}) can be rewritten as
\begin{eqnarray}
\Interference_{2}(\overline{W},p_{i}) &=& \sum_{j=i}^{\rho}
\frac{1}{r_{j}^{2}-x_{p_{i}}^{2}} ~<~
\sum_{j=1}^{i-1} \frac{4}{3 r_{j}^{2}} ~=~
\sum_{j=1}^{i-1} \frac{4}{3 \cdot 4^{2j}} ~<~ 5,
\end{eqnarray}
where the first inequality follows by the fact that
$x_{p_{i}} \leq 1/2 \cdot r_{j}$, for every $j \in \{i,...,\rho\}$.
Overall, we get that
\begin{eqnarray*}
\SINR_{\cA}(\Station_{1},p_{i}) &=& \frac{\Power_{1}}{x_{p_{i}}^{2} \cdot
(\Interference(\overline{W},p_{i})+N)} ~\geq~ \frac{\Power_{1}}{16^{i-2}
\cdot (6+N)} ~>~ \frac{\Power_{1}}{16^{\rho-1} \cdot (7+N)}\geq 1.
\end{eqnarray*}
The claim follows.
\QED
\commabsend

\commabs
\subsection{The interference function and the maximum principle}
\label{subsec:maximum_principle}
\commabsend
\commful
\dnsparagraph{The interference function and the maximum principle:}
\commfulend
\commabs
We next show an interesting property of the interference function for which
the tools presented in Section \ref{section:wires_formulation} become useful.
\commabsend
Throughout this section we consider the $2$-dimensional Euclidean plane and assume $\alpha=2$.
Let $f$ be a function defined on some connected closed subset $D$ of the Euclidean space $\R^{d}$. Let $\Boundary (D)$ denote the boundary of the domain. Then $f$ follows the \emph{maximum principle} if the maximum of $f$ in the domain $D$ is attained on its boundary $\Boundary (D)$.
In this section we establish the following theorem.
\begin{theorem}
\label{thm:max_princ}
The interference function $\Interference(S\setminus \{\Station_1\},p)$ follows the \emph{Maximum principle}.
\end{theorem}
We begin by establishing an auxiliary claim. Let $B=\Ball(q,r)$ be a ball with radius $r$ and center $q \in \R^{2}$. Let $\Station_{i}$ be a station positioned at $(x_i,y_i)$ with power $\Power_{i}$ where $\Station_{i} \notin \Ball(q,r)$. The \emph{average interference} experienced at $\Boundary (B)$ due to $\Station_{i}$ is denoted by $\varepsilon_{i}(\Boundary (B))$.
\commabs
Note that this scenario is the dual to the \emph{continuous} wire scenario. In the wire case, the boundary of the ball,  $B$,  corresponds to \emph{stations} and $q$ is the point where we evaluate interference. Here, the \emph{single} point corresponds to a transmitting \emph{station} and the circumference $B$ is where we evaluate the interference  caused by this station.
\commabsend
\begin{claim}
\label{cl:everage_inter}
$\varepsilon_{i}(\Boundary (B)) =\Power_{i}/|\dist{\Station_{i},p}-r^{2}|$.
\end{claim}
\commabs
\Proof
Without loss of generality, let $q=(0,0)$.
As before, we begin by considering the discrete case.
Consider a positive integral $\chi$. Let $\Delta_\chi=2\pi/\chi$ and let
$\thetacj=j\Delta_\chi$, for every $j\in\{0,...,\chi-1\}$.
Denote by $\Boundary (B,\chi)$ a discrete circumference of $\Ball(q,r)$, corresponding to
a collection of $\chi$ equally spaced points on $B$ given by
$\{p(r,\thetacj) \mid 0 \leq j \leq \chi-1\}$.
Note that in contrast to the construction of Claim \ref{cl:wire_inter},
the points of $\Boundary (B,\chi)$ do not correspond to stations.
The expected interference on $\Boundary (B,\chi)$ is given by
\begin{eqnarray*}
\varepsilon_{i}(\Boundary (B,\chi)) &=& \frac{\sum_{j=0} ^{\chi-1}
\Interference(\Station_i,p(r,\thetacj))}{\chi} ~=~
\frac{\Power_{i}}{2\pi} \cdot \sum_{j=0} ^{\chi-1}
\frac{\Delta_\chi}{\dist{\Station_{i},p(r,\thetacj)}^{2}}~.
\end{eqnarray*}
For the continuous case, the expected interference is given by
\begin{eqnarray*}
\label{eqn:average_inter_cont}
\varepsilon_{i}(\Boundary (B)) &=& \lim_{\chi \to \infty} \varepsilon_{i}(\Boundary (B,\chi))
\\&=&
\frac{\Power_{i}}{2\pi} \cdot \lim_{\Delta \theta \to 0} \sum_{j} ^{\chi-1}
\frac{\Delta \theta}{\dist{\Station_{i},p(r,j \cdot \Delta \theta)}^{2}}
\nonumber
\\&=&
\frac{\Power_{i}}{2\pi} \cdot \int_{0}^{2 \pi} \frac{d \theta}
{\dist{\Station_{i},p(r,\theta)}^{2}} ~=~
\frac{\Power_{i}}{|\dist{\Station_{i},p}-r^2|}~,
\end{eqnarray*}
where the last equality follows by Claim \ref{cl:wire_inter}. The claim follows.
\QED
\commabsend
We now turn to prove Theorem \ref{thm:max_princ}.
\Proof
Let $D \subseteq \R^{d}$ be a closed connected subset. We require $f$ to be continuous on $D$, and therefore $D$ is empty of interfering stations, that is, $D \cap (S \setminus \{\Station_1\})= \emptyset$. We then wish to show that
\commabs
\begin{equation*}
\max_{q \in D} \Interference(S \setminus \{\Station_1\},q) ~\leq~
\max_{q \in \Boundary D} \Interference(S \setminus \{\Station_1\},q).
\end{equation*}
\commabsend
\commful
$\max_{q \in D} \Interference(S \setminus \{\Station_1\},q) \leq \max_{q \in \Boundary D} \Interference(S \setminus \{\Station_1\},q).$
\commfulend
Assume toward contradiction that there exists an internal point $p \in D \setminus \Boundary (D)$ such that $\Interference(S \setminus \{\Station_1\},p)> \max_{q \in \Boundary (D)} \Interference(S \setminus \{\Station_1\},q)$. Let $\Ball(p,r)$ be the maximal ball  inside $D$ centered at $p$ (since $p$ is an internal point of $D$, $r >0$). Then by the maximality of $p$ and linearity of expectation
\commabs
\begin{equation*}
\Interference(S \setminus \{\Station_1\},p) ~\geq~
\sum_{i=2}^{n} \varepsilon_{i}(\Boundary (\Ball(p,r))).
\end{equation*}
\commabsend
Plugging Claim \ref{cl:everage_inter}, we have that
\begin{equation}
\label{ineq: Interference S setminus a1 > sum i=1 n}
\Interference(S \setminus \{\Station_1\},p) \geq \sum_{i=2}^{n} \frac{\Power_{i}}{|\dist{\Station_{i},p}^2-r^{2}|}~.
\end{equation}
In addition, the fact that $D\cap(S\setminus\{s_1\})=\emptyset$, implies that $\dist{s_i,p}>r$.
Combining this together with the
the definition of interference $\Interference(S \setminus \{\Station_1\},p) = \sum_{i=2}^{n} \frac{\Power_{i}}{\dist{\Station_{i},p}^2}$,
we get a contradiction to Equation (\ref{ineq: Interference S setminus a1 > sum i=1 n}),
which is contradiction to the maximality of $p$.
\QED
\commabs
\section{The fatness of the reception zones}
\label{section:Fatness}

In Section~\ref{section:Connectivity}, we showed that the reception zone $\ReceptionZone_{i}(\cA_{d+1})$ of each station $\Station_{i}$ in a \NUPN{} is hyperbolic-convex.
In this section we develop a deeper understanding of the shape of the
reception zones $\ReceptionZone_{i}$ and $\ReceptionZone_{i}(\cA_{d+1})$ by analyzing their fatness.
Consider a \NUPN{} \( \cA = \langle d=2, S, \Power, \Noise, \beta, \alpha=2
\rangle \), where \( S = \{\Station_1, \dots, \Station_{n}\} \) and
$\alpha > 0$ and $\beta > 1$ are constants.
We focus on \(\Station_1\) and assume that its location is not shared by any
other station (otherwise, its reception zone is \( \ReceptionZone_{1} =
\{\Station_1\} \)). In addition, without loss of generality, we let the minimal transmission energy be 1 and denote the maximal energy by \(\MaxPower\).

In Section~\ref{section:FatnessExplicit}, we establish explicit bounds on
the maximal and minimal radii
\(\LargeRadius(\Station_1, \ReceptionZone_{1})\) and \(\SmallRadius(\Station_1,
\ReceptionZone_{1})\) of the zone \(\ReceptionZone_{1}\).
In addition, we provide a bound on the perimeter of $\ReceptionZone_{i}$ by bounding the length of the curve $\Boundary (\ReceptionZone_{i})$.
\subsection{Explicit bounds}
\label{section:FatnessExplicit}

The goal of this section is to establish an explicit lower bound on
\(\SmallRadius(\Station_1, \ReceptionZone_{1}(\cA))\) and an explicit upper bound on
\(\LargeRadius(\Station_1, \ReceptionZone_{1}(\cA))\).
\commabsend

\commabs
To avoid cumbersome notation, we assume a two-dimensional space ($d=2$)
throughout this section;
the proof is trivially generalized to arbitrary dimensions $d$.

Fix \( \MinDist = \min \{ \dist{\Station_1, \Station_i} \mid i > 1 \} \).
For establishing a lower bound on
\(\SmallRadius(\Station_1, \ReceptionZone_{1})\), an extreme scenario (making $\SmallRadius$ as small as possible) would be to place
\(\Station_1\) at $(0, 0)$ with $\Power_1=1$ and all other \( n - 1 \) stations at $(\MinDist,
0)$ with $\Power_i = \MaxPower$ for $i \in \{2, \ldots ,n\}$. For the sake of analysis, let us replace the noise $N$ by a new imaginary station $\Station_{n+1}$ located at $(\MinDist ,0)$ whose power is $\Noise \cdot \kappa^{2}$. This introduces the \NUPN{} $\cA^{\SmallRadius} = \langle d=2, \{ (0, 0),(\MinDist, 0), \dots, (\MinDist, 0) \}, \{1, \MaxPower, \ldots, \MaxPower,N \cdot \kappa^{2}\}, 0, \beta, \alpha=2 \rangle$. Note that the energy of the new station \(\Station_{n+1}\) at point \((x, 0)\)
satisfies
(1) \( \Energy(\Station,(x,0)) > \Noise\) for all \( 0<x<\MinDist \);
(2)\( \Energy(\Station_{n+1},(x,0)) = \Noise\) for \( x = 0 \); and
(3) \( \Energy(\Station_{n+1},(x,0)) < \Noise\) for all \( x < 0 \).
Therefore, the value of \(\SmallRadius(\Station_1, \ReceptionZone_{1})\) can only get smaller by this replacement, i.e., \(\SmallRadius(\Station_1, \ReceptionZone_{1}(\cA^{\SmallRadius}))< \SmallRadius(\Station_1, \ReceptionZone_{1}(\cA))\).
The point \(q_{\SmallRadius}\) whose distance to \(\Station_1\) realizes
\(\SmallRadius(\Station_1, \ReceptionZone_{1})\) is thus located at
$(\dhat, 0)$ for some \( 0 < \dhat < \MinDist \) that satisfies
the equation $\SINR_{\mathcal{A}^{\delta}}(\Station_{1},q_{\SmallRadius})=\beta$, or,
\begin{equation*}
\frac{\dhat^{-2}}{(\MaxPower(n-1)+\Noise \cdot \MinDist^{2})
(\MinDist-\dhat)^{-2}} ~=~ \beta~.
\end{equation*}
Solving for $\dhat$ yields
\begin{equation}
\label{eq:sm_d}
\dhat ~=~ \frac{\MinDist}{\sqrt{\beta(\MaxPower(n-1) +
\Noise \cdot \MinDist^{2})}+1} ~\geq~
\frac{\MinDist}{2\sqrt{2\beta \cdot \MaxPower \cdot  n)}}~,
\end{equation}
where the inequality follows by assuming that
$\Noise \cdot \MinDist^{2} \leq \MaxPower \cdot n$.
Hence we have the following.
\begin{lemma}
\label{lem:explicit_delta_bound}
$\SmallRadius(\Station_1, \ReceptionZone_{1}(\cA)) \geq
\MinDist / \sqrt{\MaxPower \cdot n}$.
\end{lemma}
To establish an upper bound on
\(\LargeRadius(\Station_1, \ReceptionZone_{1}(\cA))\),
consider the case where $\Station_{1}$ transmits with power $\MaxPower$
while the other stations remain silent $(\Power_i=0,~ \mbox{for}~ i>1)$.
The point \(q_{\LargeRadius}\) whose distance to \(\Station_1\) realizes
\(\LargeRadius(\Station_1, \ReceptionZone_{1})\) is thus located at
$(\pm \dhat, 0)$ such that
$\dhat \leq \sqrt{\MaxPower/(\beta \cdot \Noise)}~,$
hence we get the following.
\begin{lemma}
\label{lem:explicit_Delta_bound}
$\LargeRadius(\Station_1, \ReceptionZone_{1}) \leq
\sqrt{\MaxPower / (\beta \cdot \Noise)}$.
\end{lemma}
The fatness parameter of \(\ReceptionZone_{1}(\cA)\) with respect to
\(\Station_1\) thus satisfies
\begin{equation}
\label{eq:ratio_d}
\FatnessParameter(\ReceptionZone_{1}(\cA)) ~\leq~
O\left (\frac{\MaxPower}{\MinDist}\cdot \sqrt{\frac{n}{\Noise}}\right)~ .
\end{equation}
\commabsend

\commabs
\subsection{Bounding the perimeter of $\ReceptionZone_{1}(\cA)$}
In this section we provide an upper bound on the perimeter length $\Perimeter(\ReceptionZone_{1}(\cA))$. The perimeter of a cell $\ReceptionZone_{1,i}(\cA)$ is the length of the closed curve given by $\Boundary (\ReceptionZone_{1,i}(\cA))$. The perimeter length of a zone is the sum of  the perimeters of the cells it contains.
Again we assume $d=2$ for clarity of presentation, yet the bound can be naturally extended to any dimension $d$. In the case of an \UPN, ~a bound on the perimeter of $\ReceptionZone_{1}(\cA)$ is simply given by the perimeter of the large disk of radius \(\LargeRadius(\Station_1, \ReceptionZone_{1}(\cA))\). In the case of a \NUPN, $\ReceptionZone_{1}(\cA)$ is non-convex and therefore the trivial bound of $2 \pi \cdot \LargeRadius(\Station_1, \ReceptionZone_{1}(\cA)) =O(\sqrt{\MaxPower/\Noise})$ does not hold. We begin by providing the following useful fact in this context.
\begin{fact}
\cite{Santalo04}
\label{fc:curve_length}
Let $C_{out}$ be a closed curve of length $l_{out}$. Let $C_{in}$ be a curve of length $l_{in}$ enclosed by $C_{out}$. Let $\mathbb{Y} (L,C)$ be the number of intersection points between the straight line $L$ and the curve $C$.
Then there exists a straight line $L$ such that $\mathbb{Y} (L,C_{in}) \geq 2l_{in}/l_{out}.$
\end{fact}

\begin{corollary}
\label{cor:curve_length_SINR}
$\Perimeter(\ReceptionZone_{1}(\cA)) \leq 3 \pi \cdot \LargeRadius(\Station_1, \ReceptionZone_{1}(\cA)) \cdot n^{2} $.
\end{corollary}
\Proof
We first bound from above the perimeter of a cell \(\ReceptionZone_{1,i}(\cA) \subseteq \ReceptionZone_{1} \). Let $f(L)$ be the projection of $F^{1}_{\cA}(p=(x,y))$ on the line $L=ax+b$. Then $\deg(f) \leq 2n$ (when $\Noise \neq 0$) and $\mathbb{Y} (L,\Boundary (\ReceptionZone_{1,i}(\cA))) \leq 2n$. Recall that any connected cell $\ReceptionZone_{1,i}(\cA)$ is enclosed by a disk of radius $\LargeRadius(\Station_1, \ReceptionZone_{1}(\cA))$. Combining this with Fact \ref{fc:curve_length}, we have that there exists a line $L$ such that
\begin{equation*}
\frac{2 \cdot \Perimeter(\Boundary (\ReceptionZone_{1,i}(\cA)))}{2\pi \cdot
\LargeRadius(\Station_1, \ReceptionZone_{1}(\cA))} ~\leq~
\mathbb{Y} (L,\Boundary (\ReceptionZone_{1,i}(\cA))) ~\leq~ 2n.
\end{equation*}
Hence for every $i \in \{1, \ldots, O(n^{2})\}$,
\begin{equation*}
\Perimeter(\Boundary ( \ReceptionZone_{1,i}(\cA) )) ~\leq~
2 \pi \cdot \LargeRadius(\Station_1, \ReceptionZone_{1}(\cA)) \cdot n~.
\end{equation*}
Overall, summing over the connected cells of $\Station_1$, whose number is at most $O(n^{2})$ by Theorem \ref{thm:milnor}, we have that
\begin{equation}
\label{eq:curve_bound}
\Perimeter( \Boundary(\ReceptionZone_{1}(\cA))) ~\leq~
O\left(\LargeRadius(\Station_1, \ReceptionZone_{1}(\cA)) \cdot n^{3} \right).
\end{equation}
The claim follows.
\QED
In summary, in this section we achieved the following.
\begin{theorem} \label{theorem:ExplicitBounds}
In a non-uniform energy network \( \cA = \langle d=2, S,\Power, \Noise, \beta,
\alpha=2 \rangle \), where \( S = \{\Station_1, \dots, \Station_{n - 1}\} \) and
$\alpha > 0$ and $\beta > 1$ are constants, if
\( \MinDist = \min \{ \dist{\Station_1, \Station_i} \mid i > 1 \} > 0 \),
then
\begin{eqnarray}
\Theta \left(\frac{\MinDist^{2}}{\MaxPower \cdot n} \right) &\leq&
\Area(\ReceptionZone_{1}(\cA)) ~\leq~ O \left(\frac{\MaxPower}{\Noise}\right),
\label{eq:bounds_area} \\
\Theta \left (\frac{\MinDist}{\sqrt{\MaxPower \cdot n}}\right) &\leq& \Perimeter(\ReceptionZone_{1}(\cA)) ~\leq~ O \left( n^{3} \cdot \sqrt{\frac{\MaxPower}{\Noise}}\right)~.
\label{eq:bounds_perimeter}
\end{eqnarray}
\end{theorem}
\commabsend

\section{Approximate point location}
\label{sec:point_location}
\commabs
\subsection{The Setting}
\label{subsec:the setting}
\commabsend
Consider a non-uniform power network \( \cA = \langle d=2, \{\Station_1, \dots, \Station, \Psi,\Noise, \beta, \alpha=2 \rangle \). Given some point \( p \in \Reals^{2} \), we are interested in the question: is \( \Station_1\) heard at \(p\) under the interference of $S \setminus \{ \Station_1\}$ and background noise \(\Noise\)?
One can directly compute \(\SINR_{\cA}(\Station_1, p)\) in time \(\Theta (n)\)
and answer the above question.
However, typically, this question is asked for many different points \(p\), thus linear time computations may be too expensive.
\commful
We describe a mechanism that answers some approximated variants of the above question much faster. We begin with notation. Let $\MinDist$ denote the minimal distance between two stations. We focus on \(\Station_1\) and assume that its location is not shared by any other station (otherwise, its reception zone is \( \ReceptionZone_{1} = \{\Station_1\} \)). In addition, without loss of generality, we let the minimal transmission energy be 1 and denote the maximal energy by \(\MaxPower\).
Using the techniques from \cite{Avin2009PODC}, we show that
\begin{claim}
\label{cl:explicit_bound}
$\SmallRadius(\Station_1, \ReceptionZone_{1}) \geq
\frac{\MinDist}{\sqrt{\MaxPower \cdot n}}, ~,~
\LargeRadius(\Station_1, \ReceptionZone_{1}) \leq
\sqrt{\frac{\MaxPower}{\beta \cdot \Noise}}
~~\mbox{and}~~
\FatnessParameter(\ReceptionZone_{1}) \leq
O\left (\frac{\MaxPower}{\MinDist}\cdot \sqrt{\frac{n}{\Noise}}\right)~ .$
\end{claim}
\commfulend
Our goal in this section is to provide mechanisms that answers some approximated variants of the above question much faster.
In Section~\ref{section:Colinear}, we present point location scheme for the case where all stations are aligned on a line. In Section~\ref{section:General networks}, we provide several schemes for point location for the general case where stations are embedded in $\R^{d}$.
Generally speaking, the mechanisms we present construct an efficient data structure that maintains a partition of the Euclidean plane. We consider two types of data structures. The first partitions the plane into three disjoint zones \(
\Reals^2 = \ReceptionZone_{1}^{+} \cup \ReceptionZone_{1}^{-} \cup \ReceptionZone_{1}^{?} \) such that
(1) \( \ReceptionZone_{1}^{+} \subseteq \ReceptionZone_{1} \);
(2) \( \ReceptionZone_{1}^{-} \cap \ReceptionZone_{1} = \emptyset \); and
(3) \(\ReceptionZone_{1}^{?}\) is a bounded set determined by the requested accuracy level of the algorithm. The second type partitions the plane into two disjoint zones \(
\Reals^2 = \ReceptionZone_{1}^{+} \cup \ReceptionZone_{1}^{-}\) such that the set of misclassified points is bounded.
Given a query point \( p \in \Reals^2\), \(\QuerDS_1\) answers in logarithmic time (with respect to the fatness parameter, $1/ \epsilon$ and number of stations) whether \(p\) is in \(\ReceptionZone_{1}^{+}\),
\(\ReceptionZone_{1}^{-}\), or \(\ReceptionZone_{1}^{?}\) (possible only for \(\QuerDS_1\) of first type).
We construct a separate data structure \(\QuerDS_i\) for every \( 1 \leq i \leq n \).

Recall that by Lemma \ref{cl:wv_sinr},
a point \(p\) cannot be in \(\ReceptionZone_{i}\) unless it belongs to $ \wvor_{i}(\wvorsys_{\cA})$, where $\wvor_{i}(\wvorsys_{\cA})$ is the weighted Voronoi cell of $\Station_i$ with weight $\Vweight_i =\Power_i^{1/\alpha}$.
Thus for such a point \(p\) there is no need to query the data structure
\(\QuerDS_j\) for any \( j \neq i \).

Due to \cite{AurenhammerE84}, a weighted Voronoi diagram of quadratic size for the \(n\) stations is constructed in \( O(n ^{2}) \) preprocessing time. Then given a query point \( p \in \Reals^2
\), the station \(\Station_i\) such that $p \in \wvor_{i}(\wvorsys_{\cA})$ can be identified in time \( O (\log n)
\). We then invoke the appropriate data structure \(\QuerDS_i\).

We first provide some notation and then present the common framework for any point-location schemes discussed next. For ease of notation, let the characteristic polynomial of $\ReceptionZone_{1}(\cA)$, namely, $F^{1}_{\cA}(p)$, see Eq. (\ref{eq:reception_polynomial}), be given by $F_{\beta}(p)$. In the same manner, the characteristic polynomial of $\ReceptionZone_{1}(\cA_{\beta'})$, for $\beta'\neq \beta$, is given by $F_{\beta'}(p)$. Let \(\QuerDS=\QuerDS_1\).
In addition, for station $\Station_i \in S$, define $\LargeRadiusBound_i$ as the upper bound on
\(\LargeRadius(\Station_i, \ReceptionZone_{i})\), \(\SmallRadiusBound_i\)
as the lower bound on
\(\SmallRadius(\Station_i,\ReceptionZone_{i})\) and \(\FatnessParameterBound_i\)
as the upper bound on \(\FatnessParameter(\ReceptionZone_{i})\)
(i.e., $\LargeRadiusBound_i /\SmallRadiusBound_i$).\\
\QuerDS{} is based upon imposing a $\GridSpace \in \R_{>0}$-spaced \emph{grid}, denoted by
$\Grid_{\GridSpace}$, on the Euclidean plane, $\GridSpace$ is determined later on.
The notions of grid \emph{columns}, \emph{rows}, \emph{vertices},
\emph{edges}, and \emph{cells} are defined in the natural manner.
We assume that $\Grid_{\GridSpace}$ is aligned so that the point $\Station_1$ is a grid
vertex.

The parameter $\GridSpace$ is set to be sufficiently small so that the cell
containing point $\Station_1$ is internal to the ball inscribed in $\ReceptionZone_{1}$, namely, $\Ball(\Station_1,\SmallRadiusBound_1)$.

In fact, we take $\GridSpace \leq \min \{\SmallRadiusBound_1 / (2 \sqrt{2}) \}$
so that the ball of radius $\SmallRadiusBound_1$ centered at $\Station_1$
is guaranteed to contain $\Omega ((\LargeRadiusBound_1 / \SmallRadiusBound_1)^2)$
cells (all of them are internal by definition).
The main ingredient of our algorithm is a \emph{segment testing} procedure
\cite{Avin2009PODC}, named hereafter Procedure $\SegTest$.
Given a segment $\sigma$, the segment testing procedure returns the number
of distinct intersection points of $\sigma$ and
$\Boundary(\ReceptionZone_{1}(\cA))$. The segment test is implemented to run
in time $O(n^{2})$ by employing Sturm condition \cite{BPR03} of the projection
of the polynomial $F_{\beta}(p)$ on $\sigma$ and by direct calculation
of the SINR function in the endpoints of $\sigma$.
\footnote{By applying advanced numerical techniques, segment test procedure can be implemented in $O(n\log n)$};
In particular, the segment testing allows one to decide whether
$\sigma \cap \ReceptionZone_{1}(\cA)= \emptyset$ or not.
Procedure $\SegTest$, presented formally below, is common to all
point location schemes presented later on.

\commabs
\begin{figure*}[ht]
\begin{center}
\framebox{\parbox{6in}{
{\boldmath Procedure $\SegTest$} ($\sigma,F_{\beta}(p)$)
\begin{enumerate}
\item
Employ Sturm condition on $\sigma$ for $F_{\beta}(p)$, \\
let $t$ be the number of distinct intersection points of
$\sigma$ and $F_{\beta}(p)$.
\item
Evaluate $F_{\beta}(p)$ on $\sigma$ endpoints, $p_{1}$ and $p_{2}$.
\begin{enumerate}
\item
If $t=0$
\begin{itemize}
\item
If $F_{\beta}(p_1) > 0 ~~and ~~F_{\beta}(p_2)> 0$ return $-$;
\item
Else, return $+$;
\end{itemize}
\item
Else, return $?$;
\end{enumerate}
\end{enumerate}
}}
\end{center}
\end{figure*}
\commabsend


Given a grid $\Grid_{\GridSpace}$, Procedure $\SegTest$ is invoked for each of the 4 edges for every cell $c_{i} \in \Grid_{\GridSpace}$ at distance at most $\LargeRadiusBound_1$ from $\Station_{1}$. The overall number of invocations is thus bounded by $O\left(\pi \cdot \LargeRadiusBound_1^{2} /\GridSpace^{2} \right)$. The difference between the schemes we present is in the definition of the performance parameter $\epsilon$. We conclude this section by evaluating $M_{\cA}(\QuerDS_{})$ respectively, $T_{\cA}(\QuerDS_{})$ corresponding to the memory, respectively time costs of the procedure and the schemes that use it, in terms of $\GridSpace$. Each of the schemes chooses $\GridSpace$ so that the error is controlled (where the precise notion of error is scheme-dependant). We begin with bounding the size of the data structure \(\QuerDS_{}\). Let $C_{\GridSpace}$ denote the number of cells in $\Grid_{\GridSpace}$ then due to area consideration
$C_{\GridSpace} =O \left( \left(\frac{\LargeRadiusBound_1}{\GridSpace} \right)^{2} \right)$.
It is required to keep the tag of each cell tag , therefore \(\QuerDS_{}\) is of size
\begin{equation}
\label{eq:pointlocation_generalmemory}
M_{\cA}(\QuerDS_{}) ~=~O \left( \left(\frac{\LargeRadiusBound_1}{\GridSpace} \right)^{2} \right)~.
\end{equation}
Note that it is sufficient to keep in \(\QuerDS_{}\) only cells in $\ReceptionZone_{1}^{+} \cup \ReceptionZone_{1}^{?}$. Next we bound the time complexity, $T_{\cA}(\QuerDS_{})$. The dominating step is the invocations of Procedure $\SegTest$. As the cost of a single $\SegTest$ invocation is $O(n^{2})$ and there are $O(C_{\GridSpace})$ invocations, the processing time for \(\QuerDS_{}\) construction is given by
\begin{equation}
\label{eq:pointlocation_generaltime}
T_{\cA}(\QuerDS_{}) ~=~ O \left(  \left(\frac{n \cdot \LargeRadiusBound_1}{\GridSpace} \right)^{2} \right) ~.
\end{equation}
Finally, we analyze the cost for a single point location query. This is bounded by
\begin{equation}
\label{eq:pointlocation_generalquerytime}
T^{query}_{\cA}(\QuerDS_{}) ~=~ O \left(\log C_{\GridSpace} \right)=O \left( \log  \left( \frac{\LargeRadiusBound_1}{\GridSpace} \right) \right)~,
\end{equation}
which corresponds to the time for finding the cell to which $p$ belongs. The latter can be done by preforming binary search on $C_{\GridSpace}$ cells. Recall that there is a prior step involving an access to the weighted Voronoi diagram data structure. As mentioned, that step is bounded by $O(\log n)$, which is dominated by $O(\log C_{\GridSpace}).$

\commabs

\subsection{Collinear networks}
\label{section:Colinear}

In this subsection we focus on the Euclidean plane \(\Reals^2\) and
consider a special type of \NUPN{}.
A network \( \cA = \langle d=2, \{\Station_1, \dots, \Station_{n - 1}\}, \Psi,
\Noise, \beta, \alpha=2 \rangle \) is said to be \emph{collinear} \cite{Avin2009PODC} if \(
\Station_1 = (0, 0) \) and \( \Station_i = (a_i, 0) \) for \( a_i \in \R \)
for every \( 1 \leq i \leq n - 1 \).
the point-location task is simpler for collinear networks
due to the following lemma.

\begin{lemma} \label{claim:pointlocation_colinearhyper}
Let \(\cA\) be a colinear \NUPN{}. Then \(\ReceptionZone_{1}\) is hyperbolic-convex and therefore connected.
\end{lemma}
\Proof
The proof follows immediately by Thm. \ref{thm:d+1_Connected},
in Section \ref{section:HyperbolicConvexity},
setting $d=1$. Specifically, the stations of colinear network are essentially
embedded in $\R^{1}$, and therefore their 2-dimensional reception zones
$\ReceptionZone_i(\cA_{d=2})$ are hyperbolic-convex.
\QED

Note that by Lemma \ref{claim:pointlocation_colinearhyper}, the reception zones
$\ReceptionZone_{1}$ of colinear network follow the property of
Corollary \ref{corollary:hyperbolic_ClosedShape}.
We now complete the description of the data-structure \(\QuerDS{}\).
Let $c_{i} \in \Grid_{\GridSpace}$ be a grid cell.
Procedure $\SturmCellB$ is a tagging mechanism invoked for every cell
$c_{i} \in \Grid_{\GridSpace}$ (in fact, due to symmetry it is sufficient
to restrict attention to the half space $y \geq 0$).

\begin{figure*}[ht]
\begin{center}
\framebox{\parbox{4in}{
{\boldmath Algorithm $\SturmCellB$} ($c_{i},F_{\beta}(p)$)
\begin{enumerate}
\item
For any edge $e_{j}$ of $c_{i}$
\begin{itemize}
\item
$t_{j} \gets \SegTest(e_{j},F_{\beta}(p));$
\end{itemize}
\item
If $t_{j}=~-~$ for any $j \in \{1,...,4\}$ return $-$;
\item
If $t_{j}=~+~$ for any $j \in \{1,...,4\}$ return $+$;
\item
return $?$;
\end{enumerate}
}}
\end{center}
\end{figure*}

 $\QuerDS{}$ maintains the collection of $\ReceptionZone_{1}^{?} \cup \ReceptionZone_{1}^{+}$ cells, where $c_{i} \in \ReceptionZone_{1}^{?}$ if $\SturmCellB(c_{i},\ReceptionZone_{1})$ returns $?$ and $c_{i} \in \ReceptionZone_{1}^{+}$ if $\SturmCellB(c_{+},\ReceptionZone_{1})$ returns $+$.
We begin by bounding the number of cells in $\ReceptionZone_{1}^{?}$. Let $C_{\GridSpace}$ be the number of rows and columns in $\Grid_{\GridSpace}$.
Then $C_{\GridSpace} \leq 4\pi \LargeRadiusBound_1/\GridSpace$.
Since $\deg(F_{\beta}) \leq 2n$, the number of intersection points of $F_{\beta}(p)$ with any grid row or column is at most $2n$ (see Eq. (\ref{eq:reception_polynomial} for definition). Overall, we get that the total number of intersection points of $F_{\beta}(p)$ with any of the $C_{\GridSpace}$ rows and columns in $\Grid_{\GridSpace}$
is at most \footnote{Note that by the hyperbolic convexity property of
$\ReceptionZone_{1}$ we have that the number of intersection points
of $F_{\beta}(p)$ and any vertical line (grid column) is at most 4,
see Corollary \ref{cor:hyperbolic_segment_intersection};
To keep things simple we do not take it into account.} $2n \cdot C_{\GridSpace}$.
Hence the total number of $\ReceptionZone_{1}^{?}$ cells is bounded by $2n \cdot C_{\GridSpace}$. Since the area of each cell is $\GridSpace^{2}$, it follows that
\begin{equation}
\label{eq:pointlocation_colinques}
\Area(\ReceptionZone_{1}^{?}) ~\leq~ 8\pi \cdot n\cdot \LargeRadiusBound_1 \cdot\GridSpace~.
\end{equation}
In order to guarantee that $\Area(\ReceptionZone_{1}^{?}) \leq \epsilon \cdot \Area(\ReceptionZone_{1})$, we demand that $8\pi \cdot n \cdot \LargeRadiusBound_1 \cdot\GridSpace \leq \epsilon \cdot \pi \SmallRadiusBound_1^{2}$ (this is sufficient as $\Area(\ReceptionZone_{1}) \geq \Ball(\Station_1,\SmallRadiusBound_1)$).
Therefore it is sufficient to fix
\begin{eqnarray}
\label{eq:pointlocation_colinear_gamma}
\GridSpace &=&\frac{\epsilon \SmallRadiusBound_1}{8 n \cdot
\FatnessParameter}
\end{eqnarray}

We are now ready to establish the correctness of Procedure $\SturmCellB$.

\begin{lemma}
\label{lemma:pointlocation_colinear_correctness}
(a) If $\SturmCellB(c_{i},\ReceptionZone_{1}(\cA))$ returns $+$,
then $c_{i} \subseteq\ReceptionZone_{1}(\cA)$. \\
(b) If $\SturmCellB(c_{i},\ReceptionZone_{1}(\cA))$ returns $-$,
then $c_{i} \cap \ReceptionZone_{1}(\cA)=\emptyset$. \\
(c) Let $c_{i} \subseteq \Ball(\Station_1,\Delta)$ be such that
$\SturmCellB(c_{i},H(\Station_1,\beta))$ returns $?$.
Then the total area of such $c_{i}$ cells is bounded from above
by $\epsilon \cdot \Area(\ReceptionZone_{1})$.
\end{lemma}
\Proof
(a) and (b) follow by Corollary \ref{corollary:hyperbolic_ClosedShape},
where the grid cell $c_{i}$ corresponds to a closed shape whose circumference
is tested. Finally (c) is guaranteed by the way we set $\GridSpace$.
\QED

Let $\MaxFatnessParameterBound=\max_{i=1}^{n}\{\FatnessParameterBound_{i}\}$
and $\SumFatnessSquares^{4} = \sum_{i=1}^{n}\FatnessParameterBound_{i}^{4}$.
Throughout this section we established the following theorem, by Eq.
(\ref{eq:pointlocation_generaltime}, \ref{eq:pointlocation_generalmemory},
\ref{eq:pointlocation_generalquerytime}, and
\ref{eq:pointlocation_colinear_gamma}).
\begin{theorem}
\label{theorem:pointlocation_colinear}
Given a a colinear \NUPN{} \(\cA\), it is possible to construct,
in $\widetilde{O} (n^{4} \cdot \SumFatnessSquares^{4}/\epsilon^{2})$
preprocessing time, a data structure \DataStructure{} of size
\( O \left(n^{2} \cdot \SumFatnessSquares^{4}/\epsilon^{2}\right) \) that
imposes a $(2 n + 1)$-wise partition
${\bar\ReceptionZone} = \left\langle
\ReceptionZone_{1}^{+}, \ldots, \ReceptionZone_{n}^{+},
\ReceptionZone_{1}^{?}, \ldots, \ReceptionZone_{n}^{?},
\ReceptionZone^{-} \right\rangle$
of the Euclidean plane $\Reals^{2}$
(that is, the zones in $\bar{\ReceptionZone}$ are pairwise disjoint and
$\Reals^{2} = \bigcup_{i = 1}^{n} \ReceptionZone_{i}^{+} \cup
\ReceptionZone^{-} \cup \bigcup_{i = 1}^{n} \ReceptionZone_{i}^{?}$),
such that for every \( 1 \leq i \leq n \):
\\
(1)\( \ReceptionZone_{i}^{+} \subseteq \ReceptionZone_{i} \);
\\
(2) \( \ReceptionZone^{-} \cap \ReceptionZone_{i} = \emptyset \);
\\
(3)\(\ReceptionZone_{i}^{?}\) is bounded and its area is at most an
\(\epsilon\)-fraction of the area of \(\ReceptionZone_{i}\).
Furthermore, given a query point \( p \in \Reals^{2} \),
it is possible to extract from \DataStructure{}, in time
\( O \left (\log \left(n \cdot \MaxFatnessParameterBound /\epsilon \right) \right) \),
the zone in $\bar\ReceptionZone$
to which \(p\) belongs.
\end{theorem}
\commabsend

\commabs
\subsection{Different schemes for point location in general networks}
\label{section:General networks}
In this section, we assume the general setting where stations are embedded in $\R^{d}$ and therefore their reception zones $\ReceptionZone_{i}$ are not necessarily hyperbolic convex. Specifically, we cannot assume our zones to satisfy the property of Corollary \ref{corollary:hyperbolic_ClosedShape}. We devise several approaches for point location in this setting. The key difference underlying the different approaches is in the definition of the performance measure.
\subsubsection{Scheme A}
\label{section:SchemeA}
The first algorithm we present constructs a data structure \(\QuerDS_{}\) that partitions the Euclidean plane to 2 disjoint zones \(
\Reals^2 = \ReceptionZone_{1}^{+} \cup \ReceptionZone_{1}^{-}\). The accuracy of the grid $\GridSpace$ is set such that the fraction of area on which the algorithm might give a false result is bounded by $\epsilon$. We begin by presenting the tagging procedure ($\SturmCell$) invoked on each cell $c_{i} \in \Ball(\Station_1,\LargeRadiusBound_1)$.

\commabs
\begin{figure*}[ht]
\begin{center}
\framebox{\parbox{4in}{
{\boldmath Algorithm $\SturmCell$} ($c_{i},F_{\beta}(p))$)
\begin{enumerate}
\item
For any edge $e_{j}$ of $c_{i}$
\begin{itemize}
\item
$t_{j} \gets \SegTest(e_{j},F_{\beta}(p));$
\end{itemize}
\item
If $t_{j}=~-~$ for any $j \in \{1,...,4\}$ return $-$;
\item
Else return $+$;
\end{enumerate}
}}
\end{center}
\end{figure*}
\commabsend

\begin{figure}[htb]
\begin{center}
\includegraphics[scale=0.5]{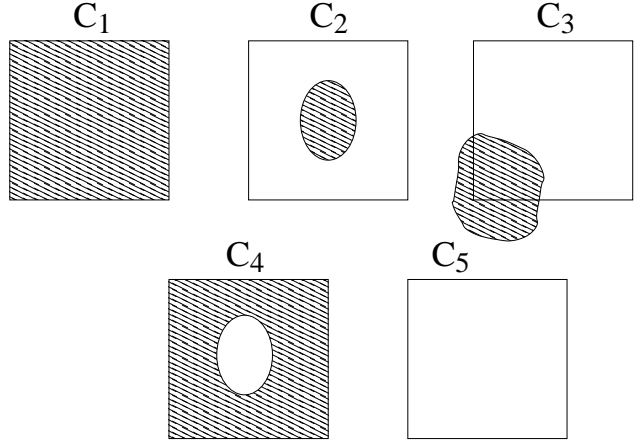}
\caption{ \label{figure:CellConfig}
\sf Cells Classification. Types of cells observed during \(\QuerDS_{}\) preparation.
Dashed area correspond to reception area of $\Station_1$. White area correspond to an area where no station is correctly received. Cells $C_{1}$ and $C_{5}$ are easy while others are hard.
}
\end{center}
\end{figure}

We say that the point location algorithm fails for a point query $q$ if it decides that $q$ hears $\Station_1$ where in fact it does not and vice-verse. We wish to bound the total area of points $q$ for which such errors might occur. We begin with some notation.
A cell in $c_{i}$ in $G_{\GridSpace}$ is referred to as \emph{easy} if either
$c_{i} \in \ReceptionZone_{1}^{+}$ or $c_{i} \cap \ReceptionZone_{1}=\emptyset$
(see cells $C_{1}, C_{5}$ in Figure \ref{figure:CellConfig}).
A \emph{non-easy} cell $c_{i}$ is referred to as \emph{hard}
(see cells $C_{2-4}$ in Figure \ref{figure:CellConfig}).
Note that a cell is hard if there exist points $p_{1},p_{2} \in c_{i}$ such that
$\SINR(\Station_{1},p_1) \geq \beta$ and $\SINR(\Station_1,p_{2}) < \beta$.
It is easy to see that easy cells are tagged correctly by Procedure $\SturmCell$ (which tags each cell based on its circumference). The algorithm might fail on a point query $q$ only if $q \in c_{i}$ where $c_{i}$ is hard. We now turn to bound the number of hard cells in $G_{\GridSpace}$. There are essentially three types of hard cells corresponding to the type of mistake the algorithm might make. Mistake type 1 (false-negative) occurs when $c_{i} \in \ReceptionZone_{1}^{-}$ but there exists $p \in c_{i}$ such that $\SINR(\Station_1,p) \geq \beta$. Let $M_{1}$ denote the number of cells for which the algorithm might make a mistake of Type 1. Since $c_{i} \subseteq \ReceptionZone_{1}^{-}$ only if $\Boundary (c_{i}) \cap \ReceptionZone_{1}= \emptyset$, it follows that the point $p$ corresponds to a connected zone of $\Station_1$ which is fully located in $c_{i}$
(see cell $C_{2}$ at Figure \ref{figure:CellConfig}).
By Corollary \ref{cor:nzones}, $\NZones_1 \leq c_{1} \cdot n^{2}$
hence $M_{1} \leq c_{1} \cdot n^{2}$, for a constant $c_1>0$.

We next bound the second type of mistake. Mistake type 2 (false-positive)
occurs when $c_{i} \in \ReceptionZone_{1}^{+}$ but there exists some point $p \in c_{i}$ such that $\SINR(\Station_1,p) < \beta$.
Within the class of type 2 mistakes there is a further division. Let $M_{2}^{1}$ be the number of cells $c_{i}$ such that $\Boundary (c_{i}) \cap \ReceptionZone_{1} =\Boundary (c_{i})$. Let $M_{2}^{2}$ be the number of cells $c_{i}$ such that $\Boundary (c_{i}) \cap \ReceptionZone_{1} \neq \Boundary (c_{i})$. We begin by bounding $M_{2}^{1}$. Since every point $q \in \Boundary (c_{i})$ is a reception point, the existence of non-reception point $p \in c_{i}$ corresponds to a connected cell of the zone $\ReceptionZone_{\emptyset}(\cA)$, where no station is received correctly, which located entirely in $c_{i}$
(see cell $C_{4}$ in Figure \ref{figure:CellConfig}).
By Corollary \ref{cor:nzones_noise}, $\NZones_{\emptyset}=O(n^{4})$, hence $M_{2}^{1}= c_{2} \cdot n^{4}$ for a constant $c_2>0$. Finally, it remains to bound the number of cells that intersect $\Boundary (\ReceptionZone_{1})$
(see cell $C_{3}$ in Figure \ref{figure:CellConfig}).
Note that these cells are exactly the $\ReceptionZone_{1}^{?}$ cells of Procedure $\SturmCellB$. The bound is then given by Eq. (\ref{eq:pointlocation_colinques}). Overall, we have that the number of hard cells for which the Algorithm $\SturmCell$ might fail is at most
\begin{eqnarray*}
M &=&M_{1}+M_{2}^{1}+M_{2}^{2} ~\leq~
c_1 \cdot n^2+ c_2 \cdot n^4+8\pi \cdot (n+1)\cdot \LargeRadiusBound_1/\GridSpace
\\ & \leq &
8\pi \cdot (c_{1}+c_{2}) \cdot n^{4} \cdot \LargeRadiusBound_1 / \GridSpace~,
\end{eqnarray*}
and their total area is at most
\begin{equation*}
\Area(M) ~\leq~ 8\pi \cdot (c_1+c_2) \cdot n^4 \cdot \LargeRadiusBound_1
\cdot \GridSpace~.
\end{equation*}
In order to guarantee that $\Area(M) \leq \epsilon \cdot \Area(\ReceptionZone_{1})$, we employ Inequality (\ref{eq:bounds_area}) and demand that $8\pi \cdot (c_{1}+c_{2}) \cdot n^{4} \cdot \LargeRadiusBound_1 \cdot \GridSpace \leq \epsilon \cdot \pi \SmallRadiusBound_1^{2}$.
Therefore it is sufficient to fix
\begin{eqnarray}
\label{eq:pointlocation_gamma_sa}
\GridSpace &=& \frac{\epsilon \SmallRadiusBound_1}{8 \cdot (c_{1}+c_{2})
\cdot n^{4} \cdot \FatnessParameter}
\end{eqnarray}

Let $\MaxFatnessParameterBound=\max_{i=1}^{n}\{\FatnessParameterBound_{i}\}$
and $\SumFatnessSquares^{4} = \sum_{i=1}^{n}\FatnessParameterBound_{i}^{4}$.
By combining Eq. (\ref{eq:pointlocation_generaltime}, \ref{eq:pointlocation_generalmemory},\ref{eq:pointlocation_generalquerytime}, and \ref{eq:pointlocation_gamma_sa}) we derive the following concluding theorem.

\begin{theorem} \label{theorem:pointlocation_sa}
It is possible to construct, in $O \left(n^{10} \cdot \SumFatnessSquares^{4}/\epsilon^{2}\right)$
preprocessing time,
a data structure \DataStructure{} of size $O \left(n^{8} \cdot \SumFatnessSquares^{4} /\epsilon^{2}\right)$ that
imposes a $(n + 1)$-wise partition
${\bar\ReceptionZone} = \left\langle
\ReceptionZone_{1}^{+}, \ldots, \ReceptionZone_{n}^{+},
\ReceptionZone^{-} \right\rangle$
of the Euclidean plane $\Reals^{2}$
(that is, the zones in $\bar{\ReceptionZone}$ are pair-wise disjoint and
$\Reals^{2} = \bigcup_{i = 1}^{n} \ReceptionZone_{i}^{+} \cup
\ReceptionZone^{-}$)
such that for every \( 1 \leq i \leq n \):
\begin{description}
\item{(1)}
\(\ReceptionZone_{i}^{+} \subseteq \ReceptionZone_{i} \);
\item{(2)} \( \ReceptionZone^{-} \cap \ReceptionZone_{i} = \emptyset \); and
\item{(3)} \(\ReceptionZone_{i}^{?}\) is bounded and its area is at most an
\(\epsilon\)-fraction of the area of \(\ReceptionZone_{i}\).
\end{description}
Furthermore, given a query point \( p \in \Reals^{2} \),
it is possible to extract from \DataStructure{}, in time \( O \left (\log \left(n \cdot \MaxFatnessParameterBound /\epsilon \right) \right) \),
the zone in $\bar\ReceptionZone$
to which \(p\) belongs.
\end{theorem}

\commabs
\commabsend

\subsubsection{Scheme B}
Using a different definition for the performance measure, we can use Procedure $\SturmCellB$ to devise a simpler (yet not as powerful) scheme. By applying Procedure $\SturmCellB$ to each cell $c_{i}$, we construct a data structure \(\QuerDS_{}\) that partitions the Euclidean plane into two disjoint zones \(
\Reals^2 = \ReceptionZone_{1}^{+} \cup \ReceptionZone_{1}^{-}\). The accuracy of the grid $\GridSpace$ is set to $\epsilon/\sqrt{2}$. We can provide the following guarantee (taking no advantage of the properties we established for \(\ReceptionZone_{1}\)). Let $p \in \R^{2}$ be a point query. Then if the algorithm (based on $\SturmCellB$ tagging) claims that $p \in \ReceptionZone_{1}$, then there exists $q \in \Ball(p,\epsilon)$ such that $\SINR(\Station_1,q) \geq \beta$. In addition, if algorithm claims that $p \notin \ReceptionZone_{1}$, then there exists some $q \in \Ball(p,\epsilon)$ such that $\SINR(\Station_1,q) < \beta$. Note that this follows simply by the fact that we impose $\Grid_{\epsilon/\sqrt{2}}$ on the Euclidean plane and evaluate the circumference of each grid cell.

\subsubsection{Scheme C}
\label{section:SchemeC}
\commabsend
Let $0 < \epsilon < 1$ be a predetermined performance parameter.
We construct in $O (\FatnessParameter^{2} \cdot n^{2}/\epsilon^{2})$ preprocessing time a data structure \(\QuerDS_{}\) of
size \( O (\FatnessParameter^{2}/\epsilon^{2}) \).
\(\QuerDS_{}\) essentially partitions the Euclidean plane into three disjoint zones \(
\Reals^2 = \ReceptionZone_{1}^{+} \cup \ReceptionZone_{1}^{-} \cup\ReceptionZone_{1}^{?} \), where
(1) \( \ReceptionZone_{1}^{+} \subseteq \ReceptionZone_{1} \);
(2) \( \ReceptionZone_{1}^{-} \cap \ReceptionZone_{1} = \emptyset \); and
(3) \( \ReceptionZone_{1}^{?} \subseteq \ReceptionZone_{1}(\betahat) \),
for $\betahat>(1-\epsilon)^{2\alpha}\cdot\beta$.
\commful 
A data structure \(\QuerDS_i\) is constructed for every station $\Station_i \in S$.
Recall that by Lemma \ref{cl:wv_sinr}(a),
a point \(p\) cannot be in \(\ReceptionZone_{i}\) unless it belongs to $ \wvor_{i}(\wvorsys_{\cA})$, where $\wvor_{i}(\wvorsys_{\cA})$ is the weighted Voronoi cell of $\Station_i$ with weight $\Vweight_i =\Power_i^{1/\alpha}$.
Thus for such a point \(p\) there is no need to query the data structure
\(\QuerDS_j\) for any \( j \neq i \).
Due to \cite{AurenhammerE84}, a weighted Voronoi diagram of quadratic size for the \(n\) stations is constructed in \( O(n ^{2}) \) preprocessing time. Then given a query point \( p \in \Reals^2
\), the station \(\Station_i\) such that $p \in \wvor(\Station_i)$ can be identified in time \( O (\log n)\). We then invoke the appropriate data structure \(\QuerDS_i\).
For ease of notation, let the characteristic polynomial of
$\ReceptionZone_{1}(\cA)$, namely, $F^{1}_{\cA}(p)$
(see Sect. \ref{section:Preliminaries})
be given by $F_{\beta}(p)$. In the same manner, the characteristic polynomial of $\ReceptionZone_{1}(\cA_{\beta'})$, for $\beta'\neq \beta$, is given by $F_{\beta'}(p)$. Let \(\QuerDS=\QuerDS_1\).
In addition, define $\LargeRadiusBound_1$ as the upper bound on
\(\LargeRadius(\Station_1, \ReceptionZone_{1})\), and \(\SmallRadiusBound_1\) as the lower bound on
\(\SmallRadius(\Station_1,\ReceptionZone_{1})\). \\
\QuerDS{} is based upon imposing a $\GridSpace \in \R_{>0}$-spaced \emph{grid}, denoted by
$\Grid_{\GridSpace}$, on the Euclidean plane, $\GridSpace$ is determined later on.
\commabs
The notions of grid \emph{columns}, \emph{rows}, \emph{vertices},
\emph{edges}, and \emph{cells} are defined in the natural manner.
We assume that $\Grid_{\GridSpace}$ is aligned so that the point $\Station_1$ is a grid
vertex.
\commabsend
The parameter $\GridSpace$ is set to be sufficiently small so that the cell
containing point $\Station_1$ is internal to the ball inscribed in $\ReceptionZone_{1}$, namely, $\Ball(\Station_1,\SmallRadiusBound_1)$. We now describe the construction of \QuerDS{}.
\\ {\bf Section 8.3.3 repeats many things said in section 8.1. In particular, procedure SturmCell appears twice, and so are the bounds ob complexity of QDS. Needs careful cleaning.}\\
The main ingredient of the algorithm is a \emph{segment testing} procedure
\cite{Avin2009PODC}, named hereafter Procedure $\SegTest$.
Given a segment $\sigma$, the segment testing procedure returns the number of distinct intersection points of $\sigma$ and $\Boundary (\ReceptionZone_{1}(\cA))$. The segment test is implemented to run in time $O(n^{2})$ by employing Sturm condition \cite{BPR03} of the projection of the polynomial $F_{\beta}(p)$ on $\sigma$ and by direct calculation of the SINR function in the endpoints of $\sigma$. In particular, the segment testing allows one to decide whether $\sigma \cap \ReceptionZone_{1}(\cA)= \emptyset$ or not
(see the code of Procedure $\SegTest$).
Given a grid $\Grid_{\GridSpace}$, Procedure $\SegTest$ is invoked for each of the 4 edges for every cell $c_{i} \in \Grid_{\GridSpace}$ at distance at most $\LargeRadiusBound_1$ from $\Station_{1}$.
\commabs
The overall number of invocations is thus bounded by $O\left (\Area(\Ball(\Station_1,\LargeRadiusBound_1))/\GridSpace^{2} \right)$.
\commabsend
We processed by describing the tagging procedure, given by Procedure $\TagCell$.
\commfulend  
Procedure $\TagCell$ tests $\Boundary (c_{i})$ for high and low $\beta$, namely, $(1+\epsilon)^{\alpha}\beta$ and $(1-\epsilon)^{\alpha} \cdot \beta$ respectively. If there exists at least one point $p_{\Boundary} \in \Boundary(c_{i})$ such that $\SINR(\Station_1,p_{\Boundary}) \geq (1+\epsilon)^{\alpha}\beta$ the entire cell is declared to be in $\ReceptionZone_{1}$. In addition, if $\SINR(\Station_1,p_{\Boundary}) < (1-\epsilon)^{\alpha}\cdot \beta$, for any point $p_{\Boundary} \in \Boundary (c_{i})$ then the cell is declared to be out of $\ReceptionZone_{1}$. Otherwise the cell $c_{i}$ is under question mark. Essentially, the question mark cells correspond to the case where $(1- \epsilon)^{2\alpha} \cdot \beta \leq \SINR(\Station_1,p_{in}) \leq (1+\epsilon)^{2\alpha}\cdot\beta$ for any $p_{in} \in c_{i}$.
For details see the code of Algorithm $\TagCell$ below.

\commabs
\begin{figure*}[ht]
\begin{center}
\framebox{\parbox{4in}{
{\boldmath Algorithm $\TagCell$} ($c_{i},F(p)$))\\
\begin{enumerate}
\item
Let $t_{1}= \SturmCell(c_{i},F_{(1+\epsilon)^{\alpha}\beta}(p));$
\item
If $t_{1} \neq ~- ~$ return $+$;
\item
Let $t_{2}=\SturmCell(c_{i},F_{(1-\epsilon)^{\alpha}\beta}(p));$
\item
If $t_{2}=~- ~$ return $-$;
\item
Else, return $?$;
\end{enumerate}
}}
\end{center}
\end{figure*}
\commabsend

Let $\GridSpace$ (grid resolution) be given by
\begin{eqnarray}
\label{eq:pointlocation_gamma_sb2}
\GridSpace &=& \frac{\epsilon \SmallRadiusBound_1 }{3 \cdot \sqrt{2}}
\end{eqnarray}
The rest of this section is dedicated for establishing the correctness of
Procedure $\TagCell$. The following lemma shows that the $\SINR$ ratio of
neighboring points within a grid cell $c_{i}$ is similar.
Let $\epsilonhat=\epsilon/3$.
\begin{lemma}
\label{lem:pointlocation_sinr_decay}
Let $\SINR(\Station_1,p)=\betahat$. Then
$\SINR(\Station_1,\widetilde{p}) \in
\left[\left(\frac{1-\epsilonhat}{1+\epsilonhat}\right)^{\alpha} \cdot \betahat,
\left(\frac{1+\epsilonhat}{1-\epsilonhat}\right)^{\alpha} \cdot \betahat \right]$
for any $\widetilde{p} \in \Ball(p,\sqrt{2} \GridSpace)$.
\end{lemma}
\commabs
\Proof
Let $\gamma'=\sqrt{2} \gamma$. Note that we are interesting in the points $p$
such that $p \notin \Ball(\Station_{i},\SmallRadiusBound_1)$
for any $\Station_{i} \in S$
(since for other points $p$, the location is determined easily).
It then follows that
\begin{eqnarray*}
\Energy(\Station_i,\widetilde{p}) &=&
\Power_i \cdot \dist{\Station_1,\widetilde{p}}^{-\alpha} ~\geq~
\Power_i \cdot (\dist{\Station_i,p}+ \GridSpace')^{-\alpha} ~=~
\Power_i \cdot (\dist{\Station_i,p}+ \epsilonhat \cdot
\SmallRadiusBound_1)^{-\alpha}
\\&\geq&
\Power_i\cdot((\epsilonhat+1)\cdot\dist{\Station_i,p})^{-\alpha} ~\geq~
\frac{1}{(\epsilonhat+1)^{\alpha}}\Energy(\Station_i,p)~,
\end{eqnarray*}
relying on the equality of $\GridSpace' =\epsilonhat \cdot \SmallRadiusBound_1$, which follows by Equation (\ref{eq:pointlocation_gamma_sb2}).
In the same manner,
\begin{eqnarray*}
\Energy(\Station_i,\widetilde{p}) &=&
\Power_i \cdot \dist{\Station_i,\widetilde{p}}^{-\alpha} ~\leq~
\Power_i \cdot (\dist{\Station_i,p}-\GridSpace')^{-\alpha} ~\leq~
\Power_i \cdot (\dist{\Station_i,p}-
\epsilonhat \SmallRadiusBound_1)^{-\alpha} \\&\leq&
\Power_i\cdot((1-\epsilonhat)\cdot\dist{\Station_i,p})^{-\alpha} ~\leq~
\frac{1}{(1-\epsilonhat)^{\alpha}}\Energy(\Station_i,p)~,
\end{eqnarray*}
obtained by using Equation (\ref{eq:pointlocation_gamma_sb2}) again.
Overall we get that
\begin{eqnarray*}
\SINR(\Station_{1},\widetilde{p}) &=&
\frac{\Energy(\Station_1,\widetilde{p})}
{\Interference(S \setminus \{\Station_1\},\widetilde{p})+N}
\\&\subseteq&
\left[\left(\frac{1-\epsilonhat}{1+\epsilonhat} \right)^{\alpha}
\cdot \frac{\Energy(\Station_1,p)}{\Interference(S \setminus \{\Station_1\},p)+
\Noise}~,~\left(\frac{1+\epsilonhat}{1-\epsilonhat}
\right)^{\alpha} \cdot \frac{\Energy(\Station_1,p)}
{\Interference(S \setminus \{\Station_1\},p)+\Noise} \right]
\\&\subseteq&
\left[\left(\frac{1-\epsilonhat}{1+\epsilonhat} \right)^{\alpha}
\cdot \SINR \left(\Station_{1},p \right)~,~\left(\frac{1+\epsilonhat}
{1-\epsilonhat} \right)^{\alpha} \cdot \SINR
\left(\Station_{1},p \right) \right]~,
\end{eqnarray*}
yielding our claim.
\QED
\commabsend

We are now turn to prove the correctness of Procedure $\TagCell$.
\begin{lemma}
\label{lem:correctness}
(a) If $\TagCell(c_{i},F_{\beta}(p))$ returns $+$, then $c_{i} \subseteq \ReceptionZone_{1}$. (b) If $\TagCell(c_{i},F_{\beta}(p))$ returns $-$, then $c_{i} \cap \ReceptionZone_{1}=\emptyset$. (c) Let $c_{i} \subseteq \Ball(\Station_1,\LargeRadiusBound_1)$ be such that $\TagCell(c_{i},F_{\beta}(p))$ returns $?$. Then $c_{i} \subseteq \ReceptionZone_{1}((1-\epsilon)^{2\alpha} \cdot \beta)$, or $\SINR(\Station_1,p) \in [(1-\epsilon)^{2\alpha}, (1+\epsilon)^{2\alpha}]$, for every $p \in c_{i}$;
\end{lemma}
\Proof
We begin with property (a). Let $c_{i}$ be such that $\TagCell(c_{i},F_{\beta}(p))$ returns $+$. That implies that there exists a point $p_{\Boundary} \in \Boundary (c_{i})$ such that $\SINR(\Station_1,p_{\Boundary}) \geq (1+\epsilon)^{\alpha}\cdot \beta$.
By Lemma \ref{lem:pointlocation_sinr_decay} it then follows that
\begin{eqnarray*}
\SINR(\Station_1,p_{in}) &\geq&
\left(\frac{1-\epsilonhat}{1+\epsilonhat}\right)^{\alpha}
(1+\epsilon)^{\alpha} \cdot \beta
\\&\geq&
\beta, ~\mbox{for every} ~p_{in} \in \Ball(p_{\Boundary},\sqrt{2} \GridSpace)~,
\end{eqnarray*}
where the last inequality follows from the choice of $\epsilonhat$.
In particular, this holds for any point $p_{in}$ in $c_{i}$,
and (a) is established. Let $c_{i}$ be such that $\TagCell(c_{i},F_{\beta}(p))=-$. That implies that $\SINR(\Station_1,p_{\Boundary}) <(1-\epsilon)^{\alpha} \cdot \beta, ~\mbox{for every} ~ p_{\Boundary} \in ~\Boundary (c_{i}).$
Assume, by the way of contradiction, that there exists some point
$p_{in} \in c_{i}$ such that
$\SINR(\Station_1,p_{in}) \geq \beta$.
Then by Lemma \ref{lem:pointlocation_sinr_decay} it must be the case that
\begin{eqnarray*}
\SINR(\Station_1,p) &\geq&
\left(\frac{1-\epsilonhat}{1+\epsilonhat}\right)^{\alpha} \cdot
\beta
\\&\geq&
(1-\epsilon)^{\alpha} \cdot \beta, ~\mbox{for every} ~ p \in
\Ball(p_{in},\sqrt{2} \GridSpace).
\end{eqnarray*}
Thus $\SturmCell(c_{i},F_{(1-\epsilon)^{\alpha} \cdot \beta}(p))$ returns $+$ and we end with contradiction which establishes (b).
Finally, it is left to prove (c).
As $\SturmCell(c_{i},F_{(1+\epsilon)^{\alpha} \cdot \beta}(p))$ does not return $+$,
$\SINR(\Station_1,p_{\Boundary}) < (1+\epsilon)^{\alpha}, ~\mbox{for every}~
p_{\Boundary} \in \Boundary (c_{i})$ and therefore
\begin{eqnarray*}
\SINR(\Station_1,p_{in}) &\leq& \left(1+\epsilon \right)^{\alpha} \cdot
\left(\frac{1+\epsilonhat}{1-\epsilonhat}\right)^{\alpha}
\\&\leq&
(1+\epsilon)^{2\alpha}\cdot \beta ~\mbox{for every}~p_{in} \in c_{i}.
\end{eqnarray*}
Next, as $\SturmCell(c_{i},F_{(1-\epsilon)^{\alpha} \cdot \beta}(p))$ does not return
$-$, there exists $p_{\Boundary} \in \Boundary (c_{i})$ such that
$\SINR(\Station_1,p_{\Boundary}) \geq (1-\epsilon)^{\alpha}$. Therefore
\begin{eqnarray*}
\SINR(\Station_1,p_{in}) &\geq&
\left (1-\epsilon\right)^{\alpha} \cdot \left(\frac{1-\epsilonhat}
{1+\epsilonhat}\right)^{\alpha}
\\&\geq&
(1-\epsilon)^{2\alpha} \cdot \beta ~\mbox{for every}~p_{in} \in c_{i}~,
\end{eqnarray*}
establishing the claim.
\QED

\commful
Next, we evaluate the memory and time costs of \(\QuerDS_{}\) construction.
The number of cells in $\Grid_{\GridSpace}$ be given by
\begin{eqnarray*}
C_{\GridSpace} &=& O \left( \left(\frac{\LargeRadiusBound_1}{\GridSpace}
\right)^{2} \right)
\end{eqnarray*}


For each cell we keep its tag (of constant size), therefore \(\QuerDS_{}\)
is of size
\begin{equation}
\label{eq:pointlocation_generalmemory}
M_{\cA}(\QuerDS_{}) ~=~ C_{\GridSpace}~.
\end{equation}

Note that it is sufficient to keep in \(\QuerDS_{}\) only cells in
$\ReceptionZone_{1}^{+} \cup \ReceptionZone_{1}^{?}$. Next, we bound the time
complexity, $T_{\cA}(\QuerDS_{})$. The dominating step is the invocations of
Procedure $\SegTest$. The cost of a single $\SegTest$ invocation is $O(n^{2})$
and since there are $O(C_{\GridSpace})$ invocations, the processing time
for \(\QuerDS_{}\) construction is given by
\begin{equation}
\label{eq:pointlocation_generaltime}
T_{\cA}(\QuerDS_{}) ~=~ O(C_{\GridSpace} \cdot n^{2})~.
\end{equation}

Finally, we analyze the cost for a single point location query.
This is bounded by
\begin{equation}
\label{eq:pointlocation_generalquerytime}
T^{query}_{\cA}(\QuerDS_{}) ~=~ O\left(\log C_{\GridSpace} \right)
\end{equation}
which corresponds to the time for finding the cell to which $p$ belongs.
The latter can be done by preforming binary search on $C_{\GridSpace}$ cells.
Recall that there is a prior step involving an access to the weighted Voronoi
diagram data structure. As mentioned, that step is bounded by $O(\log n)$,
which is dominated by $O(\log C_{\GridSpace}).$
\commfulend

Let $\MaxFatnessParameterBound=\max_{i=1}^{n}\{\FatnessParameterBound_{i}\}$
and $\SumFatnessSquares = \sum_{i=1}^{n}\FatnessParameterBound_{i}^{2}$.
\begin{theorem} \label{theorem:pointlocation_sc}
It is possible to construct, in
$O (n^{2}\cdot \SumFatnessSquares/\epsilon^{2})$ preprocessing time,
a data structure \DataStructure{} of size
\( O (\SumFatnessSquares/\epsilon^{2}) \) that imposes a $(2 n + 1)$-wise
partition
\\
${\bar\ReceptionZone} = \left\langle
\ReceptionZone_{1}^{+}, \ldots, \ReceptionZone_{n}^{+},
\ReceptionZone_{1}^{?}, \ldots, \ReceptionZone_{n}^{?},
\ReceptionZone^{-} \right\rangle$
of the Euclidean plane $\Reals^{2}$
(that is, the zones in $\bar{\ReceptionZone}$ are pair-wise disjoint and
$\Reals^{2} = \bigcup_{i = 1}^{n} \ReceptionZone_{i}^{+} \cup
\ReceptionZone^{-} \cup \bigcup_{i = 1}^{n} \ReceptionZone_{i}^{?}$)
such that for every \( 1 \leq i \leq n \):
\commabs
\begin{description}
\item{(1)}
\( \ReceptionZone_{i}^{+} \subseteq \ReceptionZone_{i} \);
\item{(2)} \( \ReceptionZone^{-} \cap \ReceptionZone_{i} = \emptyset \); and
\item{(3)} \(\ReceptionZone_{i}^{?} \subseteq \ReceptionZone_{i}((1-\epsilon)^{2\alpha} \cdot \beta)\)
\end{description}
\commabsend
\commful
(1) \( \ReceptionZone_{i}^{+} \subseteq \ReceptionZone_{i} \); (2) \( \ReceptionZone^{-} \cap \ReceptionZone_{i} = \emptyset \); and (3) \(\ReceptionZone_{i}^{?} \subseteq \ReceptionZone_{i}((1-\epsilon)^{2\alpha} \cdot \beta)\).
\commfulend
Furthermore, given a query point \( p \in \Reals^{2} \),
it is possible to extract from \DataStructure{}, in time \( O \left (\log \left(\MaxFatnessParameterBound /\epsilon \right) \right) \),
the zone in $\bar\ReceptionZone$ to which \(p\) belongs.
\end{theorem}

\def\APPENDW{
%
%
Due to Lemma \ref{lem:correctness}, it remains to evaluate the memory and time costs of \(\QuerDS_{}\) construction. The number of cells in $\Grid_{\GridSpace}$ be given by
\begin{eqnarray*}
C_{\GridSpace} &=&
O\left( \left(\frac{\LargeRadiusBound_1}{\GridSpace} \right)^{2} \right)
\end{eqnarray*}
For each cell we keep its tag (of constant size), therefore \(\QuerDS_{}\)
is of size
{\bf third appearance}
\begin{equation}
\label{eq:pointlocation_generalmemory}
M_{\cA}(\QuerDS_{}) ~=~ O\left(C_{\GridSpace}\right)~.
\end{equation}
Note that it is sufficient to keep in \(\QuerDS_{}\) only cells in $\ReceptionZone_{1}^{+} \cup \ReceptionZone_{1}^{?}$. Next, we bound the time complexity, $T_{\cA}(\QuerDS_{})$. The dominating step is the invocations of Procedure $\SegTest$. The cost of a single $\SegTest$ invocation is $O(n^{2})$ and since there are $O(C_{\GridSpace})$ invocations, the processing time for \(\QuerDS_{}\) construction is given by
\begin{equation}
\label{eq:pointlocation_generaltime}
T_{\cA}(\QuerDS_{}) ~=~O\left(n^{2} \cdot C_{\GridSpace}\right)~.
\end{equation}
Finally, we analyze the cost for a single point location query. This is bounded by
\begin{equation}
\label{eq:pointlocation_generalquerytime}
T^{query}_{\cA}(\QuerDS_{}) ~=~ O \left(\log C_{\GridSpace} \right)
\end{equation}
which corresponds to the time for finding the cell to which $p$ belongs. The latter can be done by preforming binary search on $C_{\GridSpace}$ cells. Recall that there is a prior step involving an access to the weighted Voronoi diagram data structure. As mentioned, that step is bounded by $O(\log n)$, which is dominated by $O(\log C_{\GridSpace}).$
By combining Eq. (\ref{eq:pointlocation_gamma_sb2} \ref{eq:pointlocation_generaltime}, \ref{eq:pointlocation_generalmemory} and \ref{eq:pointlocation_generalquerytime}),
the theorem follows.
\QED
}
\commful
We note that the hyperbolic convexity property of zones in $\R^{d+1}$ can also be utilized to devise a different scheme for point location (following \cite{Avin2009PODC}). In this scheme the total area of question marked cells is bounded by $\epsilon \cdot \Area(\ReceptionZone_{1})$. Section \ref{section:Colinear} describes such a scheme for $d=1$.
\commfulend

\clearpage
\def\thepage{}
{\small
\bibliographystyle{plain}
\bibliography{SINR_v7}
}

\end{document}